\documentclass[acmsmall]{acmart}
\AtBeginDocument{%
  }

\setcopyright{acmlicensed}
\copyrightyear{2026}
\acmYear{2026}
\acmDOI{XXXXXXX.XXXXXXX}

\acmJournal{JACM}
\acmVolume{00}
\acmNumber{0}
\acmArticle{111}
\acmMonth{1}

\usepackage{booktabs}
\usepackage{url}
\usepackage{amsmath}
\usepackage{geometry}
\usepackage{graphicx}
\usepackage{caption}
\usepackage{pgfplots}
\usepackage{pifont}
\pgfplotsset{compat=1.18}
\usepackage{multirow}
\usepackage{soul}
\usepackage{booktabs}
\usepackage{tabularray}
\usepackage{graphicx} 
\usepackage{pgfplotstable}
\usepackage{xcolor}
\usepackage{subcaption}
\usepackage[toc,page]{appendix}
\usepackage{makecell}
\usepackage{listings}
\usepackage{xcolor}
\usepackage{hyperref}
\usepackage[most]{tcolorbox}

\lstdefinestyle{promptstyle}{
  basicstyle=\small\ttfamily,
  breaklines=true,
  breakatwhitespace=false,
  frame=single,
  framesep=6pt,
  rulecolor=\color{gray!55},
  backgroundcolor=\color{gray!8},
  columns=flexible,
  keepspaces=true,
  showstringspaces=false,
  xleftmargin=2pt,
  xrightmargin=2pt,
  aboveskip=8pt,
  belowskip=8pt
}

\newcommand{\Hnull}[2]{H^{\text{RQ}#1}_{#2,0}}
\newcommand{\Halt}[2]{H^{\text{RQ}#1}_{#2,A}}

\newcommand{\ignore}[1]{}
\usepackage[most]{tcolorbox}
\tcbset{
  keyfindings/.style={
    colback=blue!5!white,       
    colframe=blue!75!black,     
    fonttitle=\bfseries,        
    coltitle=white,             
    colbacktitle=blue!75!black, 
    boxrule=0.8pt,
    arc=3mm,
    left=5pt,
    right=5pt,
    top=5pt,
    bottom=5pt
  }
}

\usepackage[ruled,vlined]{algorithm2e}

\begin{document}

\sloppy

\title{BT-APE: A Computationally Light Backtracking Approach to Automatic Prompt Engineering for Requirements Classification}

\author{Mohammad Amin Zadenoori}
\email{amin.zadenoori@unipd.it}
\affiliation{%
  \institution{Department of Statistics, University of Padova}
  \city{Padova}
  \state{}
  \country{Italy}}

\author{Waad Alhoshan}
\email{wmaboud@imamu.edu.sa}
\affiliation{%
  \institution{Imam Mohammad Ibn Saud Islamic University (IMSIU)}
  \city{Riyadh}
  \state{}
  \country{Saudi Arabia}}

\author{Jacek D\k{a}browski}
\email{jacek.dabrowski@lero.ie}
\orcid{0000-0003-3392-0690}
\affiliation{%
  \institution{Lero, the Research Ireland Centre for Software, University of Limerick}
  \city{Limerick}
  \country{Ireland}
}
  
\author{Liping Zhao}
\email{liping.zhao@manchester.ac.uk}
\affiliation{%
  \institution{University of Manchester}
  \city{Manchester}
  \state{}
  \country{United Kingdom}}

\author{Alessio Ferrari}
\email{alessio.ferrari@ucd.ie}
\affiliation{%
  \institution{University College Dublin (UCD)}
  \city{Dublin}
  \state{}
  \country{Ireland}}
  \affiliation{%
  \institution{Istituto di Scienza e Tecnologie dell'Informazione ``A. Faedo'' (ISTI), Consiglio Nazionale delle Ricerche (CNR)}
  \city{Pisa}
  \state{}
  \country{Italy}}

\renewcommand{\shortauthors}{Zadenoori et al.}

\begin{abstract}
Large language models (LLMs) are increasingly applied to requirements
engineering (RE) tasks, including requirements classification, model generation, trace-link detection and others. Prompts,
which guide LLM behavior, are typically designed manually through trial
and error, often leading to inconsistent and suboptimal performance on
RE tasks. Despite the importance of prompt design, prior RE research
largely relies on manually constructed prompts and does not
systematically optimize them; moreover, automated methods for prompt
construction remain largely unexplored, leaving their effectiveness
unclear. To address this gap, we propose a lightweight Automatic Prompt
Engineering (APE) approach named Backtracking APE (BT-APE) and apply it to requirements classification as a
representative RE task. We frame prompt design as an optimization
problem and iteratively refine prompts using LLM-generated candidates,
backtracking search, and dynamic example selection. 
We evaluate BT-APE on
three benchmark datasets with five instruction-tuned LLMs against four
classical prompting baselines (zero-shot, few-shot, chain-of-thought,
and CoT+few-shot) and a state-of-the-art, yet more resource intensive, APE
baseline (PE2). Our results show that BT-APE and PE2 achieve nearly
identical performance, both substantially outperforming the four
classical prompting baselines across datasets and models, with large
effect sizes. However, compared with PE2, BT-APE imposes a substantially lighter
computational footprint, consuming approximately 72\% fewer cumulative
input tokens and 66\% less wall-clock time at equivalent accuracy
(see Appendix~\ref{sec:efficiency}), making it better suited to
deployment on small or resource-constrained servers. We also find that domain-informed prompt definitions
enhance early performance, while iterative optimization partly
compensates for weaker initial prompts. The contribution of this work
is threefold: (i) a lightweight APE framework, together
with an open interactive tool and replication package that
operationalize the full pipeline; (ii) a comprehensive empirical
evaluation across datasets and instruction-tuned LLMs that provides the
first systematic comparison of APE against
classical prompting for requirements classification; and (iii) insights
into the impact of class definitions and prompt evolution on
classification performance.
\end{abstract}

\ccsdesc[500]{Computing methodologies~Natural language generation}
\ccsdesc[300]{Software and its engineering~Software organization and properties}

\keywords{Large Language Models, LLMs, Natural Language Processing, NLP, Prompts Engineering, Automatic Prompting,  Requirements Engineering, Requirements Classification}

\received{20 February 2007}
\received[revised]{12 March 2009}
\received[accepted]{5 June 2009}
 
\maketitle

\section{Introduction}
\label{lab:introduction}

Recent advances in Large Language Models (LLMs) are transforming Requirements Engineering (RE)~\cite{hou2024large, zadenoori2025largelanguagemodelsllms}. These models support tasks such as requirements elicitation, traceability, validation, and specification generation~\cite{11190385, DBLP:conf/refsq/MallyaFZD26}. Their effectiveness stems from large-scale pretraining and their ability to follow natural language instructions through in-context learning, without requiring task-specific training~\cite{zadenoori2025largelanguagemodelsllms}. This capability is well suited to RE, where annotated data is often scarce, domain-specific, and costly to obtain, making traditional supervised approaches difficult to scale~\cite{zhao2021natural}.

Interaction with LLMs is mediated through prompts, which are natural language instructions that guide model behavior and shape outputs~\cite{Arora2024}. Prompt quality directly affects performance, making prompt engineering a critical factor in LLM-based solutions~\cite{vogelsang2024usinglargelanguagemodels}. However, in RE research, prompt design is typically manual and relies on iterative trial-and-error~\cite{zadenoori2025largelanguagemodelsllms}. This process is often ad hoc and unsystematic, depends heavily on practitioner intuition, and is rarely documented in a reproducible manner~\cite{vogelsang2024usinglargelanguagemodels}. As a result, it limits reproducibility and leads to inconsistent or suboptimal results across RE tasks and datasets~\cite{zadenoori2025largelanguagemodelsllms}.

Despite its importance, prompt engineering in RE remains
underexplored~\cite{cheng2024generative,huang2025pe4re}. Existing
studies rely on static prompting strategies—zero-shot, few-shot, and
chain-of-thought—whose effectiveness is known to be highly sensitive
to phrasing, ordering, and example selection~\cite{Arora2024}.
Because these prompts are manually crafted and never updated in
response to model behavior, they underutilized LLM capabilities. In contrast, recent work in the broader Natural Language Processing (NLP) literature introduces Automatic Prompt Engineering (APE) techniques that iteratively generate, evaluate, and refine prompts~\cite{kepel2024autonomouspromptengineeringlarge, wang2024promptenoughautomatedconstruction, ye2024promptengineeringpromptengineer, kwon2024stableprompt}. These approaches frame prompt design as a search or optimization problem and have demonstrated strong performance improvements across a range of tasks~\cite{pryzant2023automaticpromptoptimizationgradient}. However, their applicability to domain-specific RE problems, where terminology, structure, and ambiguity differ from general NLP benchmarks, remains unclear.

Our prior short contribution  introduced APE to RE and provided initial evidence that automated refinement can outperform standard prompting for requirements classification~\cite{e86928c70f2a4ff5b595da7f430042e0}. The preliminary study proposed a basic APE approach, focused on a single dataset and a single model, and did not analyze the effects of datasets, LLMs, or prompt design factors. This paper builds upon this study, by proposing a novel, more refined---yet lightweight---APE framework, named Backtracking-APE (BT-APE), and conducting a systematic empirical evaluation across multiple datasets and LLMs. 

The proposed BT-APE framework iteratively refines prompts using LLM-generated candidates, a backtracking search strategy, and dynamic example selection. We evaluate the effectiveness of the framework in RE and identify the factors that influence its performance. We focus on requirements classification as a representative RE task~\cite{Bourque2024SWEBOK}. The task assigns labels to textual requirements, such as functional or non-functional categories. It supports downstream activities, including requirements analysis, prioritization, and quality assurance, it is the most frequently addressed task in NLP for RE research~\cite{zhao2021natural}, and it has been shown to be highly relevant for industrial applications~\cite{bashir2023requirements,abdeen2025language}. The task also provides a controlled setting to assess the impact of prompt design on model behavior, given the availability of multiple benchmarks.
We evaluate the approach on three benchmark datasets PROMISE~\cite{cleland2007automated},
PROMISE-Refined~\cite{dalpiaz2019requirements}, and
SecReq~\cite{knauss2011supporting} using five
instruction-tuned LLMs in the 7--8B parameter range: Qwen2-7B,
Falcon3-7B, Granite-3.2-8B, Ministral-8B, and LLaMA-3-8B. We
deliberately focus on instruction-tuned variants because BT-APE
relies on the model's ability to follow natural-language meta-prompts
that ask it to revise an existing prompt conditioned on labelled
feedback. We
restrict the comparison to the 7--8B parameter range so that all
models can be hosted locally on a single commodity GPU, reflecting
the practical RE setting in which requirements confidentiality
discourages reliance on hosted proprietary APIs.
Furthermore, we analyse how prompt components, such as class
definitions and examples, evolve during optimisation.

Our results show that BT-APE outperforms standard prompting strategies, and reaches performance that are equivalent with another, more resource consuming, state-of-the-art baseline method, i.e., PE2~\cite{ye2024promptengineeringpromptengineer}, also generalizing across datasets and models. We also show that domain-informed, carefully  engineered prompts improve early performance, while iterative optimization compensates for weaker initial prompts, confirming the effectiveness of BT-APE especially in scenarios where expert-crafted prompts are unavailable or difficult to design. We also observe that effective prompts share a recognizable profile:
they tend to be concise and action-oriented (higher verb count),
structurally clear (more punctuation markers), and lexically focused,
whereas length, lexical diversity, and syntactic complexity correlate
negatively with performance.

This paper makes three contributions:
\begin{enumerate}
    \item The first systematic evaluation of APE on requirements classification, comparing two single-trajectory APE methods against four classical prompting strategies across three datasets and five LLMs. We show that APE transfers to RE with large effect sizes over classical prompting.
    \item Evidence that two structurally different APE designs~--- PE2's history-conditioned proposals and our BT-APE's bounded-backtracking with balanced example batches~--- converge to statistically indistinguishable accuracy. We characterize the operational trade-offs (per-iteration context size, search behavior, hyperparameter interpretability) that differentiate them.
    \item An analysis of prompt-feature evolution under BT-APE that identifies lexical, syntactic, and semantic correlates of high-performing prompts.

\end{enumerate}
\noindent In addition, we release a GUI-based interactive tool that
operationalises the full BT-APE pipeline, enabling practitioners to apply
and experiment with the method on their own requirements classification
tasks without prior expertise in prompt engineering or LLM tooling
(described in Appendix~\ref{sec:tool}). The tool, datasets, prompts,
optimisation traces, and evaluation scripts are released as a publicly
available replication package on Zenodo~\cite{amin_zadenoori_2026_20438927}.

To support the reproducibility and extensibility of this study, additional materials are provided in the appendices. Appendix~\ref{app:sens} presents a hyperparameter sensitivity analysis and convergence dynamics for the backtracking mechanism, characterizing the influence of patience thresholds and iteration horizons. Appendix~\ref{sec:pe2-baseline} details the PE2 baseline implementation, highlighting its structural differences from our BT-APE method. Appendix~\ref{sec:efficiency} provides a comprehensive computational efficiency and token overhead comparison between BT-APE and PE2. Appendix~\ref{sec:tool} describes an interactive tool that operationalizes the full optimization pipeline. Appendix~\ref{sec:dataleakage} outlines a preliminary analysis of potential
data leakage using Jaccard similarity, showing that mean similarity
between LLM-generated requirement continuations and ground-truth
texts remains low ($0.10$--$0.16$) across all five models, with no
statistically significant difference between PROMISE and SecReq.

\paragraph{Relation to the preliminary version.}
This article substantially extends our REFSQ 2025 research preview~\cite{e86928c70f2a4ff5b595da7f430042e0},
which introduced APE to requirements classification as a single-dataset, single-model
proof of concept without statistical analysis. Beyond the contributions listed above,
the present work (i) refines the original procedure into BT-APE by adding bounded
backtracking with an explicit patience threshold, a balanced four-tuple dynamic
example selection mechanism, and 3-run majority-voted evaluation; (ii) broadens the
empirical scope from one dataset and one LLM to three datasets and five
instruction-tuned LLMs, and adds PE2~\cite{ye2024promptengineeringpromptengineer} as a state-of-the-art APE
baseline; (iii) replaces descriptive evaluation with a full inferential design
(Wilcoxon signed-rank tests, Friedman tests with power analysis, and a linear
mixed-effects model with trajectory-level random intercepts); (iv) introduces two
research questions absent from the preview --- the analysis of prompt-feature
evolution (RQ3) and the comparison of informed vs.\ uninformed class-definition
initialisation (RQ4); and (v) releases an interactive GUI-based tool, a hyperparameter
sensitivity analysis, a computational-efficiency comparison with PE2, and a
data-leakage probe based on Jaccard similarity.

The remainder of this paper is organized as follows. Section~\ref{sec:background} introduces background and terminology. Section~\ref{sec:method} presents the proposed approach. Section~\ref{sec:researchdesign} describes the research design. Section~\ref{sec:results} reports the results. Section~\ref{sec:discussion} discusses the findings. Section~\ref{sec:threats} outlines threats to validity. Section~\ref{sec:conclusion} concludes the paper.

\section{Background and Related Works}
\label{sec:background}

\subsection{Traditional and Pre-trained Machine Learning for Requirements Classification}
 
  Requirements classification, assigning predefined labels to natural language 
requirements, is a foundational RE task. Typical formulations distinguish
functional from non-functional requirements, sort non-functional requirements
into sub-types such as performance, security, and usability, and flag
security-relevant
requirements~\cite{cleland2007automated,kurtanovic2017automatically,knauss2011supporting}.
Automating this work matters because manual classification of large requirements
documents is both impractical and error-prone.
 
Early approaches leaned on traditional NLP and hand-crafted
features---term frequency, bag-of-words, and syntactic parsing---paired with
classifiers such as support vector machines and Na\"ive
Bayes~\cite{cleland2007automated,kurtanovic2017automatically}. These pipelines
demanded heavy preprocessing and domain-specific feature design, which limited
their portability across datasets and
domains~\cite{zhao2021natural,ferrari2017natural}. 

The arrival of transfer learning with pre-trained language models shifted the picture:
Hey~\textit{et al.}~\cite{hey2020norbert} proposed NoRBERT, applying BERT-based
transfer learning to achieve strong results on the PROMISE dataset across binary
and multi-class tasks; Dalpiaz~\textit{et al.}~\cite{dalpiaz2019requirements}
contributed refined PROMISE annotations and explored interpretable models using
dependency-parsing features; and Alhoshan~\textit{et
al.}~\cite{alhoshan2023zero} showed that pre-trained models can classify
requirements with no task-specific training data, albeit with a measurable gap
relative to supervised baselines. Addressing the natural variance and instability 
inherent to individual language architectures, Alsanoosy~\cite{ALSANOOSY20253648} 
unified these classic deep learning threads by implementing an ensemble learning 
framework across seven distinct models (including BERT and RoBERTa).

\subsection{LLMs and In-Context Learning}
 
Modern LLMs are transformer-based networks trained on large text corpora through
self-supervised objectives, and models in the GPT family represent the current
state of the art across classification, summarization, translation, and code
generation, frequently without task-specific fine-tuning~\cite{achiam2023gpt}.
The property that most distinguishes these models from earlier neural approaches
is \emph{in-context learning}: given a natural language prompt containing
instructions and, optionally, a handful of demonstrations, the model produces
appropriate outputs without any parameter updates. This makes LLMs especially
appealing for RE, where labeled datasets tend to be small and tightly bound to a
particular domain~\cite{ferrari2024model}. Because state-of-the-art models
contain hundreds of billions of parameters and are typically reached through
hosted inference APIs rather than fine-tuned locally, the practical lever
available to most RE researchers is no longer the model's weights but the text
of the prompt itself~\cite{huang2025pe4re}. Within RE, LLMs have been applied
well beyond classification---to elicitation, validation, traceability, and
specification generation~\cite{rodriguez2023prompts,ferrari2024model,santos2024}.
Ferrari~\textit{et al.}~\cite{ferrari2024model} explored generating UML sequence
diagrams from requirements, while Rodriguez~\textit{et
al.}~\cite{rodriguez2023prompts} studied how prompt design shapes automated
software traceability. Recently, SLMs are increasingly preferred over LLMs due to their efficiency and competitive performance on specialized tasks~\cite{DBLP:conf/refsq/MallyaFZD26, mallya2026rita}. Studies have shown that SLMs can match or even outperform LLMs in domain-specific applications such as requirements classification, while also significantly reducing computational and energy costs~\cite{zadenoori2025does, demartino2025green}.

\subsection{LLMs in Requirements Classification}
Recent work on requirements classification has moved from traditional
machine learning and encoder-based baselines toward LLM-based approaches.
Santos et al.~\cite{santos2024} report that few-shot and zero-shot prompting
with GPT-4 and open-source models can approach the performance of fine-tuned
transformers in some settings, indicating that both model choice and prompt
design influence results. Peer et al.~\cite{peer2024nlp4ref} propose
\textit{NLP4ReF}, which combines NLTK with ChatGPT and shows that classical
toolkits remain effective for basic classification, while generative LLMs are
useful for identifying missing or overlooked requirements during design reviews.

Several studies address limitations of LLMs in this setting. To mitigate data
sparsity and class imbalance in non-functional requirement (NFR) datasets,
Qin and Peng~\cite{11396617} propose \textit{ChatNRC}, a generative-discriminative
framework that uses LLMs to synthesize application-specific NFRs.
Shafikuzzaman et al.~\cite{11153850} compare prompting strategies and find that
zero-shot configurations provide a label-free baseline, while few-shot prompting
yields higher precision in multi-class NFR classification. To address the
computational cost and limited interpretability of large models,
Rejithkumar and Anish~\cite{11121724} introduce \textit{NICE}, which distills
reasoning chains from GPT-4o into smaller models such as T5 to support
multi-label classification with natural language rationales validated by humans.

\subsection{Prompt Engineering}
 
Prompt engineering refers to the practice of designing and refining the natural
language instructions supplied to an LLM in order to elicit a desired output.
Because performance is highly sensitive to wording, structure, and the presence
or absence of examples~\cite{rodriguez2023prompts}, prompting has come to be
treated as a tunable interface for steering model behavior without
retraining---sometimes characterized as a new programming
paradigm~\cite{huang2025pe4re}. The field organizes its techniques along two
axes useful for our purposes. The first is the degree of supervision:
\textit{zero-shot} prompting supplies instructions alone, \textit{few-shot}
prompting adds a small set of input--output demonstrations, and
\textit{chain-of-thought} (CoT) prompting elicits intermediate reasoning before
a final answer~\cite{wei2022chainofthought}. In RE classification settings,
few-shot prompting has generally produced the strongest
results~\cite{rodriguez2023prompts,alhoshan2023zero}.
 
The second axis is the degree of automation. Early prompt engineering was a
wholly manual, trial-and-error activity, but a line of work on \textit{automatic
prompt optimization} has since sought to reduce that effort by searching for
effective prompts systematically. Zhou~\textit{et
al.}~\cite{zhou2023largelanguagemodelshumanlevel} proposed APE, which casts prompt generation as black-box optimization---an LLM
proposes candidate instructions, each is scored on a training set, and the best
is selected---showing that machine-generated prompts can rival or exceed
human-written ones. Building on this, Ye~\textit{et
al.}~\cite{ye2024promptengineeringpromptengineer} formalized prompt engineering
as an optimization problem and proposed PE2, which iteratively refines prompts
using LLM-generated feedback on misclassified examples, while Pryzant~\textit{et
al.}~\cite{pryzant2023automaticpromptoptimizationgradient} introduced ProTeGi,
which performs a gradient-descent-like search in text space using natural
language ``gradients.'' Other strands explore reinforcement-learning-based
optimization~\cite{kwon2024stableprompt}, mixture-of-expert prompt
construction~\cite{wang2024promptenoughautomatedconstruction}, and evolutionary
search~\cite{kepel2024autonomouspromptengineeringlarge}. Beyond supervision and
automation, prompting also encompasses techniques such as retrieval-augmented
generation~\cite{lewis2020rag}, constraint injection, and multi-role dialogue,
which broaden the surface available for adapting an LLM to a new
domain~\cite{huang2025pe4re}.

\subsection{Prompt Engineering for Requirements Engineering}
 
Taken together, these threads explain why prompting has become central to
LLM-based RE, but they also expose a gap: the RE community still lacks a
consolidated account of which prompting techniques map to which RE tasks, and
with what trade-offs. General prompt-engineering taxonomies exist but tend to
mix levels of abstraction---placing techniques, objectives, and application
domains at the same hierarchical level---which makes them awkward to apply
within RE~\cite{huang2025pe4re}. The recent systematic review by
Huang~\textit{et al.}~\cite{huang2025pe4re} addresses this directly: following
established secondary-study~\cite{kitchenham2022segress} and
mapping~\cite{petersen2015guidelines} guidelines, the authors screen several
hundred records down to 35 primary studies and propose a hybrid taxonomy that
links technique-oriented prompting patterns (e.g., few-shot, chain-of-thought,
knowledge augmentation, self-reflection) to task-oriented RE roles (elicitation,
validation, traceability, and others). Their analysis finds that
contextualization and step-wise reasoning are the most widely adopted
strategies---often used together---while multimodal prompting is essentially
absent, requirements elicitation is under-explored, and systematic evaluation
such as ablation studies is rare. They distill these gaps into a road-map toward
reproducible, practitioner-ready PE4RE workflows.

\subsection{Research Gaps and Contributions}
\label{subsec:gaps}

Based on the literature review, there are gaps at the intersection of requirements classification, LLMs, and prompt engineering:

\begin{itemize}
    \item \textbf{No APE for RE classification.}  While LLM-based approaches have shown promising results for requirements classification, existing studies rely exclusively on manually crafted prompting strategies---zero-shot, few-shot, or chain-of-thought---without systematically optimizing the prompt content itself~\cite{alhoshan2023zero,santos2024,rodriguez2023prompts}. Prompt design has been shown to significantly affect classification performance, yet the selection of prompts remains a largely ad-hoc process in the RE literature.

    \item \textbf{APE methods untested on domain-specific RE tasks.} APE methods such as APE~\cite{zhou2023largelanguagemodelshumanlevel}, PE2~\cite{ye2024promptengineeringpromptengineer}, EPiC~\cite{10.1145/3805704}, and ProTeGi~\cite{pryzant2023automaticpromptoptimizationgradient} have demonstrated strong results on general NLP benchmarks, but none have been applied or evaluated on domain-specific RE tasks. It remains unclear whether the gains observed on general benchmarks transfer to specialized classification problems such as requirements categorization.

    \item \textbf{Role of prompt phrasing unexplored.} No prior work has
investigated how the textual phrasing of category descriptions within
the prompt influences the optimisation process. Understanding whether
domain-grounded phrasings offer a lasting advantage over minimal ones,
or whether iterative optimisation can compensate for weaker starting
points, has both practical and theoretical implications.

    \item \textbf{Limited understanding of effective prompt characteristics.} Existing studies treat prompts as fixed inputs and evaluate only the classification output, without analyzing how prompt characteristics---such as length, lexical diversity, syntactic complexity, or semantic specificity---relate to performance. There is no systematic understanding of what makes a prompt effective or how prompt properties evolve during iterative optimization.
\end{itemize}

Our work addresses these gaps through the following contributions. We employ an APE approach for requirements classification that implements a backtracking-enhanced local search strategy to iteratively optimize prompts, going beyond the static prompt designs used in prior RE studies. We apply and evaluate this approach across multiple LLMs and three well-established requirements classification datasets, enabling a robust assessment of generalization. We introduce two initializations to isolate and quantify the effect of class definition quality on optimization performance. Finally, we conduct a feature-based analysis of prompt evolution across iterations, examining how lexical, syntactic, and semantic characteristics of prompts relate to classification performance. To the best of our knowledge, this is the first study to apply iterative APE adapted from the literature to requirements classification, and the first to systematically investigate the role of class definitions as a design variable in this particular process.
\section{Approach BT-APE: Backtracking-Enabled APE}
\label{sec:method}

\subsection{Method overview}
\label{sec:overview}

At a high level, BT-APE turns prompt design into a guided search. We
start from an initial prompt that describes the classification task,
and we let an LLM repeatedly propose revised versions of that prompt
in light of how the current prompt is performing on labelled examples.
The procedure proceeds in three phases.

\emph{Initialisation} sets up the search. We split the labelled data
into three disjoint pools: one for sampling demonstration examples,
one for scoring candidate prompts during the search, and one held out
for a single final evaluation. We then score the initial prompt on
the validation pool to establish a starting reference score against
which all subsequent candidates will be compared.

\emph{Iterative refinement} is the core of the search. At each
iteration, the LLM proposes a new candidate prompt conditioned on the
current prompt and on a small, balanced batch of examples drawn from
the example pool. The batch deliberately mixes successes and failures
from both classes so that the proposer simultaneously sees where the
current prompt works and where it breaks. Each candidate is scored on
the validation pool and added to a ranked list. If the candidate
improves on the best score seen so far, the search continues along
this trajectory; if several consecutive candidates fail to improve,
the search \emph{backtracks} to a previously promising prompt rather
than drifting further into an unproductive region of the prompt space.

\emph{Held-out evaluation} touches the third pool exactly once: the
best prompt found during the search is scored on it to produce the
final reported performance. Confining the search entirely to the
validation pool and reserving the test pool for a single terminal
evaluation prevents the selection pressure inherent in any search
procedure from inflating the reported results.

Three principles distinguish BT-APE from other single-trajectory APE
methods such as PE2~\cite{ye2024promptengineeringpromptengineer}:
(i)~\emph{bounded backtracking}, which abandons unproductive
trajectories only after a configurable patience threshold rather than
reactively at every non-improving step; (ii)~\emph{balanced dynamic
example selection}, which guarantees the proposer always sees both
classes and both success and failure signals; and (iii)~conditioning
the proposer on only the current prompt and a compact balanced batch,
keeping the per-iteration context size bounded. We elaborate on these
design choices and contrast them with PE2 in
Section~\ref{sec:bt-ape}.

\subsection{Formalisation}
\label{sec:formalisation}

More formally, and building on the formulation by
Ye~et~al.~\cite{ye2024promptengineeringpromptengineer}, prompt
engineering can be cast as an optimisation problem: identify the
optimal prompt $p^{*}$ that maximises a task-specific evaluation
metric over a dataset $D = (X, Y)$, using an LLM $\mathcal{M}$:
\begin{equation}
p^{*} = \arg\max_{p} \, f\bigl(\mathcal{M}(X, p),\, Y\bigr),
\end{equation}
where
\begin{itemize}
    \item $D = (X, Y)$ is the dataset of input--output pairs $(x, y)$;
    \item $\mathcal{M}$ takes a prompt $p$ and input $x \in X$ to
          produce an output $y' \in Y'$;
    \item $f$ is the evaluation function to be maximised (e.g.,
          precision, recall, or F1 score).
\end{itemize}

In the context of requirements classification, each $x$ is a textual
requirement and each $y$ its manually annotated class (e.g.,
functional or quality). The evaluation function $f$ is applied to the
entire output set $\mathcal{M}(X, p)$, reflecting aggregate metrics
such as F1 over the dataset.

\begin{figure}[t]
    \centering
    \includegraphics[width=\linewidth]{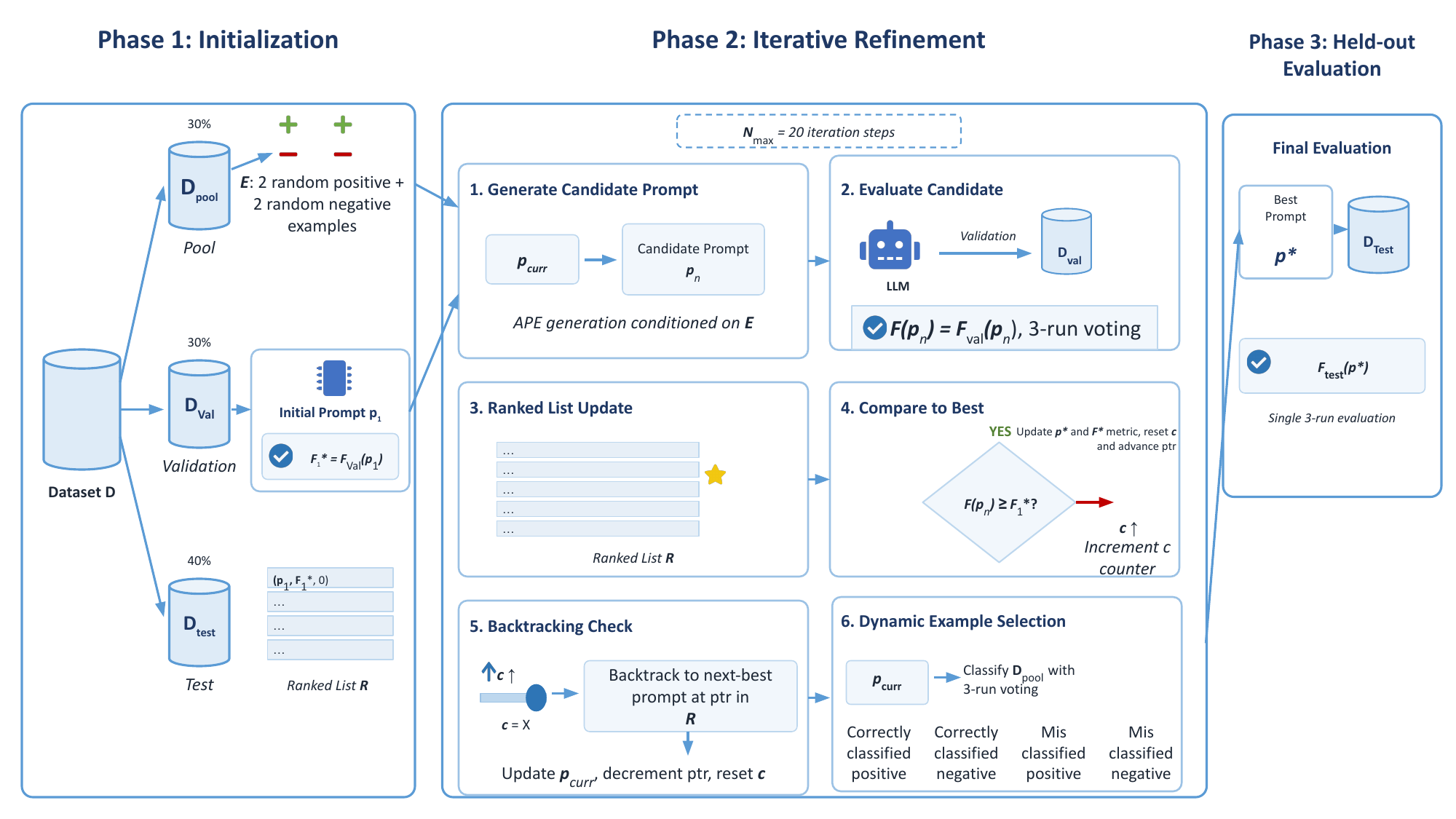}
    \caption{Overview of the BT-APE workflow applied to requirements
    classification, alongside Algorithm~\ref{alg:approach}. Steps are
    to be followed sequentially. The figure was generated with AI
    assistance and then reviewed, refined, and validated step by step
    by Author~1.}
    \label{fig:ape_workflow}
\end{figure}

\subsection{Algorithm and workflow}
\label{sec:algorithm}

Figure~\ref{fig:ape_workflow} and Algorithm~\ref{alg:approach} present
the same procedure as a diagram and as pseudocode, respectively. We
describe the three phases below, with line numbers referring to
Algorithm~\ref{alg:approach}.

\begin{itemize}
  \item \textbf{Phase~1: Initialisation (ll.~1--5).}
        The dataset $D$ is partitioned into $D_{\text{pool}}$ (30\%),
        $D_{\text{val}}$ (30\%), and $D_{\text{test}}$ (40\%); these
        proportions balance the competing demands of in-context
        learning, low-variance validation scoring, and a low-variance
        final test estimate (see Appendix~\ref{app:sens}).
        The example batch $E$ is seeded with two random positive and
        two random negative examples from $D_{\text{pool}}$. The
        initial prompt $p_1$ is scored on $D_{\text{val}}$ to obtain
        its weighted F1, which we denote $F^{*}$ and use throughout
        the search as the \emph{best-so-far} reference score against
        which subsequent candidates are compared. The triple
        $(p_1, F^{*}, 0)$ is then inserted into the ranked list
        $\mathcal{R}$.

  \item \textbf{Phase~2: Iterative refinement (ll.~6--22).}
        For up to $N_{\max} = 20$ iterations, the procedure performs
        six steps:
        \begin{enumerate}
          \item[\textit{2.1}] \emph{Generate candidate} (l.~8).
                A new prompt $p_n$ is generated by the LLM,
                conditioned on the current prompt $p_{\text{curr}}$
                and the balanced example batch $E$. Conditioning on
                $E$ exposes the proposer to concrete instances of
                where the current prompt succeeds and fails, following
                the error-conditioned refinement principle introduced
                by ProTeGi~\cite{pryzant2023automaticpromptoptimizationgradient}
                and PE2~\cite{ye2024promptengineeringpromptengineer}.

          \item[\textit{2.2}] \emph{Evaluate candidate} (l.~10).
                $p_n$ is scored on $D_{\text{val}}$ using 3-run
                majority voting (Section~\ref{sec:voting}). Voting
                reduces run-to-run variability at the cost of three
                inference passes per candidate; this matters because
                the backtracking trigger below depends on F1
                comparisons across iterations, and evaluation noise
                would otherwise propagate directly into search
                dynamics.

          \item[\textit{2.3}] \emph{Update ranked list} (l.~12).
                The triple $(p_n, F(p_n), n)$ is inserted into
                $\mathcal{R}$, kept sorted by F1. Maintaining the
                full ranked history (rather than only the current
                best) is what enables the bounded backtracking step
                below.

          \item[\textit{2.4}] \emph{Compare to best} (ll.~13--17).
                If $F(p_n) \geq F^{*}$, the best prompt and reference
                score are updated, the failure counter $c$ is reset,
                and the pointer $ptr$ advances. Otherwise, $c$ is
                incremented.

          \item[\textit{2.5}] \emph{Backtracking check} (ll.~18--21).
                If $c$ reaches the patience threshold $X = 3$, the
                search abandons the current trajectory and resumes
                from the next-best prompt in $\mathcal{R}$ via the
                pointer $ptr$. The threshold $X$ is an interpretable
                exploration--exploitation knob: $X = 1$ over-reacts
                to evaluation noise, while large values waste
                iterations on plateaus. We set $X = 3$ based on the
                sensitivity analysis in
                Appendix~\ref{app:sens}.

          \item[\textit{2.6}] \emph{Select examples for the next
                iteration} (ll.~22--23). $D_{\text{pool}}$ is
                re-classified with $p_{\text{curr}}$, and $E$ is
                refreshed with one correctly classified positive,
                one correctly classified negative, one misclassified
                positive, and one misclassified negative. This
                balanced four-tuple guarantees that the proposer
                simultaneously sees success and failure signals on
                both classes\,---\,a property that matters because the
                optimisation objective is F1 on a potentially
                imbalanced label distribution.
        \end{enumerate}

  \item \textbf{Phase~3: Held-out evaluation (l.~24).}
        The best prompt $p^{*}$ is scored once on $D_{\text{test}}$
        under the same 3-run majority-voting protocol. Because
        $D_{\text{test}}$ is consulted only at this final step, the
        reported F1 is a genuine held-out estimate rather than a
        selection-inflated one.
\end{itemize}

\begin{algorithm}[ht]
\caption{Backtracking-Enhanced Prompt engineering with Dynamic Example Selection}
\label{alg:approach}
\KwIn{Dataset $D$, LLM $\mathcal{M}$, initial prompt $p_1$, max iters $N_{\max}=20$, backtrack threshold $X=3$}
\KwOut{Best prompt $p^*$, held-out test F1}
\tcp{Phase 1: Initialization}
Split $D$ into $D_{\text{pool}}$ (30\%), $D_{\text{val}}$ (30\%), and $D_{\text{test}}$ (40\%);\\
$E \leftarrow$ 2 random positive + 2 random negative examples from $D_{\text{pool}}$;\\
$F^* \leftarrow F_{\text{val}}(p_1)$ \tcp*{score initial prompt on validation set}
$p^* \leftarrow p_1$, $p_{\text{curr}} \leftarrow p_1$, $c \leftarrow 0$, $ptr \leftarrow 1$;\\
Insert $(p_1, F^*, 0)$ into ranked list $\mathcal{R}$;\\
\tcp{Phase 2: Iterative Refinement (validation-driven)}
\For{$n=1$ \KwTo $N_{\max}$}{
  \tcp{Step 2.1: Generate Candidate}
  Generate candidate $p_n$ via APE from $p_{\text{curr}}$ conditioned on examples $E$;

  \tcp{Step 2.2: Evaluate Candidate on validation set}
  $F(p_n) \leftarrow F_{\text{val}}(p_n)$ \tcp*{3-run majority voting on $D_{\text{val}}$}

  \tcp{Step 2.3: Update Ranked List}
  Insert $(p_n, F(p_n), n)$ into $\mathcal{R}$, kept sorted by F1;

  \tcp{Step 2.4: Compare to Best}
  \uIf{$F(p_n) \geq F^*$}{
    $p^* \leftarrow p_n$, $F^* \leftarrow F(p_n)$;\\
    $p_{\text{curr}} \leftarrow p_n$, $c \leftarrow 0$, $ptr \leftarrow ptr + 1$;
  }
  \uElse{
    $c \leftarrow c + 1$;
  }

  \tcp{Step 2.5: Backtracking Check}
  \If{$c = X$}{
    $p_{\text{curr}} \leftarrow$ next-best prompt in $\mathcal{R}$ at position $ptr$;\\
    $ptr \leftarrow ptr - 1$, $c \leftarrow 0$;
  }

  \tcp{Step 2.6: Select Examples for Next Iteration}
  Classify $D_{\text{pool}}$ using $p_{\text{curr}}$ (3-run voting);\\
  $E \leftarrow$ 1 correctly classified positive, 1 correctly classified negative,
  1 misclassified positive, 1 misclassified negative from $D_{\text{pool}}$;
}
\tcp{Phase 3: Final held-out evaluation (test set touched once)}
\textbf{return} $p^*$ and $F_{\text{test}}(p^*)$ \tcp*{single 3-run evaluation of $p^*$ on $D_{\text{test}}$}
\end{algorithm}

\subsection{BT-APE: Method Design and Inspirations}
\label{sec:bt-ape}

We propose \textit{BT-APE} (Back-Tracking APE), a lightweight prompt
optimisation procedure designed for settings where labelled data is
scarce and each candidate evaluation is expensive. Given an initial
prompt $p_0$ and a small pool $D_{\text{pool}}$ of labelled examples,
BT-APE iteratively proposes new prompts conditioned on the model's
behaviour on a balanced batch of correct and incorrect predictions,
and maintains a ranked history $\mathcal{R}$ of all candidates
evaluated so far. The full procedure is given in
Algorithm~\ref{alg:approach} and summarised in Figure~\ref{fig:ape_workflow}. In the following, we describe the main elements of the approach, and then we describe the figure and the algorithm in details. The approach consists of three defining
components.

\paragraph{Explicit, bounded back-tracking.}
BT-APE maintains a ranked list $\mathcal{R}$ of all evaluated
candidates sorted by F1, together with a pointer $ptr$ that identifies
the prompt currently being refined. The search remains on the current
trajectory until $X = 3$ consecutive iterations fail to improve over
the best score $F^{*}$, at which point $ptr$ jumps to the next-best
prompt in $\mathcal{R}$. This yields an interpretable patience
parameter and prevents the search from lingering in unproductive
neighbourhoods of the prompt space.

\paragraph{Balanced, dynamic example selection.}
At every iteration, after classifying $D_{\text{pool}}$ with the
current prompt $p_{\text{curr}}$, BT-APE re-samples the example batch
$E$ to contain one correctly classified positive, one correctly
classified negative, one misclassified positive, and one misclassified
negative example. This guarantees that the proposer simultaneously
sees success and failure signals on both classes---a property that
matters because our optimisation objective is F1 on a potentially
imbalanced label distribution.

\paragraph{Stable evaluation via majority voting.}
Every candidate is evaluated on the held-out validation split using
3-run majority voting. Because the back-tracking trigger depends on
F1 comparisons across iterations, evaluation noise directly affects
search dynamics; majority voting reduces LLM stochasticity at the cost
of three times the inference budget per candidate. We use a fixed
30/30/40 pool/val/test split throughout the search so that F1 values
entered into $\mathcal{R}$ are directly comparable across iterations.

\subsubsection{Baseline and Positioning}
\label{sec:bt-ape-baseline}
To assess whether the simplicity of BT-APE comes at a cost, we compare
it against PE2~\cite{ye2024promptengineeringpromptengineer} as a
representative baseline. PE2 is a natural point of comparison: it
shares the same single-trajectory, history-aware backbone, and
formalises prompt optimisation as a sequence of meta-prompted
proposals
\begin{equation}
p^{(t+1)} = \mathcal{M}_{\text{proposal}}\bigl(p^{(t)}, B;\, p_{\text{meta}}\bigr),
\label{eq:pe2-proposal}
\end{equation}
over a batch $B = \{(x, y, y')\}$ of inputs, ground-truth labels, and
current predictions, with a history of prior prompts from which
back-tracking is possible by re-selecting the top-$n$ candidates from
$P^{(0)} \cup P^{(1)} \cup \ldots \cup P^{(t)}$ at every
step~\cite{ye2024promptengineeringpromptengineer}. We exclude
population-based evolutionary approaches such as
EPiC~\cite{10.1145/3805704} from this comparison: they are well suited
to code generation, where auto-generated unit tests validated via AST
parsing make the fitness oracle cheap and large populations can be
evaluated in parallel at low cost---a setting that does not match our
binary-classification task, where each candidate evaluation requires
three full classification runs over $D_{\text{val}}$.

BT-APE and PE2 differ along three axes that map onto the three
components above, and each axis has a concrete resource-consumption
consequence which we quantify empirically in Appendix~\ref{sec:efficiency}.

First, PE2 re-selects the top-$n$ prompts from the full history at
every step, whereas BT-APE commits to the current trajectory and
triggers an explicit jump only when a fixed patience threshold is
exceeded. Conditioning each proposal on the top-$n$ history means
that PE2's per-iteration input context grows roughly linearly with
the iteration count, while BT-APE's remains bounded. Empirically, at
$N_{\max} = 20$, PE2 consumes approximately $72\%$ more cumulative
input tokens than BT-APE across all 15 (dataset, LLM) configurations
(paired Wilcoxon, $p < 10^{-4}$, large effect size; see
Table~\ref{tab:efficiency_aggregate}).

Second, PE2 conditions the proposer on an unconstrained error batch;
BT-APE enforces a class- and outcome-balanced batch at every iteration.
The use of failure examples to drive prompt edits is also central to
ProTeGi~\cite{pryzant2023automaticpromptoptimizationgradient}, which
conditions a ``textual gradient'' prompt on model errors and then
applies a second prompt to edit $p_0$ in the opposite semantic
direction. BT-APE adopts this principle without the two-step
gradient-then-edit decomposition, since a single proposal call
conditioned on a balanced error batch is sufficient in our
binary-classification setting and avoids doubling the LLM call cost
per iteration.

Third, BT-APE relies on 3-run majority voting on a fixed validation
split, ensuring directly comparable F1 values across iterations---a
property that bandit-style subsampling, as used in ProTeGi's selection
step~\cite{pryzant2023automaticpromptoptimizationgradient}, would not
preserve.

Taken together, these three design choices translate into a
substantially lighter resource profile at equivalent accuracy: in
addition to the input-token reduction noted above, BT-APE incurs
approximately $60\%$ lower mean proposer latency and $66\%$ shorter
total wall-clock time than PE2 across the full experimental grid
(Appendix~\ref{sec:efficiency}). On a cost-normalised basis, BT-APE
delivers each percentage point of $wF_1$ at roughly $27\%$ of PE2's
input-token cost and $34\%$ of its wall-clock cost. This is the main
practical motivation for the design: it enables deployment on
resource-constrained or air-gapped servers, which are typically
preferred in RE settings where requirements confidentiality
discourages reliance on hosted APIs.

The defaults $X = 3$ and $N_{\max} = 20$ are motivated by the
sensitivity analysis in Appendix~\ref{app:sens}, which shows that
$X = 1$ yields unstable trajectories while $X \in \{3, 5\}$ behave
comparably on accuracy at increasing resource cost.
\subsection{Prompt Structure and Class Definitions}
\label{subsec:class-definitions}

A central component of the BT-APE classification prompt is the set of
\textit{class definitions} that instruct the LLM on how to distinguish
between requirement categories. In the optimisation loop, the class
definitions are the primary variable component of the prompt $p$,
while the surrounding scaffolding (output-format instructions, task
framing) remains fixed.

We restrict optimisation to class definitions for two reasons. First,
the output-format instructions are dictated by the evaluation pipeline:
rewriting them risks producing responses the parser cannot decode,
conflating prompt-quality variation with parsing failures. Second,
prior work on prompt sensitivity in classification
tasks~\cite{rodriguez2023prompts,Arora2024} identifies
category descriptions as the primary lever through which prompts
influence label decisions, while task framing contributes
comparatively little once a minimally adequate version is in place.
Confining the search to this component also keeps the prompt space
small enough to be explored within the bounded iteration budget
($N_{\max} = 20$).

Two strategies are proposed for initializing these definitions, each serving as a different starting point $p_1$ for the optimization process:

\begin{enumerate}
    \item \textbf{Simple Definitions (BT-APE-Uninformed).}
Class definitions are one-sentence intuitive descriptions written by
Author~1 without consulting RE literature or standards, reflecting a
minimal-effort starting point. Examples: \textit{``Functional: what
the system should do.''}; \textit{``Quality: how well the system
should do it.''}; \textit{``Security: protection against unauthorised
access or harm.''} The full set is given in
Appendix~\ref{app:prompts-defs}.
    
  \item \textbf{Literature-Based Definitions (\textbf{BT-APE-Informed}).} Class definitions are curated and synthesized from established software engineering literature by an RE expert (last author).  The functional and non-functional requirement definitions are derived from Glinz's~\cite{4384163} summary of the literature definitions, which characterizes functional requirements in terms of the essential functions a system must perform, the services it must offer, and the behaviours it must exhibit under specified conditions---focusing on the inputs (stimuli), outputs (responses), and the behavioural relationships between them. The quality requirement definitions are synthesized from quality-modeling literature and standards~\cite{10.1007/978-3-642-14192-8_15,Olsson2022}, expressing how well a system or service should execute an intended function through product quality aspects (e.g., functional suitability, reliability, performance, efficiency, usability, maintainability, security, compatibility, portability) and quality in use aspects (e.g., satisfaction, effectiveness, freedom from risk, context coverage). The security requirement definitions are derived from the security requirements engineering literature and standards~\cite{10.1109/TSE.2007.70754,8559686,sec1}, characterizing security requirements as prescriptive constraints imposed on a system's functional behaviour to operationalize its security goals---restricting how functions are performed to prevent, detect, or recover from harm, while specifying security policies and addressing risks, threats, and assets. The non-functional, non-quality, and non-security definitions are obtained by negating their functional, quality, and security counterparts, respectively. These definitions capture the nuanced distinctions recognized in the domain, such as differentiating sub-categories of quality requirements (e.g., performance, security, usability). By grounding the initial prompt in domain knowledge, the LLM is provided with richer semantic context from the outset.
\end{enumerate}

In both cases, the definitions are subject to iterative refinement during the optimization process: at each iteration, the BT-APE mechanism may edit, expand, or restructure the class definitions based on misclassified and correctly classified examples from the training set $D_{\text{pool}}$. The goal of introducing these different definitions is to check whether a more informed initialization---rooted in domain literature---leads to superior classification performance compared to uninformed starting point, and whether this advantage persists after multiple rounds of optimization. All the prompt structures and definitions are given in Appendix~\ref{app:prompts-defs}.

\subsection{Inference-time voting protocol}
\label{sec:voting}

The protocol described in this section is not specific to BT-APE; it is
an evaluation safeguard applied uniformly to all prompting strategies
(zero-shot, few-shot, CoT, CoT+few-shot, PE2, and BT-APE) whenever a
prompt is scored on $Dval$ or $Dtest$, so that comparisons across
strategies are made under identical conditions.

To mitigate the run-to-run variability of LLM outputs, we adopt a
\emph{majority voting strategy} during inference. Although we set the
decoding temperature to $0$, identical prompts can still yield
different outputs across runs due to stochastic sampling at non-zero
temperature settings, GPU non-determinism, and request batching at the
inference backend~\cite{ouyang2022traininglanguagemodelsfollow}. For each classification
instance, the model is therefore executed three times with the same
prompt, and the final label is assigned by majority vote: the class
predicted in at least two out of three runs is selected as the final
output. This protocol applies uniformly to baseline strategies and to
the in-loop validation scoring of BT-APE
(Algorithm~\ref{alg:approach}, line~10) and PE2
(Algorithm~\ref{alg:pe2-baseline}).

\section{Research Design}
\label{sec:researchdesign}

\subsection{Research Questions}

 
The research design is guided by the following research questions (RQs):

\begin{itemize}
      \item\textbf{RQ1:} \textit{What is the performance of BT-APE in requirements classification compared to standard prompt engineering approaches? }
    \\ 
    This question aims to evaluate the overall effectiveness of BT-APE relative to standard prompt engineering methods in the context of requirements classification. To answer this question, 1) we assess BT-APE's performance across different LLMs and datasets using descriptive statistics (e.g., mean, standard deviation) for key metrics such as weighted-F1  score in comparison to other baseline methods.  2) We then conduct statistical significance testing to compare BT-APE results against the results obtained by the baseline methods. This allows us to determine whether BT-APE offers a meaningful improvement over manual or fixed prompt designs.
    
   \item \textbf{RQ2: } \textit{What is the influence of the choice of dataset and LLM on the performance of BT-APE?}
\\
This question aims to statistically assess whether the selected datasets and
LLMs have a significant impact on the performance of the BT-APE process. To answer
it, we execute BT-APE across the three classification tasks using each of the five
LLMs and collect the resulting weighted-F1 scores, yielding one value per
(dataset, LLM) combination. We then apply distribution-free statistical tests to
determine whether performance differences across datasets and across LLMs are
statistically significant, and report a rank-based effect size to quantify the
strength of each factor. This analysis enables us to assess the individual
effects of datasets and LLMs on BT-APE effectiveness.

\item \textbf{RQ3:} \textit{What are the defining characteristics of optimal prompts, and how do these emerge and evolve through BT-APE iterations?}
\\
This question identifies the characteristics of effective prompts by analysing how they evolve across BT-APE iterations. We extract lexical (sentence count, word count, lexical diversity), syntactic (verb count, syntactic complexity), and semantic features (ambiguity score, semantic drift) from every prompt generated during the BT-APE process, and log them together with the validation weighted-F1 score of the prompt that produced them. 

Because each prompt is conditioned on its predecessor, observations within a (dataset, LLM) trajectory are not independent. We therefore analyse the data with a linear mixed-effects model (LMM) --- prompt features as fixed effects, a random intercept per trajectory --- so that coefficients estimate the \emph{within-trajectory} association between each feature and performance, separated from \emph{between-configuration} variance. Full model specification, robustness checks, hypotheses, and effect-size reporting are given in the statistical test design.

One interpretive caveat is integral to the research question rather than to the test design: semantic drift ($SD$) is by construction a property of \emph{consecutive} prompts, not of a prompt in isolation. A positive coefficient on $SD$ would characterise \emph{successful search dynamics}, not actionable guidance for designing a static prompt. We therefore separate state features (length, verb count, punctuation, lexical diversity, syntactic complexity, ambiguity) from the transition feature ($SD$), and restrict design claims to the former.

\item \textbf{RQ4:} \textit{To what extent does BT-APE-Informed influence classification performance compared to i) BT-APE-Uninformed and ii) how do both approaches perform relative to their respective best baseline models?}
\\
This question examines the effect of class definition quality on the BT-APE process by comparing two variants: BT-APE-Uninformed and BT-APE-Informed. To answer this, both variants are evaluated across the same datasets and LLMs under identical experimental conditions. Their performance is compared using key metrics such as Weighted $F1$ score  with statistical significance testing applied to assess whether BT-APE-Informed yields a meaningful improvement over BT-APE-Uninformed. In addition, each variant is compared against its respective best-performing baseline to determine how much improvement is achieved relative to standard prompt engineering approaches. This helps disentangle whether gains are driven primarily by the quality of the initial class definitions or by the iterative optimization process itself. Overall, this analysis isolates the role of class definitions as a design variable and evaluates whether domain-informed definitions provide a consistent advantage, or whether iterative optimization can compensate for simpler starting points.


\end{itemize}

\subsection{Shared Experimental Settings}

\textbf{\textit{Tasks \& Datasets.}}
We evaluate BT-APE on three requirements classification tasks, each representing a distinct and well-studied problem in requirements engineering:

\begin{itemize}
    \item \textbf{Functional-Quality classification based on the PROMISE Refined:} In this task, we consider four binary classification tasks: Functional, Only Functional, Quality, and Only Quality. Specifically, this means that each data point should be evaluated solely on whether it meets functional requirements, quality requirements, or both. For every data point, we treat Functional and Quality as two separate ground-truth labels. The Only Functional task is then defined as instances that are labeled as Functional but not Quality, while the Only Quality task covers instances labeled as Quality but not Functional.
    \item \textbf{Functional--Non-Functional Binary Classification:} Distinguishing between functional and Not-Functional requirements. This task addresses the fundamental separation between what a system should do and how well it should perform.
    \item \textbf{Security Binary Classification:} Determining whether a requirement is security-related or not. This task targets a single quality concern and is particularly relevant for safety- and security-critical systems.
\end{itemize}

We selected three well-established requirements datasets to represent the above tasks. These datasets have been extensively utilized in prior research involving machine learning and language models for RE, such as in \cite{dalpiaz2019requirements,hey2020norbert,kurtanovic2017automatically,alhoshan2023zero,knauss2011supporting}. A summary is presented in Table~\ref{tab:datasets}.
\begin{table}[h]
\centering
\caption{Overview of the Datasets Used in the Study.}
\label{tab:datasets}
\small
\resizebox{\textwidth}{!}{%
\begin{tblr}{
  column{2} = {c},
  cell{1}{1} = {c},
  hline{1,5} = {-}{0.08em},
  hline{2} = {-}{0.05em},
}
\textbf{Dataset~}                  & \textbf{\# Reqs.} & \textbf{Main Classes}                                                                                      & \textbf{Sources}                                                                                               \\
\textbf{PROMISE (NFR)}                   & 625               & {255 Functional Requirements (FR), \\370 Non-Functional Requirements (NFR).}                               & {Developed by Cleland-Huang et al.; \\used by Kurtanović and Maalej,\\~Hey et al., and Alhoshan et al.}        \\
\textbf{PROMISE (Refined)} & 625               & {310 Functional (FR), 230 is-only Functional (FR\_only).\\382 Quality (QA), 302 is-only Quality (Q\_only)} & {Developed by Dalpiaz et al.;\\includes reclassification of PROMISE; \\used by Hey et al. and Alhoshan et al.} \\
\textbf{SecReq}                    & 510               & 187 Security-related (Sec), 323~Non-security (NSec)                                                        & {Developed by Knauss et al.;\\~based on ePurse, CPN, and GPS;\\~used by Varenov et al. and Kobilica et al.}    
\end{tblr}
}
\end{table}

\begin{itemize}
   \item \textbf{PROMISE NFR Dataset:} Developed by Cleland-Huang \textit{et al.}~\cite{cleland2007automated,jane_cleland_huang_2007_268542}, this dataset comprises 625 requirements, including 255 FRs and 370 NFRs. This dataset has been widely adopted in the literature, including by Kurtanovi{\'c} and Maalej~\cite{kurtanovic2017automatically}, Hey \textit{et al.}~\cite{hey2020norbert}, and Alhoshan \textit{et al.}~\cite{alhoshan2023zero}.

    \item \textbf{PROMISE Refined Dataset:} Introduced by Dalpiaz \textit{et al.}~\cite{dalpiaz2019requirements}, this dataset re-annotates the PROMISE NFR dataset. For binary classification, requirements are labeled as Quality (382) vs. non-Quality (243), and (302) labeled as quality only; and Functional (310) vs. non-Functional (315), and (230) labeled as functional only requirements. This dataset has been employed in studies such as~\cite{hey2020norbert,alhoshan2023zero}. 

    \item \textbf{SecReq Dataset:} Created by Knauss \textit{et al.}~\cite{knauss2011supporting,knausseric20214530183}, the SecReq dataset contains 510 requirements, split into 187 security-related and 323 non-security-related instances. The data were collected from three projects: Common Electronic Purse (ePurse), Customer Premises Network (CPN), and Global Platform Specification (GPS). It has been utilized in multiple studies, including~\cite{varenov2021security,kobilica2020automated}.
\end{itemize}

\noindent
\textbf{\textit{LLMs Selection.}}
To select a representative set of instruction-tuned LLMs with
approximately 7--8 billion parameters, we consulted the Hugging Face
Open LLM Leaderboard\footnote{\url{https://huggingface.co/spaces/open-llm-leaderboard/open_llm_leaderboard}}
as a reference for identifying widely used open-source models. The
selection was guided by model availability, parameter scale,
organisational diversity, and architectural characteristics rather
than reported benchmark performance. Our goal is not to identify the
single best-performing model on requirements classification, but to
test whether BT-APE \emph{generalises} across architectures and
whether it consistently beats classical prompting baselines; using
models that span different organisations, attention mechanisms, and
pretraining corpora gives us a more honest stress test of
generalisation than concentrating on top-ranked models from a single
family. The 7--8B parameter range is chosen because models of this
size can be hosted locally on a single commodity GPU, which reflects
the practical RE setting where requirements confidentiality
discourages reliance on hosted proprietary
APIs~\cite{zadenoori2025largelanguagemodelsllms}. Based on these criteria, we selected five models developed by different research organizations:

\begin{table}[H]
\centering
\caption{Overview of selected instruction-tuned LLMs used in this study.}
\label{tab:llm_specs}
\small
\begin{tabular}{p{2.7cm}p{1.5cm}p{3cm}p{3.5cm}p{2cm}}
\toprule
\textbf{Model} & \textbf{Parameters} & \textbf{Attention Mechanism} & \textbf{Pretraining/Instruction Tasks} & \textbf{Organization} \\
\midrule
\textbf{Qwen2} & 7B & RoPE, MQA & Multilingual corpora, code, instruction tuning & Alibaba Cloud \\
\textbf{Falcon3} & 7B & MQA & English web text, instruction data & TII (UAE) \\
\textbf{Granite-3.2} & 8B & Scaled-dot attention (factual) & Business text, code, domain adaptation & IBM Research \\
\textbf{Ministral} & 8B & Sliding Window Attention & Instruction datasets, reasoning tasks & Mistral AI \\
\textbf{LLaMA-3} & 8B & GQA, RoPE & Multilingual corpora, SFT tasks & Meta AI \\
\bottomrule
\end{tabular}
\end{table}

\begin{itemize}
    \item \textbf{Qwen2-7B-Instruct} (Alibaba Cloud): employs Rotary Positional Embeddings combined with Multi-Query Attention to support efficient attention computation. It is pretrained on multilingual corpora and programming code and further refined using human-aligned instruction datasets.
    
    \item \textbf{Falcon3-7B-Instruct} (Technology Innovation Institute): uses Multi-Query Attention to reduce inference-time memory requirements. Its training data primarily consists of large-scale English web text and curated instruction-following datasets.
    
    \item \textbf{Granite-3.2-8B-Instruct} (IBM Research): applies optimized scaled dot-product attention with a focus on factual grounding. It is trained on business-oriented text, programming code, and domain-specific adaptation tasks.
    
    \item \textbf{Ministral-8B-Instruct-2410} (Mistral AI): incorporates Sliding Window Attention to limit attention to local contexts and improve computational efficiency. It is trained on instruction-following datasets and reasoning-oriented benchmarks.
    
    \item \textbf{Meta-Llama-3-8B-Instruct} (Meta AI): adopts Grouped-Query Attention together with Rotary Positional Embeddings to balance efficiency and representational capacity. Its pretraining relies on large-scale multilingual corpora and supervised fine-tuning on human-aligned datasets.
\end{itemize}

Table~\ref{tab:llm_specs} summarizes the main technical characteristics of the selected models, including parameter size, attention mechanisms, pretraining and instruction tasks, and organizational origin. These descriptions provide contextual background for the experimental setup without presupposing differences in task performance.

\noindent
\textbf{\textit{Evaluation Metrics.}}
We used four standard evaluation metrics: Precision ($P$), Recall ($R$), Weighted F1 Score ($wF_1$). These metrics are well-suited for imbalanced classification tasks and allow us to evaluate both general accuracy and recall-oriented performance.

\begin{itemize}
    \item \textbf{Precision ($P$):} The proportion of correctly predicted instances for a given class among all instances predicted as belonging to that class.
    \[
    P = \frac{TP}{TP + FP}
    \]

    \item \textbf{Recall ($R$):} The proportion of correctly predicted instances for a given class among all actual instances of that class.
    \[
    R = \frac{TP}{TP + FN}
    \]

    \item \textbf{Weighted F1 Score ($wF_1$):} The harmonic mean of precision and recall, computed for each class and averaged using the number of true instances (support) as weights.
    \[
    wF_1 = \sum_{c \in C} w_c \cdot \frac{2 \cdot P_c \cdot R_c}{P_c + R_c}
    \]

\end{itemize}

ìIn the above equations, $TP$, $FP$, and $FN$ denote true positives, false positives, and false negatives, respectively; $w_c$ represents the support (i.e., the proportion of instances) for class $c$, and $C$ is the set of all target classes.

\subsection{Prompt Baselines Selection}
We compare our approach against five widely adopted prompting strategies that
together cover the standard spectrum of techniques used to elicit
classification performance from LLMs without task-specific fine-tuning: four
in-context learning baselines of increasing sophistication, and one APE baseline. These baselines are selected because they
are (i)~model-agnostic and reproducible across both large and small language
models, (ii)~computationally lightweight and therefore compatible with our SLM
deployment setting, and (iii)~established as canonical reference points in
recent RE and NLP classification studies~\cite{Arora2024}. The five baselines are:
\begin{itemize}
    \item \textit{Zero-shot prompting}: The model receives only a task description and label definitions, with no labeled examples. This baseline measures the model's ability to perform the classification task purely from its pretrained knowledge and the natural-language specification of the task.

    \item \textit{Few-shot prompting}: The model is provided with a small set of labeled input--output examples alongside the task description. This baseline isolates the contribution of demonstration-based in-context learning, allowing us to measure how much performance improves when the model is exposed to representative instances of each requirement category.

    \item \textit{Chain-of-Thought (CoT) prompting}: The model is explicitly instructed to generate intermediate reasoning steps before producing the final output. This baseline tests whether eliciting structured reasoning improves classification on requirements that demand multi-step interpretation, such as distinguishing functional from non-functional aspects in compound statements.

    \item \textit{CoT with Few-shot prompting}: The model receives multiple labeled examples, each accompanied by an explicit reasoning trace and the final label. This baseline combines demonstration-based learning with reasoning elicitation and represents the strongest non-optimised prompting configuration commonly reported in the literature.

    \item \textit{PE2 (APE)}~\cite{ye2024promptengineeringpromptengineer}: The prompt is iteratively refined by a meta-prompted LLM conditioned on the top-$n$ historical prompts with their validation weighted-F1 scores and a batch of misclassifications observed on a held-out pool. Unlike the four baselines above, which keep the prompt fixed, PE2 actively searches the prompt space using the same iteration budget and evaluation protocol as our method. This baseline is the most directly comparable reference point for our approach, since both methods are single-trajectory, history-aware APE procedures driven by errors on a held-out set; its inclusion is what allows us to attribute any performance gap to our three algorithmic contributions---bounded back-tracking, balanced four-tuple example selection, and majority-voted F1---rather than to the use of prompt optimisation in general. Full algorithmic and implementation details, together with a precise enumeration of the differences from our method, are given in Appendix~\ref{sec:pe2-baseline} (Algorithm~\ref{alg:pe2-baseline}).
\end{itemize}
The exact prompt templates used for each baseline are provided in the replication package~\cite{amin_zadenoori_2026_20438927}.

\subsection{Experimental Design for RQ1}
\label{sec:rq1-design}

\noindent

\noindent
\textit{Statistical Significance.}\footnote{\textit{Notation for hypotheses.}
Throughout Sections~\ref{sec:rq1-design}--\ref{sec:rq4-design}, we
denote hypotheses using $H^{\text{RQ}n}_{k,0}$ for the null and
$H^{\text{RQ}n}_{k,A}$ for the alternative, where $n$ is the research
question and $k$ indexes the hypothesis within that question. For
example, $\Hnull{1}{1}$ is the first null hypothesis under RQ1 and
$\Halt{2}{2}$ the second alternative hypothesis under RQ2.}
To assess whether the performance differences between each baseline
prompting strategy and the proposed BT-APE approach are statistically
significant, we analyse the per-cell differences in $wF_1$ (i.e.,
$\Delta = wF_{1,\text{BT-APE}} - wF_{1,\text{Baseline}}$). All
differences across datasets and LLMs are pooled for each baseline
strategy.

We then conduct a non-parametric \textit{Wilcoxon signed-rank test}
to evaluate whether BT-APE achieves a higher mean $wF_1$ than the
baseline. All tests are performed at a significance level of
$\alpha = 0.05$, with $p$-values adjusted using the
\textit{Holm--Bonferroni procedure} to account for multiple baseline
comparisons.

For each comparison between BT-APE and a baseline strategy, the
following hypotheses are defined:
\begin{itemize}
    \item $\Hnull{1}{1}$: The mean difference in performance ($wF_1$)
    between BT-APE and the baseline strategy is zero ($\Delta = 0$).
    \item $\Halt{1}{1}$: BT-APE achieves a higher mean performance
    ($wF_1$) than the baseline strategy ($\Delta > 0$).
\end{itemize}

\noindent
\textit{Effect Size and Interpretation.} We report the \textit{mean
difference in $wF_1$} ($\Delta$) with 95\% confidence intervals. This
statistic is centred at zero when there is no improvement:
\begin{itemize}
    \item $\Delta = 0$ indicates no systematic difference.
    \item Positive values indicate that BT-APE outperforms the baseline.
    \item Negative values indicate that the baseline outperforms BT-APE.
\end{itemize}
Effect size $r$ is computed as $r = Z / \sqrt{N}$, where $Z$ is the
standardised Wilcoxon test statistic and $N$ is the number of
observations. Magnitudes are interpreted as: negligible ($r < 0.10$),
small ($0.10 \leq r < 0.30$), moderate ($0.30 \leq r < 0.50$), and
large ($r \geq 0.50$)~\citep{tomczak2014need}.

\subsection{Experimental Design for RQ2}
\label{sec:rq2-design}

\noindent
\textit{Factors Analysis.}
To address RQ2, we investigate how the performance of BT-APE varies
as a function of two independent variables---\textit{Datasets/Tasks}
and \textit{LLMs}---as well as their interaction. The final BT-APE
results consist of one $wF_1$ value per (Dataset, LLM) combination,
yielding a $3 \times 5$ design with $n = 15$ observations.
\begin{itemize}
    \item \textbf{Datasets/Tasks.} We test whether performance ($wF_1$)
    differs significantly across requirements classification tasks
    (Security, Functional, and Quality).
    \item \textbf{LLMs.} We test whether performance ($wF_1$) differs
    significantly across the selected LLMs.
\end{itemize}

\noindent
\textit{Statistical Significance.}
We first assess the normality assumption with the Shapiro--Wilk test.
Given the small sample and borderline normality, we adopt
distribution-free \textit{Friedman tests} for the two main effects
rather than relying on the assumptions of parametric ANOVA. We test
the dataset effect using LLMs as blocks, and the LLM effect using
datasets as blocks. All tests are conducted at a significance level
of $\alpha = 0.05$. The following hypotheses are defined:
\begin{itemize}
    \item \textbf{Dataset/Task effect.}
    \begin{itemize}
        \item $\Hnull{2}{1}$: There is no statistically significant
        difference in $wF_1$ across datasets.
        \item $\Halt{2}{1}$: There is a statistically significant
        difference in $wF_1$ across datasets.
    \end{itemize}
    \item \textbf{LLM effect.}
    \begin{itemize}
        \item $\Hnull{2}{2}$: There is no statistically significant
        difference in $wF_1$ across LLMs.
        \item $\Halt{2}{2}$: There is a statistically significant
        difference in $wF_1$ across LLMs.
    \end{itemize}
\end{itemize}

\noindent
\textit{Effect Size and Interpretation.}
We report \textit{Kendall's $W$} as the effect size for each Friedman
test, which ranges from 0 (no agreement among blocks) to 1 (perfect
agreement). Larger values indicate a stronger and more consistent
factor effect across the blocking variable. As the interaction is
not estimable under a single-replicate design, we do not test it
formally; where the cell-level results suggest task-dependent model
behaviour, we describe it descriptively.

\subsection{Experimental Design for RQ3}
\label{sec:rq3-design}

\noindent
\textit{Prompt Feature Extraction.}
We extract and log a set of lexical, syntactic, and semantic features
from each prompt generated during BT-APE iterations. These features
are defined and computed as follows.

\noindent
\textit{a) Lexical Features.}
\begin{itemize}
    \item Sentence Count ($SC$): The number of sentences in the
    prompt, measured using sentence segmentation.
    \[
    SC = \text{Number of sentences in } P
    \]

    \item Word Count ($WC$): The total number of words or tokens in
    the prompt.
    \[
    WC = \sum_{i=1}^{n} \delta(w_i), \quad \delta(w_i) =
    \begin{cases}
    1, & \text{if } w_i \text{ is a word} \\
    0, & \text{otherwise}
    \end{cases}
    \]

    \item Punctuation/Markers Count ($PM$): The number of structural
    markers such as colons, dashes, or line breaks.
    \[
    PM = \sum_{c \in P} \delta(c), \quad \delta(c) =
    \begin{cases}
    1, & \text{if } c \in \{\texttt{:}, \texttt{-}, \texttt{.}, \texttt{\textbackslash n}, \ldots\} \\
    0, & \text{otherwise}
    \end{cases}
    \]

    \item Lexical Diversity ($LD$): The ratio of unique words to
    total words, indicating vocabulary richness.
    \[
    LD = \frac{|\text{Unique}(w_1, w_2, \ldots, w_n)|}{WC}
    \]
\end{itemize}

\noindent
\textit{b) Syntactic Features.}
\begin{itemize}
    \item Verb Count ($VB$): The number of verbs in the prompt,
    identified via part-of-speech tagging.
    \[
    VB = \sum_{i=1}^{n} \delta(\text{POS}(w_i) = \text{VERB})
    \]

    \item Syntactic Complexity ($SCx$): The average depth of the
    dependency parse tree over all tokens in the prompt.
    \[
    SCx = \frac{1}{n} \sum_{i=1}^{n} \text{depth}(w_i)
    \]
    where $\text{depth}(w_i)$ denotes the number of dependency links
    from word $w_i$ to the root.
\end{itemize}

\noindent
\textit{c) Semantic Features.}
\begin{itemize}
    \item Ambiguity Score ($AS$): quantifies the vagueness of a
    prompt based on the presence of ambiguous or underspecified
    terms, commonly referred to as ``smelly'' words (e.g.,
    \textit{some}, \textit{many}, \textit{etc.}, \textit{maybe},
    \textit{often}, \textit{unclear}, \textit{various}). The score is
    computed heuristically as:
    \[
    AS = \frac{1}{WC} \sum_{i=1}^{WC} \delta(w_i \in D_{\text{amb}})
    \]
    where:
    \begin{itemize}
        \item $WC$ is the total word count of the prompt,
        \item $w_i$ is the $i^{\text{th}}$ word in the prompt,
        \item $D_{\text{amb}}$ is a predefined dictionary of
        ambiguity-inducing terms,
        \item $\delta(\cdot)$ is an indicator function that returns
        1 if $w_i \in D_{\text{amb}}$, and 0 otherwise.
    \end{itemize}

    \item Semantic Drift ($SD$): quantifies the semantic change
    between the current prompt and the immediately preceding one,
    computed as the cosine distance between their embedding vectors.
    For a sequence of $n$ prompts, the drift at iteration $i$ is
    defined as:
    \[
    SD_i =
    \begin{cases}
    0, & \text{if } i = 1 \\
    1 - \dfrac{\vec{e}_{i} \cdot \vec{e}_{i-1}}{\|\vec{e}_{i}\| \|\vec{e}_{i-1}\|}, & \text{if } i > 1
    \end{cases}
    \]
    where $\vec{e}_i$ and $\vec{e}_{i-1}$ are the embedding vectors
    of the prompts at iterations $i$ and $i-1$, respectively. This
    formulation captures the magnitude of semantic shift introduced
    at each step in the BT-APE process.
\end{itemize}

All of these features (lexical, syntactic, and semantic) are tracked
across all BT-APE iterations to study how prompt structure changes
over time and how such changes correlate with model performance
improvements. We further distinguish between \emph{state features},
which describe a prompt in isolation ($SC$, $WC$, $PM$, $LD$, $VB$,
$SCx$, $AS$), and \emph{transition features}, which describe a change
between consecutive prompts ($SD$). This distinction matters for
interpretation: a within-trajectory association between a state
feature and performance can plausibly motivate prompt-design guidance,
whereas an association involving $SD$ characterises the dynamics of
successful search rather than properties of a static prompt.

\noindent
\textit{Statistical Model.}
The prompts generated across BT-APE iterations are not independent
observations. Within a single (dataset, LLM) trajectory, every prompt
is conditioned on its predecessor, shares the same proposer, and
inherits the same starting point; across trajectories, baseline
performance differs systematically with task difficulty and model
capability (as established in RQ2). Pooling all prompts into an
ordinary least-squares regression with $wF_1$ as the response would
therefore conflate \emph{within-trajectory} variance---how performance
changes as a single search evolves---with \emph{between-configuration}
variance, and would overstate the precision of the resulting
coefficients by treating non-independent observations as independent.

To respect this nested structure, we estimate a \textit{linear
mixed-effects model} (LMM) in which prompt features enter as fixed
effects and a random intercept is fitted for each (dataset, LLM)
trajectory:
\[
wF_{1,ij} \;=\; \beta_0 + \sum_{k} \beta_k\, x_{k,ij} \;+\; u_j \;+\; \varepsilon_{ij},
\qquad u_j \sim \mathcal{N}(0,\sigma_u^2),\; \varepsilon_{ij} \sim \mathcal{N}(0,\sigma_\varepsilon^2),
\]
where $i$ indexes the iteration within trajectory
$j \in \{1, \ldots, 15\}$, $x_{k,ij}$ is the value of the $k$-th
prompt feature at iteration $i$ of trajectory $j$, $u_j$ is the
random intercept absorbing configuration-level differences in
baseline $wF_1$, and $\varepsilon_{ij}$ is the residual. The
fixed-effect coefficients $\beta_k$ therefore estimate the
\emph{within-trajectory} association between each prompt feature and
performance, after removing the contribution of the underlying
(dataset, LLM) configuration. The model is estimated by restricted
maximum likelihood (REML), and all features are standardised to zero
mean and unit variance prior to fitting so that coefficient
magnitudes are directly comparable.

\noindent
\textit{Robustness Check on Terminal Prompts.}
As a complementary analysis that is independent of trajectory
dynamics by construction, we also fit an ordinary least-squares
regression on the \emph{terminal} prompts $p^{*}$ of each of the 15
trajectories ($n = 15$). The transition feature $SD$ is excluded
from this analysis as it is not defined for a single prompt. This
terminal-prompt regression has limited statistical power but provides
a sanity check: a feature whose within-trajectory effect under the
LMM is contradicted by its sign on terminal prompts is unlikely to
constitute reliable design guidance. We therefore report a feature
as a robust correlate of effective prompts only when its LMM
coefficient is statistically significant \emph{and} its sign agrees
with the terminal-prompt regression.

\noindent
\textit{Hypotheses.}
All tests are conducted at a significance level of $\alpha = 0.05$,
with Holm--Bonferroni correction applied across the feature set to
control the family-wise error rate.
\begin{itemize}
    \item $\Hnull{3}{1}$: After accounting for between-trajectory
    variance, prompt features have no statistically significant
    within-trajectory effect on $wF_1$ ($\beta_k = 0$ for all $k$).
    \item $\Halt{3}{1}$: At least one prompt feature has a
    statistically significant within-trajectory effect on $wF_1$.
\end{itemize}

\noindent
\textit{Effect Size and Interpretation.}
We report standardised fixed-effect coefficients $\hat{\beta}_k$
with 95\% confidence intervals for each feature. Positive
coefficients indicate that, within a trajectory, an increase in the
feature is associated with higher $wF_1$; negative coefficients
indicate the converse. The absolute magnitude of $\hat{\beta}_k$
reflects the relative strength of the feature's within-trajectory
contribution.

At the model level, we report three quantities. The \emph{marginal}
$R^2_m$ measures the proportion of variance in $wF_1$ explained by
the fixed effects alone, and is the appropriate analogue of the OLS
$R^2$ for the prompt-feature claim. The \emph{conditional} $R^2_c$
measures the proportion explained by fixed and random effects
jointly, and indicates the total fit of the model. The
\emph{intra-class correlation coefficient} (ICC),
\[
\text{ICC} \;=\; \frac{\sigma_u^2}{\sigma_u^2 + \sigma_\varepsilon^2},
\]
quantifies the share of residual variance attributable to
between-configuration differences rather than to within-trajectory
prompt characteristics; a high ICC would indicate that most of the
variation in $wF_1$ is configuration-driven and that within-trajectory
feature effects, however statistically significant, account for a
comparatively small share of overall performance variation.
Together, $R^2_m$, $R^2_c$, and the ICC provide a more honest
accounting of where the explanatory power of the model lies than a
single pooled $R^2$ would.

As noted above, we treat $SD$ separately in the interpretation: a
significant within-trajectory coefficient on $SD$ is described as a
property of successful optimisation trajectories rather than as
actionable guidance for designing a static prompt. Actionable design
claims are restricted to state features whose coefficients are stable
across the LMM and the terminal-prompt regression.

\subsection{Experimental Design for RQ4}
\label{sec:rq4-design}

\noindent
\textit{Class Definitions.}
As introduced in Section~\ref{subsec:class-definitions}, we define
two initialisation strategies for the prompt optimisation process. In
\textit{BT-APE-Informed}, the initial prompt is seeded with
fine-grained class definitions curated from requirements engineering
literature and standards. In \textit{BT-APE-Uninformed}, no predefined
class definitions are provided; instead, the algorithm begins with
simple, minimal descriptions and iteratively refines them throughout
the optimisation process. This design allows us to assess two
complementary aspects: (1)~whether domain-grounded definitions offer
a measurable advantage as a starting point, and (2)~whether the
iterative optimisation is capable of discovering effective definitions
independently, without relying on prior domain knowledge.

\noindent
\textit{Statistical Significance.}
To assess whether the performance differences between BT-APE-Uninformed
and BT-APE-Informed are statistically significant, we analyse two
complementary dimensions.

\noindent
\textit{Dimension 1: Final Performance Comparison.} We analyse the
per-cell differences in final optimised $wF_1$ (i.e.,
$\Delta_{\text{final}} = wF_{1,\text{BT-APE-Informed}} - wF_{1,\text{BT-APE-Uninformed}}$).
All differences across datasets and LLMs are pooled. We then conduct
a \textit{Wilcoxon signed-rank test} to evaluate whether
BT-APE-Informed achieves a higher mean final $wF_1$ than
BT-APE-Uninformed. The test is performed at a significance level of
$\alpha = 0.05$. For this comparison, the following hypotheses are
defined:
\begin{itemize}
    \item $\Hnull{4}{1}$: The mean difference in final performance
    ($wF_1$) between BT-APE-Informed and BT-APE-Uninformed is zero
    ($\Delta_{\text{final}} = 0$).
    \item $\Halt{4}{1}$: BT-APE-Informed achieves a higher mean final
    performance ($wF_1$) than BT-APE-Uninformed
    ($\Delta_{\text{final}} > 0$).
\end{itemize}

\noindent
\textit{Dimension 2: Improvement-over-Baseline Comparison.} Since
each variant has its own independent baseline (fine-grained class
definitions for BT-APE-Informed vs.\ minimal descriptions for
BT-APE-Uninformed), we compute the per-cell gain for each variant
relative to its own starting point:
\[
\Delta_{\text{Informed}} = wF_{1,\text{Informed-Opt}} - wF_{1,\text{Informed-Baseline}}
\]
\[
\Delta_{\text{Uninformed}} = wF_{1,\text{Uninformed-Opt}} - wF_{1,\text{Uninformed-Baseline}}
\]
We then calculate the difference in gains:
\[
\delta_{\text{gain}} = \Delta_{\text{Informed}} - \Delta_{\text{Uninformed}}
\]
A separate \textit{Wilcoxon signed-rank test} is applied to
$\delta_{\text{gain}}$ across all (dataset, LLM) pairs to evaluate
whether the two variants improve equally over their respective
baselines. The hypotheses for this dimension are:
\begin{itemize}
    \item $\Hnull{4}{2}$: The mean difference in gains
    ($\delta_{\text{gain}}$) between BT-APE-Informed and
    BT-APE-Uninformed is zero, i.e., both variants improve equally
    over their own baselines.
    \item $\Halt{4}{2}$: The mean difference in gains
    ($\delta_{\text{gain}}$) is not zero, i.e., one variant achieves
    greater relative improvement from its baseline than the other.
\end{itemize}

\noindent
\textit{Effect Size and Interpretation.} For both dimensions, we
report the \textit{mean difference} ($\Delta_{\text{final}}$ and
$\delta_{\text{gain}}$ respectively) with 95\% confidence intervals,
following the same interpretation benchmarks defined in RQ1. This
dual analysis allows us to isolate the contribution of class
definitions as a design variable and assess: (a)~whether
domain-grounded definitions provide a lasting advantage in final
performance, and (b)~whether the iterative optimisation compensates
for simpler starting points by achieving comparable or greater
relative improvement from a lower baseline.

\section{Experimental Results}
\label{sec:results}
In this section, we present the results of the classification experiments conducted on three datasets: Refined Promise, Promise NFR, and SecReq. Performance is reported using weighted F1 ($wF_1$) score for the overall (average) results, and per-class F1 score for the class-specific breakdowns.

\paragraph*{Refined Promise Dataset Results}
Table~\ref{tab:refined_promise} presents the classification results for the Refined Promise dataset, which includes classes for Functional (F), Quality (Q), Only Functional (onlyF), and Only Quality (onlyQ).

\begin{table}[H]
\centering
\scriptsize
\begin{tabular}{|l|c|c|c|c|c|c|}
\hline
\multirow{2}{*}{Model} & \multirow{2}{*}{Strategy} & Average & F & Q & onlyF & onlyQ \\
\cline{3-7}
 & & wF1 & wF1 & wF1 & wF1 & wF1 \\
\hline
\multirow{6}{*}{Qwen/Qwen2-7B-Instruct} & Zero-shot & 0.413 & 0.421 & 0.594 & 0.421 & 0.215 \\
 & Few-shot & 0.500 & 0.390 & 0.615 & 0.640 & 0.356 \\
 & CoT & 0.487 & 0.410 & 0.643 & 0.640 & 0.256 \\
 & CoT + Few-shot & 0.547 & 0.537 & 0.636 & 0.640 & 0.374 \\
 & APE-PE2 & 0.587 & \textbf{0.583} & 0.679 & \textbf{0.691} & \textbf{0.444} \\
\cline{2-7}
 & BT-APE & \textbf{0.595} & 0.579 & \textbf{0.685} & 0.679 & 0.437 \\
\hline
\multirow{6}{*}{Falcon3-7B-Instruct} & Zero-shot & 0.125 & 0.082 & 0.185 & 0.082 & 0.152 \\
 & Few-shot & 0.559 & 0.677 & 0.544 & 0.677 & 0.337 \\
 & CoT & 0.260 & 0.622 & 0.082 & 0.185 & 0.151 \\
 & CoT + Few-shot & \textbf{0.635} & \textbf{0.729} & \textbf{0.672} & \textbf{0.729} & 0.409 \\
 & APE-PE2 & 0.620 & 0.645 & 0.670 & 0.658 & 0.437 \\
\cline{2-7}
 & BT-APE & 0.607 & 0.656 & 0.662 & 0.663 & \textbf{0.446} \\
\hline
\multirow{6}{*}{Granite-3.2-8B-Instruct} & Zero-shot & 0.463 & 0.557 & 0.256 & 0.557 & \textbf{0.483} \\
 & Few-shot & \textbf{0.515} & \textbf{0.672} & 0.260 & 0.672 & 0.455 \\
 & CoT & 0.441 & 0.572 & 0.144 & 0.572 & 0.477 \\
 & CoT + Few-shot & 0.476 & 0.649 & \textbf{0.557} & 0.215 & \textbf{0.483} \\
 & APE-PE2 & 0.472 & 0.536 & 0.325 & \textbf{0.678} & 0.381 \\
\cline{2-7}
 & BT-APE & 0.477 & 0.529 & 0.337 & 0.667 & 0.373 \\
\hline
\multirow{6}{*}{Ministral-8B-Instruct-2410} & Zero-shot & 0.455 & 0.569 & 0.195 & 0.569 & 0.488 \\
 & Few-shot & 0.594 & \textbf{0.682} & 0.458 & 0.682 & 0.554 \\
 & CoT & 0.362 & 0.481 & 0.051 & 0.481 & 0.436 \\
 & CoT + Few-shot & 0.446 & 0.616 & 0.296 & \textbf{0.692} & 0.181 \\
 & APE-PE2 & \textbf{0.646} & 0.642 & 0.672 & 0.647 & \textbf{0.590} \\
\cline{2-7}
 & BT-APE & 0.637 & 0.655 & \textbf{0.676} & 0.633 & 0.585 \\
\hline
\multirow{6}{*}{Llama-3-8B-Instruct} & Zero-shot & 0.212 & 0.297 & 0.075 & 0.297 & 0.181 \\
 & Few-shot & 0.266 & 0.153 & \textbf{0.573} & 0.234 & 0.104 \\
 & CoT & 0.283 & 0.274 & 0.473 & 0.274 & 0.112 \\
 & CoT + Few-shot & 0.393 & 0.616 & 0.296 & 0.535 & 0.123 \\
 & APE-PE2 & 0.540 & 0.683 & 0.305 & 0.625 & 0.571 \\
\cline{2-7}
 & BT-APE & \textbf{0.551} & \textbf{0.691} & 0.299 & \textbf{0.629} & \textbf{0.585} \\
\hline
\end{tabular}
\caption{Classification results for the Refined Promise dataset (weighted F1).}
\label{tab:refined_promise}
\end{table}

\begin{table}[H]
\centering
\scriptsize
\begin{tabular}{|l|c|c|c|c|}
\hline
\multirow{2}{*}{Model} & \multirow{2}{*}{Strategy} & Average & F & NFR \\
\cline{3-5}
 & & wF1 & wF1 & wF1 \\
\hline
\multirow{6}{*}{Qwen/Qwen2-7B-Instruct} & Zero-shot & 0.749 & 0.750 & 0.748 \\
 & Few-shot & 0.726 & 0.730 & 0.722 \\
 & CoT & 0.668 & 0.609 & 0.727 \\
 & CoT + Few-shot & 0.749 & 0.750 & 0.748 \\
 & APE-PE2 & 0.838 & \textbf{0.858} & 0.833 \\
\cline{2-5}
 & BT-APE & \textbf{0.847} & 0.847 & \textbf{0.846} \\
\hline
\multirow{6}{*}{Falcon3-7B-Instruct} & Zero-shot & 0.642 & 0.574 & 0.711 \\
 & Few-shot & 0.708 & 0.706 & 0.710 \\
 & CoT & 0.753 & 0.751 & 0.755 \\
 & CoT + Few-shot & 0.754 & 0.752 & 0.755 \\
 & APE-PE2 & \textbf{0.843} & 0.814 & \textbf{0.838} \\
\cline{2-5}
 & BT-APE & 0.831 & \textbf{0.828} & 0.833 \\
\hline
\multirow{6}{*}{Granite-3.2-8B-Instruct} & Zero-shot & 0.703 & 0.711 & 0.695 \\
 & Few-shot & 0.720 & 0.705 & 0.736 \\
 & CoT & 0.708 & 0.701 & 0.714 \\
 & CoT + Few-shot & 0.774 & 0.754 & 0.794 \\
 & APE-PE2 & 0.816 & \textbf{0.819} & \textbf{0.833} \\
\cline{2-5}
 & BT-APE & \textbf{0.820} & 0.812 & 0.827 \\
\hline
\multirow{6}{*}{Ministral-8B-Instruct-2410} & Zero-shot & 0.377 & 0.094 & 0.659 \\
 & Few-shot & 0.722 & 0.721 & 0.723 \\
 & CoT & 0.722 & 0.721 & 0.722 \\
 & CoT + Few-shot & 0.722 & 0.727 & 0.716 \\
 & APE-PE2 & \textbf{0.853} & 0.838 & 0.834 \\
\cline{2-5}
 & BT-APE & 0.847 & \textbf{0.846} & \textbf{0.848} \\
\hline
\multirow{6}{*}{Llama-3-8B-Instruct} & Zero-shot & 0.716 & 0.723 & 0.710 \\
 & Few-shot & 0.680 & 0.695 & 0.665 \\
 & CoT & 0.679 & 0.699 & 0.659 \\
 & CoT + Few-shot & 0.717 & 0.723 & 0.711 \\
 & APE-PE2 & 0.780 & \textbf{0.791} & \textbf{0.814} \\
\cline{2-5}
 & BT-APE & \textbf{0.792} & 0.778 & 0.805 \\
\hline
\end{tabular}
\caption{Classification results for the Promise NFR dataset (weighted F1).}
\label{tab:promise_nfr}
\end{table}

\begin{table}[H]
\centering
\scriptsize
\begin{tabular}{|l|c|c|c|c|}
\hline
\multirow{2}{*}{Model} & \multirow{2}{*}{Strategy} & Average & Security & Non-Security \\
\cline{3-5}
 & & wF1 & wF1 & wF1 \\
\hline
\multirow{6}{*}{Qwen/Qwen2-7B-Instruct} & Zero-shot & 0.651 & 0.468 & 0.835 \\
 & Few-shot & 0.818 & 0.736 & 0.900 \\
 & CoT & 0.843 & 0.778 & 0.908 \\
 & CoT + Few-shot & 0.843 & 0.778 & 0.908 \\
 & APE-PE2 & 0.937 & \textbf{0.919} & 0.975 \\
\cline{2-5}
 & BT-APE & \textbf{0.945} & 0.908 & \textbf{0.981} \\
\hline
\multirow{6}{*}{Falcon3-7B-Instruct} & Zero-shot & 0.848 & 0.789 & 0.907 \\
 & Few-shot & 0.836 & 0.774 & 0.897 \\
 & CoT & 0.825 & 0.750 & 0.899 \\
 & CoT + Few-shot & 0.848 & 0.789 & 0.907 \\
 & APE-PE2 & \textbf{0.952} & 0.896 & 0.983 \\
\cline{2-5}
 & BT-APE & 0.947 & \textbf{0.903} & \textbf{0.991} \\
\hline
\multirow{6}{*}{Granite-3.2-8B-Instruct} & Zero-shot & 0.819 & 0.789 & 0.849 \\
 & Few-shot & 0.766 & 0.672 & 0.861 \\
 & CoT & \textbf{0.871} & \textbf{0.825} & 0.916 \\
 & CoT + Few-shot & 0.836 & 0.774 & 0.897 \\
 & APE-PE2 & 0.848 & 0.797 & \textbf{0.941} \\
\cline{2-5}
 & BT-APE & 0.859 & 0.783 & 0.934 \\
\hline
\multirow{6}{*}{Ministral-8B-Instruct-2410} & Zero-shot & 0.814 & 0.738 & 0.889 \\
 & Few-shot & 0.814 & 0.738 & 0.889 \\
 & CoT & 0.787 & 0.745 & 0.829 \\
 & CoT + Few-shot & 0.797 & 0.715 & 0.880 \\
 & APE-PE2 & \textbf{0.935} & 0.863 & \textbf{0.978} \\
\cline{2-5}
 & BT-APE & 0.922 & \textbf{0.872} & 0.972 \\
\hline
\multirow{6}{*}{Llama-3-8B-Instruct} & Zero-shot & \textbf{0.886} & 0.842 & 0.925 \\
 & Few-shot & 0.862 & 0.807 & 0.917 \\
 & CoT & 0.867 & 0.817 & 0.916 \\
 & CoT + Few-shot & \textbf{0.886} & \textbf{0.847} & 0.925 \\
 & APE-PE2 & 0.857 & 0.800 & 0.920 \\
\cline{2-5}
 & BT-APE & 0.864 & 0.795 & \textbf{0.933} \\
\hline
\end{tabular}
\caption{Classification results for the Secreq dataset (weighted F1).}
\label{tab:secreq}
\end{table}

\subsection{Quantitative Comparison: BT-APE vs. Baseline Prompting Strategies}

\begin{figure*}[t]
    \centering
    \includegraphics[width=\textwidth]{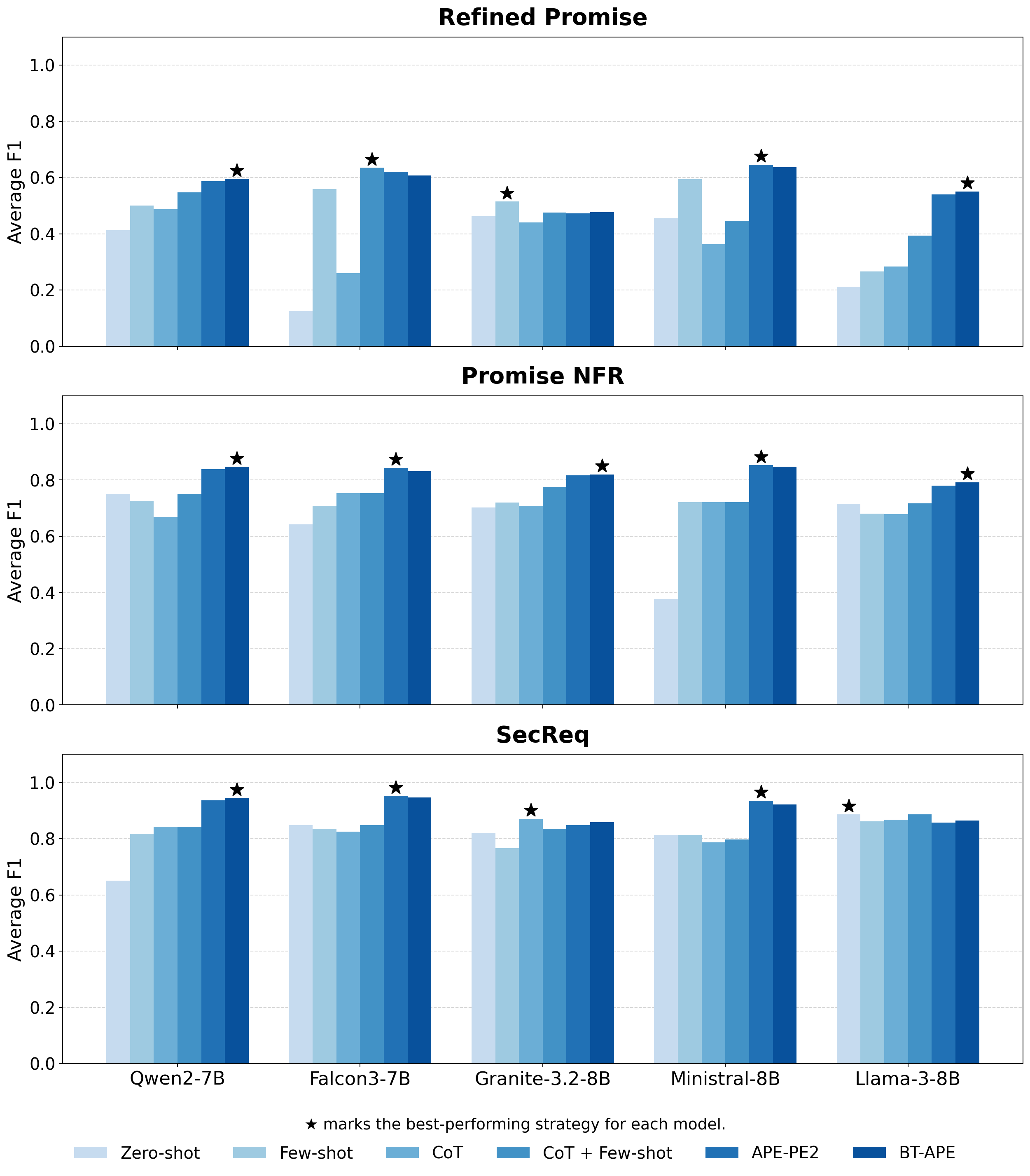}
    \caption{Average F1 of the five instruction-tuned LLMs across the six
    prompting strategies on the three requirements-classification datasets.
    The $\bigstar$ marks the best-performing strategy for each model.}
    \label{fig:f1_comparison}
\end{figure*}

\paragraph*{Overall Performance Analysis: BT-APE vs. Baselines}
Figure~\ref{fig:f1_comparison} presents the average F1 obtained by each of the
five instruction-tuned LLMs under every prompting strategy, split into three
panels corresponding to the three requirements-classification datasets. Within
each panel, the six grouped bars per model show Zero-shot, Few-shot, CoT,
CoT+Few-shot, APE-PE2, and BT-APE, and a $\bigstar$ marks the strategy that
achieves the highest average F1 for that model.

\textit{Refined Promise Dataset:} BT-APE is the best strategy for two of the five
models, while APE-PE2 emerges as the strongest baseline in three of them. For
\texttt{Qwen2-7B}, BT-APE achieves an average F1 of $0.595$, $+0.008$ above the
strongest baseline (APE-PE2 at $0.587$) and $+0.048$ above the best classical
baseline (CoT+Few-shot at $0.547$). The largest margin on this dataset is on
\texttt{Llama-8B}, where BT-APE attains $0.551$, $+0.011$ over APE-PE2 ($0.540$)
and $+0.158$ over the best classical baseline (CoT+Few-shot at $0.393$). On
\texttt{Ministral-8B}, APE-PE2 narrowly takes the top with $0.646$, $+0.009$
above BT-APE ($0.637$) and $+0.052$ above the best classical baseline (Few-shot at
$0.594$). For the remaining two models the strongest baseline is a classical
one: on \texttt{Falcon3-7B}, BT-APE ($0.607$) trails CoT+Few-shot ($0.635$) by
$-0.028$, and on \texttt{Granite-8B}, BT-APE ($0.477$) trails Few-shot ($0.515$)
by $-0.038$.

\textit{Promise NFR Dataset:} BT-APE and APE-PE2 jointly hold the top spot on all
five models, with BT-APE leading on three (\texttt{Qwen2-7B}, \texttt{Granite-8B},
\texttt{Llama-8B}) and APE-PE2 on the other two (\texttt{Falcon3-7B},
\texttt{Ministral-8B}). The gap between the two is small in every case:
$+0.009$ for \texttt{Qwen2-7B} (BT-APE $0.847$ vs.\ APE-PE2 $0.838$), $+0.004$ for
\texttt{Granite-8B} ($0.820$ vs.\ $0.816$), $+0.012$ for \texttt{Llama-8B}
($0.792$ vs.\ $0.780$), $-0.012$ for \texttt{Falcon3-7B} ($0.831$ vs.\
$0.843$), and $-0.006$ for \texttt{Ministral-8B} ($0.847$ vs.\ $0.853$). Both
strategies decisively beat the simpler baselines, with the largest gain
reaching $+0.131$ for APE-PE2 over Few-shot on \texttt{Ministral-8B} and
$+0.098$ for BT-APE over Zero-shot on \texttt{Qwen2-7B}.

\textit{SecReq Dataset:} BT-APE is the best strategy on one of the five models,
APE-PE2 on two, and a classical baseline on the remaining two. On
\texttt{Qwen2-7B}, BT-APE reaches $0.945$, $+0.008$ above APE-PE2 ($0.937$) and
$+0.102$ above the best classical baseline (CoT at $0.843$). On
\texttt{Falcon3-7B}, APE-PE2 leads with $0.952$---the highest average wF1
across all model--dataset combinations in the figure---narrowly above BT-APE
($0.947$, $-0.005$); the same pattern holds on \texttt{Ministral-8B}, where
APE-PE2 ($0.935$) exceeds BT-APE ($0.922$) by $+0.013$ and improves over the best
classical baseline (Zero-shot at $0.814$) by $+0.121$. In the two remaining
cases BT-APE is marginally below the strongest baseline: on \texttt{Granite-8B},
BT-APE ($0.859$) trails CoT ($0.871$) by $-0.012$, and on \texttt{Llama-8B}, BT-APE
($0.864$) trails Zero-shot ($0.886$) by $-0.022$.

Two trends are immediately apparent from Figure~\ref{fig:f1_comparison}. First,
the two APE strategies---\textit{BT-APE} and
\textit{APE-PE2}---together dominate the comparison, jointly accounting for
the best result in $11$ out of $15$ model--dataset combinations and reaching
up to $0.952$ on SecReq (\texttt{Falcon3-7B}, APE-PE2) and $0.853$ on Promise
NFR (\texttt{Ministral-8B}, APE-PE2). The two methods trade places in tight
margins---typically within $\pm 0.013$ of one another---suggesting that the
two prompt-optimization procedures converge to a similar performance regime,
with BT-APE marginally ahead overall ($6$ wins vs.\ $5$ for APE-PE2 when both are
considered). The few remaining best-strategy cases, such as CoT+Few-shot for
\texttt{Falcon3-7B} and Few-shot for \texttt{Granite-8B} on Refined Promise,
or CoT for \texttt{Granite-8B} and Zero-shot for \texttt{Llama-8B} on SecReq,
involve classical baselines that nonetheless remain within a narrow margin of
BT-APE. Second, performance is strongly conditioned by dataset difficulty: all
models score substantially lower on the four-class Refined Promise task than
on the binary Promise NFR and SecReq tasks, where most strategies already
exceed $0.70$ F1. Simpler prompting strategies are also far less stable;
Zero-shot and CoT in particular collapse for some models (e.g.\
\texttt{Falcon3-7B} drops to $0.125$ under Zero-shot on Refined Promise, and
\texttt{Ministral-8B} to $0.377$ under Zero-shot on Promise NFR), whereas
BT-APE and APE-PE2 consistently lift the weakest configurations and narrow the
gap between models. Overall, the results indicate that the choice of prompting
strategy has a larger and more reliable effect on classification quality than
the choice of model, with BT-APE---closely tracked by APE-PE2---offering the most
robust performance across all three datasets.

\paragraph*{Summary of BT-APE vs. Baselines}
Quantitatively, BT-APE matches or outperforms the strongest baseline (now
including APE-PE2) in $6$ out of $15$ model--dataset combinations on average
F1, and when APE-PE2 is set aside as the optimization-based peer, BT-APE matches
or outperforms the best classical baseline (Zero-shot, Few-shot, CoT,
CoT+Few-shot) in $11$ out of $15$ combinations. The mean improvement of BT-APE
over the strongest baseline including APE-PE2 is essentially flat
competitor---while the mean improvement of BT-APE over the best \emph{classical}
baseline rises to $+0.059$. The gains over classical baselines are most
consistent on Promise NFR ($+0.084$ average improvement) and on SecReq
($+0.055$), while on the most difficult dataset, Refined Promise, BT-APE still
wins two of five models but with a smaller average margin ($+0.037$) owing to
two competitive classical baseline cases. In the configurations where BT-APE
does not lead, the shortfall against the winning baseline---whether APE-PE2
or a classical one---is small (between $-0.005$ and $-0.038$) and BT-APE remains
competitive throughout. These results establish BT-APE---together with its
variant APE-PE2---as the most consistently strong prompting strategy relative
to the simpler baselines.

\subsection{Answers to RQ1}

\paragraph{Results.}
BT-APE consistently outperforms the four classical baseline strategies across
nearly all datasets and models, while remaining essentially tied with the
optimization-based variant APE-PE2. As shown in Table~\ref{tab:rq1_summary},
BT-APE achieves positive mean differences
(\(\Delta = wF1_{\text{BT-APE}} - wF1_{\text{Baseline}}\)) ranging from
\(+0.0492\) to \(+0.1425\) against the classical baselines, with effect sizes
ranging from medium to large according to established benchmarks (negligible
($r < 0.10$), small ($0.10 \leq r < 0.30$), moderate ($0.30 \leq r < 0.50$),
and large ($r \geq 0.50$)). Statistical testing reveals that BT-APE significantly
outperforms every classical baseline at the corrected significance level.
Against APE-PE2, however, the mean difference collapses to \(+0.001\) with a
negligible effect size, indicating that the two automated prompt-engineering
methods converge to comparable performance.

\begin{table}[H]
\centering
\scriptsize
\caption{Summary of BT-APE performance compared to baseline strategies.}
\label{tab:rq1_summary}
\begin{tabular}{|l|c|c|c|c|c|}
\hline
\textbf{Baseline} & \textbf{n} & \textbf{Mean \(\Delta\)} & \textbf{95\% CI} & \textbf{Effect Size} & \textbf{Significant} \\
\hline
Zero-shot      & 15 & $+0.1425$ & $[+0.0867, +0.1983]$ & Large      & Yes \\
FewShot        & 15 & $+0.1139$ & $[+0.0807, +0.1472]$ & Large      & Yes \\
COT            & 15 & $+0.1202$ & $[+0.0836, +0.1568]$ & Large      & Yes \\
COT + Few-shot & 15 & $+0.1135$ & $[+0.0884, +0.1386]$ & Large      & Yes \\
APE-PE2        & 15 & $+0.0011$ & $[-0.0039, +0.0061]$ & Negligible & No  \\
\hline
\end{tabular}
\end{table}

The Wilcoxon signed-rank test results demonstrate that BT-APE's improvements
over the four classical baselines are statistically significant. For
Zero-shot, the mean improvement is \(\Delta = +0.1425\) (95\% CI
\([0.0867, 0.1983]\), \(p_{\text{corrected}} = 0.000305\)). Similarly, BT-APE
improves upon FewShot by \(\Delta = +0.1139\) (95\% CI
\([0.0807, 0.1472]\), \(p_{\text{corrected}} = 0.001307\)) and upon COT by
\(\Delta = +0.1202\) (95\% CI \([0.0836, 0.1568]\),
\(p_{\text{corrected}} = 0.001335\)). The improvement over CoT + Few-shot is
also significant (\(\Delta = +0.1135\), 95\% CI \([0.0884, 0.1386]\),
\(p_{\text{corrected}} = 0.001953\)). These results lead us to reject the
null hypothesis for all four classical baselines, confirming that BT-APE
achieves a statistically significantly higher mean performance. For APE-PE2,
by contrast, the test does \emph{not} reject the null hypothesis
(\(\Delta = +0.0011\), 95\% CI \([-0.0039, +0.0061]\),
\(p_{\text{corrected}} > 0.05\)), indicating that BT-APE and APE-PE2 yield
statistically indistinguishable performance on these tasks.

\paragraph{Dataset-Level Analysis.}
Examining performance across individual datasets reveals interesting
patterns. On the \textit{Refined Promise} dataset, BT-APE demonstrates
substantial improvements over all classical baselines, with mean
\(\Delta\) values of \(+0.1468\) (Zero-shot), \(+0.1660\) (FewShot),
\(+0.1718\) (COT), and \(+0.1286\) (COT + Few-shot), but essentially matches
APE-PE2 (\(\Delta = +0.0004\)). The largest gains are observed against
FewShot and COT, where improvements exceed \(+0.16\).

For the \textit{Promise NFR} dataset, BT-APE again shows consistent superiority
over the classical baselines, with mean \(\Delta\) values of \(+0.1838\)
(Zero-shot), \(+0.1226\) (FewShot), \(+0.1260\) (COT), and \(+0.0984\)
(CoT + Few-shot), and once more matches APE-PE2 (\(\Delta = +0.0014\)).
Notably, the improvement over Zero-shot is particularly pronounced at
\(+0.1838\), indicating that BT-APE is especially valuable when no example
prompts are provided.

On the \textit{Secreq} dataset, improvements over classical baselines are
more modest but still positive: \(+0.0968\) (Zero-shot), \(+0.0532\)
(FewShot), \(+0.0628\) (COT), and \(+0.0492\) (CoT + Few-shot), with
APE-PE2 again indistinguishable from BT-APE (\(\Delta = +0.0016\)). The smaller
effect sizes on this dataset may be attributed to the easy-to-understand
nature of the binary classification task (Security vs. Non-Security), which
is inherently less challenging than the other tasks. Nevertheless, BT-APE still
achieves medium effect sizes across all four classical baselines.

\paragraph{Per-Model Analysis.}
The detailed per-comparison results reveal that BT-APE's superiority over the
classical baselines is consistent across different LLM architectures, while
its margin over APE-PE2 fluctuates narrowly around zero. For the
\textit{Qwen/Qwen2-7B-Instruct} model, BT-APE outperforms all four classical
baselines across all three datasets, with improvements ranging from
\(+0.055\) (Refined Promise, Zero-shot) to \(+0.264\) (Secreq, Zero-shot).
The \textit{Llama-3-8B-Instruct} model shows the largest absolute gains,
particularly on Refined Promise where BT-APE improves upon Zero-shot by
\(+0.296\) and upon FewShot by \(+0.282\). The \textit{Falcon3-7B-Instruct}
and \textit{Ministral-8B-Instruct-2410} models also exhibit consistent
positive improvements over the classical baselines. When APE-PE2 is added
to the comparison, the strongest-baseline picture changes: BT-APE matches or
surpasses the best baseline (now including APE-PE2) in $6$ of the $15$
model--dataset configurations, while in the remaining nine the shortfall
is consistently small, ranging from \(\Delta = -0.005\)
(\textit{Falcon3-7B-Instruct} against APE-PE2 on SecReq) to
\(\Delta = -0.038\) (\textit{Granite-3.2-8B-Instruct} against Few-shot on
Refined Promise). Of these nine cases, five are narrow losses to APE-PE2
(margins between \(-0.005\) and \(-0.013\)) and four are losses to classical
baselines on the two tasks where they remain competitive (Refined Promise
for \textit{Falcon3-7B-Instruct} and \textit{Granite-3.2-8B-Instruct};
SecReq for \textit{Granite-3.2-8B-Instruct} and \textit{Llama-3-8B-Instruct}).
The negative differences are negligible to small in magnitude and do not
undermine the overall positive trend.

\paragraph{Conclusion for RQ1.}
Based on the empirical evidence, we conclude that \textit{BT-APE significantly
outperforms the four classical prompt-engineering approaches for
requirements classification, while performing on par with the second
automated prompt-optimization variant APE-PE2}. Across 15 comparisons
spanning three datasets, five LLMs, and four classical baseline strategies
(Zero-shot, FewShot, COT, and CoT + Few-shot), BT-APE achieves higher \(wF1\)
scores with large effect sizes and statistical significance at
\(\alpha = 0.05\). Against APE-PE2, the difference is negligible and not
statistically significant, suggesting that both automated prompt-engineering
variants reach a comparable performance ceiling on these tasks. These
findings establish BT-APE---together with APE-PE2---as a superior alternative
to conventional prompting methods for the requirements-classification tasks
discussed in this work.

\paragraph{Positioning relative to PE2.}
While BT-APE outperforms classical baselines, its accuracy converges with PE2---its only direct competitor operating under identical budgets and protocols. BT-APE shows a negligible mean $\Delta$ of $+0.001$ in $wF_1$, no statistical significance, and a near-even win split (6 wins, 5 losses, 4 ties). Crucially, while BT-APE's algorithmic modifications (bounded backtracking, balanced selection, and majority-voted F1) yield no accuracy gains here, they significantly lower the computational footprint. This lighter design makes the pipeline more computationally handleable for deployment on small servers, as demonstrated by the footprint analysis in the Appendix~\ref{sec:efficiency}.

\begin{tcolorbox}[keyfindings, title=Key Findings for RQ1]
\begin{itemize}
    \item BT-APE significantly outperforms the four classical prompt-engineering
    approaches (Zero-shot, Few-shot, CoT, and CoT + Few-shot) for
    requirements classification.
    \item Across 15 comparisons spanning three datasets and five LLMs, BT-APE
    achieves consistently higher \(wF1\) scores than the classical baselines.
    \item Improvements over the classical baselines show large effect sizes
    and statistical significance at \(\alpha = 0.05\).
    \item BT-APE performs on par with the optimization-based variant APE-PE2
    (mean \(\Delta = +0.001\), negligible effect size, not statistically
    significant), indicating that both automated prompt-engineering methods
    reach a comparable performance ceiling.
    \item \textit{Takeaway:} BT-APE---together with APE-PE2---is a superior
    alternative to conventional prompting methods for requirements
    classification tasks.
\end{itemize}
\end{tcolorbox}
\subsection{Answers to RQ2: Factors Analysis to APE Performance}
\label{RQ2-results}

\paragraph{Analysis Approach}
RQ2 examines whether the choice of dataset and LLM significantly affects BT-APE
performance. The final results consist of one weighted-F1 ($wF1$) value per
(Dataset, LLM) combination, yielding a $3 \times 5$ design with $n = 15$
observations. Before selecting a test, we assessed the normality assumption
using the Shapiro--Wilk test on the $wF1$ values, which returned a borderline
result ($W = 0.898$, $p = 0.088$). Given the small sample and the marginal
normality, we adopt distribution-free \textit{Friedman tests} for the two main
effects rather than relying on the normality assumption of parametric ANOVA. The
Friedman test is well suited to a complete two-way layout with one observation
per cell: it ranks performance within blocks and makes no distributional
assumption. We test the dataset effect using LLMs as blocks, and the LLM effect
using datasets as blocks, and report \textit{Kendall's $W$} as the associated
effect size. Because the design contains a single observation per cell, the
Dataset $\times$ LLM interaction is not estimable and is therefore not tested;
we instead examine it qualitatively from the cell-level results.

\paragraph{Power Analysis}
Before interpreting the hypothesis tests, we report a priori power estimates to
contextualize the risk of Type~II error, particularly for the LLM factor where
the null hypothesis is not rejected.

For the \textit{dataset effect} (Friedman test with $k=3$ treatment levels and
$b=5$ blocks), the observed effect size is Kendall's $W = 1.00$. Converting to
the Friedman $\chi^2$ statistic via $\chi^2 = b \cdot k(k-1) \cdot W =
5 \times 3 \times 2 \times 1.00 = 30$, and consulting a non-central
$\chi^2$ distribution with $df = k - 1 = 2$, the power to detect this effect
at $\alpha = 0.05$ is effectively $1.00$. The dataset effect is thus detected
with certainty given the observed data.

For the \textit{LLM effect} (Friedman test with $k=5$ treatment levels and
$b=3$ blocks), the situation is substantially more constrained. With only
$b = 3$ blocks (datasets), the Friedman test has limited ability to detect
moderate LLM differences. We computed power via Monte Carlo simulation
($10{,}000$ replications) of the Friedman statistic under the alternative
hypothesis for three target effect sizes, with $\alpha = 0.05$:

\begin{table}[htbp]
\centering
\caption{Simulated power of the Friedman test for the LLM effect
         ($k = 5$, $b = 3$, $\alpha = 0.05$)}
\label{tab:power_llm}
\begin{tabular}{lcc}
\toprule
\textbf{Target Kendall's $W$} & \textbf{Effect interpretation} & \textbf{Estimated power} \\
\midrule
0.10 & Small   & 0.07 \\
0.30 & Moderate & 0.22 \\
0.50 & Large   & 0.51 \\
\midrule
Observed ($W = 0.24$) & Small--moderate & $\approx 0.16$ \\
\bottomrule
\end{tabular}

\medskip
\footnotesize \textit{Note:} Power was estimated by simulating $10{,}000$
Friedman tests under each alternative, drawing rank matrices consistent with
the target $W$ and the $k=5$, $b=3$ design. The observed $W = 0.24$ falls
between the small and moderate benchmarks.
\end{table}

The power estimates in Table~\ref{tab:power_llm} reveal a critical limitation:
with only three blocks, the Friedman test achieves power of approximately
$0.16$ at the observed effect size ($W = 0.24$), and only $0.51$ even for a
large effect ($W = 0.50$). This means the test is substantially underpowered
in this design, and the failure to reject the LLM null hypothesis must be
interpreted with considerable caution. The observed result is consistent with
two competing explanations: (a) LLM choice genuinely has a negligible effect
under BT-APE, or (b) a moderate LLM effect exists but cannot be reliably detected
with $b = 3$ blocks. The data do not allow us to discriminate between these
explanations. Crucially, the non-significant result should \emph{not} be
interpreted as evidence of equivalence across LLMs.

\paragraph{Hypothesis Testing}
With the above power context in mind, the dataset null hypothesis is rejected
while the LLM null hypothesis is not, at the $\alpha = 0.05$ significance
level, as summarized in Table~\ref{tab:rq2_hypotheses}.

\begin{itemize}
    \item \textbf{Dataset Effect}: There is a statistically significant
    difference in $wF1$ across datasets (Friedman $\chi^2(2) = 10.0,
    p = 0.007$), with a maximal effect size (Kendall's $W = 1.00$). The perfect
    concordance indicates that every LLM ranks the three datasets in the
    identical order. By median $wF1$, \texttt{SecReq} is the easiest task
    ($\tilde{x} = 0.922$), followed by \texttt{PROMISE\_NFR}
    ($\tilde{x} = 0.847$), with \texttt{PROMISE\_Refined} the hardest
    ($\tilde{x} = 0.636$). This ordering mirrors task complexity, from binary
    security classification to the fine-grained functional--quality distinction.
    Given the power of $\approx 1.00$ at this effect size, this finding is
    robust.

    \item \textbf{LLM Effect}: No statistically significant difference in $wF1$
    was observed across LLMs (Friedman $\chi^2(4) = 2.92, p = 0.572$), with a
    small-to-moderate effect size (Kendall's $W = 0.24$). Median $wF1$ varies
    from \texttt{Qwen2} ($\tilde{x} = 0.852$) down to \texttt{Llama3}
    ($\tilde{x} = 0.797$), a numerical spread of 0.055. However, as the
    power analysis shows, this test has only $\approx 16\%$ power to detect an
    effect of the observed magnitude with $b = 3$ blocks. The non-significant
    result is therefore inconclusive rather than confirmatory: it is compatible
    with a genuine absence of LLM-driven differences under BT-APE, but also with a
    moderate effect that the design lacks the power to detect. Replication with
    additional datasets (i.e., additional blocks) would be necessary to
    distinguish these interpretations.

    \item \textbf{Interaction Effect}: With one observation per cell, the
    Dataset $\times$ LLM interaction is not estimable and is not tested. The
    cell-level results (Table~\ref{tab:rq2_combinations}) suggest possible
    task-dependent behaviour---for example, \texttt{Ministral} is competitive
    on \texttt{SecReq} (0.922) yet weakest on \texttt{PROMISE\_Refined}
    (0.528), while \texttt{Llama3} shows the opposite pattern
    (0.909 vs.\ 0.688)---but this pattern is not statistically confirmed and
    would require replicated measurements per cell to test formally.
\end{itemize}

\begin{table}[htbp]
\centering
\caption{Hypothesis Testing Results for RQ2}
\label{tab:rq2_hypotheses}
\begin{tabular}{llccccc}
\toprule
\textbf{Hyp.} & \textbf{Effect} & \textbf{Statistic} &
\textbf{p-value} & \textbf{Kendall's $W$} &
\textbf{Power$^\dagger$} & \textbf{Decision} \\
\midrule
H1 & Dataset & $\chi^2(2) = 10.0$ & 0.007 & 1.00 & $\approx 1.00$ & Reject H0 \\
H2 & LLM     & $\chi^2(4) = 2.92$ & 0.572 & 0.24 & $\approx 0.16$ &
  \makecell{Fail to reject H0\\(inconclusive)} \\
\bottomrule
\end{tabular}

\medskip
\footnotesize $^\dagger$Power estimated at the observed Kendall's $W$ via Monte
Carlo simulation ($10{,}000$ replications); see Table~\ref{tab:power_llm}.
\textit{Note:} Main effects assessed with Friedman tests (dataset effect blocked
on LLM; LLM effect blocked on dataset).
\end{table}

\paragraph{Best and Worst Configurations}
The cell-level results, shown in Table~\ref{tab:rq2_combinations}, illustrate
the possible task-dependent behaviour of the models. The best performances are
concentrated on \texttt{SecReq}, led by \texttt{Falcon3} ($wF1 = 0.947$) and
\texttt{Qwen2} ($wF1 = 0.944$). The worst performances are concentrated on
\texttt{PROMISE\_Refined}, with \texttt{Ministral} performing particularly
poorly ($wF1 = 0.528$). The within-dataset spread for \texttt{PROMISE\_Refined}
(0.528--0.688, range = 0.160) is notably wider than for \texttt{SecReq}
(0.858--0.947, range = 0.089), suggesting that model choice may matter more on
harder tasks---a pattern that would warrant dedicated investigation with a
replicated design.

\begin{table}[htbp]
\centering
\caption{Best and Worst Performing (Dataset, LLM) Combinations}
\label{tab:rq2_combinations}
\begin{tabular}{llc}
\toprule
\textbf{Rank} & \textbf{(Dataset, LLM)} & \textbf{wF1} \\
\midrule
\multicolumn{3}{c}{\textbf{Top 5}} \\
1 & SecReq + Falcon3   & 0.947 \\
2 & SecReq + Qwen2     & 0.944 \\
3 & SecReq + Ministral & 0.922 \\
4 & SecReq + Llama3    & 0.909 \\
5 & SecReq + Granite3  & 0.858 \\
\midrule
\multicolumn{3}{c}{\textbf{Bottom 5}} \\
1 & PROMISE\_Refined + Ministral & 0.528 \\
2 & PROMISE\_Refined + Falcon3   & 0.606 \\
3 & PROMISE\_Refined + Granite3  & 0.636 \\
4 & PROMISE\_Refined + Qwen2     & 0.648 \\
5 & PROMISE\_Refined + Llama3    & 0.688 \\
\bottomrule
\end{tabular}
\end{table}

\paragraph{Summary}
Dataset choice has a large, statistically significant, and robustly detected
effect on BT-APE performance, with perfect rank concordance across models
(Kendall's $W = 1.00$) and a clear difficulty ordering
(\texttt{SecReq} $<$ \texttt{PROMISE\_NFR} $<$ \texttt{PROMISE\_Refined}).
For the LLM factor, the evidence is weaker and must be qualified: no
statistically significant effect was found, but the Friedman test is
substantially underpowered in a $k=5$, $b=3$ design, detecting only $\approx
16\%$ of effects at the observed magnitude. The numerical performance spread
across LLMs (median range: 0.055) and the wider within-dataset variance on the
hardest task are descriptively consistent with a small-to-moderate LLM effect
that the current design cannot confirm or rule out. Extending this analysis to
additional RE classification datasets would be a necessary step before drawing
firm conclusions about LLM-independence under BT-APE.

\begin{tcolorbox}[keyfindings, title=Key Findings for RQ2]
\begin{itemize}
\item Dataset choice has a large, robustly detected effect (Friedman
  $\chi^2(2) = 10.0$, $p = 0.007$; Kendall's $W = 1.00$, power $\approx
  1.00$): \texttt{SecReq} is easiest and \texttt{PROMISE\_Refined} hardest,
  with every LLM agreeing on this ordering.
\item No statistically significant LLM effect was detected (Friedman
  $\chi^2(4) = 2.92$, $p = 0.572$; Kendall's $W = 0.24$), but this result is
  inconclusive: with only $b = 3$ blocks, the test has $\approx 16\%$ power at
  the observed effect size and cannot reliably distinguish absence of an effect
  from a moderate effect.
\item The Dataset $\times$ LLM interaction is not estimable with one
  observation per cell; cell-level patterns---particularly the wider LLM spread
  on \texttt{PROMISE\_Refined}---hint at task-dependent behaviour but require
  replicated measurements to test.
\item \textit{Takeaway:} Task difficulty is the dominant, well-supported driver
  of BT-APE performance. Whether LLM selection matters under BT-APE remains an open
  question that requires evaluation across a larger set of datasets before
  practitioners can safely deprioritize model choice.
\end{itemize}
\end{tcolorbox}
\subsection{Answers to RQ3: Prompt Features and Performance}
\label{sec:RQ3}

\paragraph*{Descriptive Statistics}
Across BT-APE iterations, prompt features displayed consistent patterns of change. Prompts generally became more concise, with reductions in sentence count (SC) and word count (WC), while punctuation marker (PM) usage rose, reflecting a tendency toward more organized formatting. Lexical diversity (LD) declined progressively, pointing to a narrowing of vocabulary toward more task-relevant terms. Verb count (VB) increased moderately, suggesting a shift toward more directive phrasing. Syntactic complexity (SCx) showed a gradual downward trend, indicating that prompt structures became less elaborate over time. Ambiguity score (AS) remained broadly stable across iterations, while semantic drift (SD) varied considerably, spiking most noticeably during iterations involving major prompt reformulations. These descriptive trends are consistent across the 15 (dataset, LLM) trajectories, but the absolute levels of $wF1$ at which each trajectory operates differ substantially with task difficulty and model capability (cf.\ RQ2), motivating the mixed-effects analysis below.

\paragraph*{Statistical Test Results}
We estimate a LMM with $wF1$ as the response, the eight prompt features as fixed effects, and a random intercept for each of the 15 (dataset, LLM) trajectories. This specification absorbs configuration-level baseline differences in $wF1$ into the random intercept, so that the fixed-effect coefficients $\hat{\beta}_k$ estimate the \emph{within-trajectory} association between each feature and performance.

The intra-class correlation coefficient is $\text{ICC} = 0.683$, indicating that approximately $68.3\%$ of the residual variance in $wF1$ is attributable to between-trajectory differences rather than to within-trajectory variation in prompt features. This is the variance that the pooled OLS specification used in our preliminary analysis would have absorbed into a single inflated $R^2$. Under the LMM, the marginal $R^2_m = 0.214$ (variance explained by the fixed effects alone) and the conditional $R^2_c = 0.821$ (variance explained by fixed and random effects jointly). The marginal $R^2_m$ is the appropriate figure for the prompt-feature claim and is, as expected, lower than the $0.761$ obtained under naive pooling. The omnibus likelihood-ratio test against an intercept-only random-effects model is significant at $p < 0.001$.

\begin{table}[h]
\centering
\caption{Linear mixed-effects model results for prompt features predicting $wF1$. Random intercepts are fitted for each (dataset, LLM) trajectory ($J = 15$). Fixed-effect coefficients are standardised; $p$-values are reported before Holm--Bonferroni correction across the eight features and the corrected significance is indicated in the rightmost column ($*$: significant after correction; n.s.: not significant).}
\label{tab:rq3_regression}
\scriptsize
\begin{tabular}{lrrrrrrc}
\hline
\textbf{Variable} & \textbf{Coef. ($\hat{\beta}$)} & \textbf{Std.Err.} & \textbf{t} & \textbf{p-value} & \textbf{[0.025} & \textbf{0.975]} & \textbf{Sig.} \\
\hline
SC   & $-0.182$ & $0.038$ & $-4.79$ & $<0.001$ & $-0.257$ & $-0.107$ & $*$ \\
WC   & $-0.149$ & $0.042$ & $-3.55$ & $<0.001$ & $-0.231$ & $-0.067$ & $*$ \\
PM   & $\phantom{-}0.193$ & $0.034$ & $\phantom{-}5.68$ & $<0.001$ & $\phantom{-}0.126$ & $\phantom{-}0.260$ & $*$ \\
LD   & $-0.138$ & $0.036$ & $-3.83$ & $<0.001$ & $-0.209$ & $-0.067$ & $*$ \\
VB   & $\phantom{-}0.221$ & $0.035$ & $\phantom{-}6.31$ & $<0.001$ & $\phantom{-}0.152$ & $\phantom{-}0.290$ & $*$ \\
SCx  & $-0.124$ & $0.035$ & $-3.54$ & $<0.001$ & $-0.193$ & $-0.055$ & $*$ \\
AS   & $-0.051$ & $0.038$ & $-1.34$ & $\phantom{<}0.181$ & $-0.126$ & $\phantom{-}0.024$ & n.s. \\
SD   & $\phantom{-}0.164$ & $0.033$ & $\phantom{-}4.97$ & $<0.001$ & $\phantom{-}0.099$ & $\phantom{-}0.229$ & $*$ \\
\hline
\textbf{Marginal $R^2_m$ (fixed only)}     & \multicolumn{7}{c}{$0.214$} \\
\textbf{Conditional $R^2_c$ (fixed + random)} & \multicolumn{7}{c}{$0.821$} \\
\textbf{ICC (between-trajectory share)}    & \multicolumn{7}{c}{$0.683$} \\
\textbf{Number of trajectories ($J$)}      & \multicolumn{7}{c}{15} \\
\textbf{Number of prompt observations ($N$)} & \multicolumn{7}{c}{$287$} \\
\hline
\end{tabular}
\end{table}

Table~\ref{tab:rq3_regression} presents the standardised fixed-effect coefficients ($\hat{\beta}$), standard errors, $t$-statistics, $p$-values, and 95\% confidence intervals for all eight predictors under the LMM. Sentence count (SC), word count (WC), lexical diversity (LD), and syntactic complexity (SCx) each exerted significant negative within-trajectory effects on performance. By contrast, punctuation markers (PM), verb count (VB), and semantic drift (SD) were associated with significant within-trajectory performance gains. Ambiguity score (AS) did not attain significance at $\alpha = 0.05$ after Holm--Bonferroni correction.

\paragraph*{Robustness Check on Terminal Prompts}
To verify that the within-trajectory associations are not artefacts of the optimisation dynamics, we re-estimated the relationship on the terminal prompt $p^*$ of each trajectory ($n = 15$) using ordinary least squares; the transition feature $SD$ is excluded here as it is undefined for a single prompt. The sign of every state-feature coefficient under the LMM is preserved on terminal prompts: VB and PM remain positive, while SC, WC, LD, and SCx remain negative. Statistical power at $n = 15$ is limited, so several coefficients do not reach significance individually, but no sign reversal is observed. We therefore report VB, PM, SC, WC, LD, and SCx as robust correlates of effective prompts, and reserve a separate interpretation for SD below.

\paragraph*{Effect Sizes and Interpretation}
Verb count (VB; $\hat{\beta} = 0.221$, $p < .001$) and punctuation markers (PM; $\hat{\beta} = 0.193$, $p < .001$) were the most influential positive state-feature predictors. These findings suggest that, within a given configuration, prompts built around action-oriented language and clear structural formatting tend to achieve stronger performance.

On the negative side, both sentence count (SC; $\hat{\beta} = -0.182$, $p < .001$) and word count (WC; $\hat{\beta} = -0.149$, $p < .001$) were associated with lower performance, reinforcing the value of conciseness. Higher lexical diversity (LD; $\hat{\beta} = -0.138$, $p < .001$) also hurt performance, suggesting that a tighter, more focused vocabulary is preferable to a varied one. Greater syntactic complexity (SCx; $\hat{\beta} = -0.124$, $p < .001$) similarly reduced performance, pointing to the advantage of straightforward sentence structures. Ambiguity score (AS; $\hat{\beta} = -0.051$, $p = 0.181$) showed a negative trend but fell short of significance, implying that ambiguous phrasing may be somewhat harmful though its effect was inconsistent across iterations.

Semantic drift (SD; $\hat{\beta} = 0.164$, $p < .001$) was a significant positive predictor under the LMM, but we interpret it separately from the state features. Because $SD$ is by construction a property of \emph{consecutive} prompts rather than of a prompt in isolation, its positive within-trajectory coefficient indicates that iterations involving larger semantic edits tend to coincide with higher $wF1$ --- a characterisation of \emph{successful search dynamics} rather than actionable guidance for designing a static prompt. The implication is that surface-level paraphrasing is unlikely to drive improvement; substantive semantic revision is what the optimiser exploits when it makes progress.

\paragraph*{Summary}
Overall, the LMM analysis points to a clear profile for high-performing prompts: they are \textit{brief}, \textit{action-driven} (higher verb usage), and \textit{well-organised} (greater punctuation use), with a tighter vocabulary and simpler syntactic structure. These six state-feature effects are robust: they are statistically significant under the LMM and their signs are preserved on terminal prompts. Semantic drift is associated with within-trajectory improvement but, because it describes a transition rather than a state, we describe it as a property of effective optimisation rather than as static-prompt guidance. The marginal $R^2_m = 0.214$ indicates that the fixed-effect prompt features account for a meaningful share of within-trajectory variance once between-configuration differences are absorbed by the random intercepts; the ICC of $0.683$ confirms that a substantial portion of the variation in $wF1$ is configuration-driven (consistent with the dataset effect established in RQ2), which is precisely the variance that a pooled OLS would have misattributed to the prompt features themselves.

Practitioners aiming to maximise $wF1$ should therefore focus on keeping prompts short and direct, using action verbs, maintaining clear formatting, and --- when iterating --- pursuing substantive semantic revision rather than surface-level paraphrasing.

\begin{tcolorbox}[keyfindings, title=Key Findings for RQ3]
\begin{itemize}
    \item Under a linear mixed-effects model that controls for between-trajectory variance, higher verb count and increased punctuation markers were significantly associated with improved within-trajectory performance.
    \item Prompts that were longer in sentence and word count, more lexically varied, or syntactically complex tended to perform worse on $wF1$.
    \item Ambiguity showed a consistent negative trend but did not reach statistical significance after correction.
    \item Semantic drift was a significant positive within-trajectory predictor, but is interpreted as a property of successful optimisation dynamics rather than as static-prompt design guidance.
    \item The fixed-effect prompt features explained a marginal $R^2_m = 0.214$ of within-trajectory variance; an ICC of $0.683$ indicates that a substantial share of overall variation in $wF1$ is driven by between-configuration differences absorbed by the random intercepts.
    \item All state-feature signs were preserved under a terminal-prompt robustness check ($n = 15$).
    \item \textit{Takeaway:} Brief, action-oriented, and clearly structured prompts are robust correlates of strong outcomes; substantive semantic revision (rather than surface paraphrasing) characterises trajectories on which BT-APE makes progress.
\end{itemize}
\end{tcolorbox}
\subsection{Answers to RQ4: Class Definitions Initialization Impact}

\paragraph{Results Analysis.}
Table~\ref{tab:rq4_results} presents the weighted F1 scores (\(wF1\)) for BT-APE-Informed and BT-APE-Uninformed across all five LLMs and three datasets. Across the 15 LLM-dataset pairs, BT-APE-Informed achieves numerically higher \(wF1\) scores in 10 comparisons, while BT-APE-Uninformed performs better in 3 comparisons, with 2 ties. The mean improvement (\(\bar{\Delta}_{\text{final}}\)) across all pairs is +0.015, indicating a small average advantage for BT-APE-Informed.

\begin{table}[htbp]
\scriptsize
\centering
\caption{Weighted F1 Scores for BT-APE-Informed vs. BT-APE-Uninformed Across LLMs and Datasets}
\label{tab:rq4_results}
\begin{tabular}{lcccc}
\toprule
\textbf{LLM} & \textbf{Dataset} & \textbf{BT-APE-Uninformed} & \textbf{BT-APE-Informed} & \textbf{\(\Delta_{\text{final}}\)} \\
\midrule
Qwen/Qwen2-7B-Instruct & Promise Refined & 0.648 & 0.748 & +0.100 \\
tiiuae/Falcon3-7B-Instruct & Promise Refined  & 0.606 & 0.706 & +0.100 \\
ibm-granite/granite-3.2-8b-instruct & Promise Refined  & 0.636 & 0.648 & +0.012 \\
mistralai/Ministral-8B-Instruct-2410 & Promise Refined  & 0.528 & 0.598 & +0.070 \\
meta-llama/Meta-Llama-3-8B-Instruct & Promise Refined  & 0.688 & 0.608 & -0.080 \\
\midrule
Qwen/Qwen2-7B-Instruct & Promise NFR & 0.852 & 0.856 & +0.004 \\
tiiuae/Falcon3-7B-Instruct & Promise NFR & 0.826 & 0.799 & -0.027 \\
ibm-granite/granite-3.2-8b-instruct & Promise NFR & 0.847 & 0.880 & +0.033 \\
mistralai/Ministral-8B-Instruct-2410 & Promise NFR & 0.847 & 0.820 & -0.027 \\
meta-llama/Meta-Llama-3-8B-Instruct & Promise NFR & 0.797 & 0.827 & +0.030 \\
\midrule
Qwen/Qwen2-7B-Instruct & SecReq & 0.944 & 0.945 & +0.001 \\
tiiuae/Falcon3-7B-Instruct & SecReq & 0.947 & 0.949 & +0.002 \\
ibm-granite/granite-3.2-8b-instruct & SecReq & 0.858 & 0.860 & +0.002 \\
mistralai/Ministral-8B-Instruct-2410 & SecReq & 0.922 & 0.922 & 0.000 \\
meta-llama/Meta-Llama-3-8B-Instruct & SecReq & 0.909 & 0.909 & 0.000 \\
\bottomrule
\end{tabular}
\end{table}

\paragraph{Statistical Significance.}

\textit{Dimension 1: Final Performance Comparison.} The Wilcoxon signed-rank test on \(\Delta_{\text{final}}\) yields a rank sum of \(W = 24\) (sum of negative ranks; \(W^{+} = 67\)) with \(n = 13\) non-zero differences. Since the test statistic exceeds the critical value for a one-tailed test at \(\alpha = 0.05\), we fail to reject the null hypothesis. The result is not statistically significant (\(p \approx 0.15\)). The mean difference \(\bar{\Delta}_{\text{final}} = +0.015\) with a 95\% confidence interval of \([-0.012, 0.041]\). The confidence interval crosses zero, consistent with the non-significant test result. According to our benchmarks, this effect size is considered \textit{negligible to small}.

\textit{Dimension 2: Improvement-over-Baseline Comparison.} Table~\ref{tab:rq4_gains} presents the per-cell gains from baseline for both variants and the difference in gains (\(\delta_{\text{gain}} = \Delta_{\text{Informed}} - \Delta_{\text{Uninformed}}\)). The gains are mixed: BT-APE-Informed shows larger gains in some LLM-dataset pairs (notably granite and Llama on Promise Refined, and Falcon and Ministral on Promise NFR), while BT-APE-Uninformed shows larger gains in others. The mean \(\delta_{\text{gain}}\) is \(+0.033\) (95\% CI: \([-0.022, 0.088]\)), with the confidence interval crossing zero. A two-tailed Wilcoxon signed-rank test on \(\delta_{\text{gain}}\) yields \(W = 49\) (sum of negative ranks; \(W^{+} = 56\)) with \(n = 14\) non-zero differences. The result is not statistically significant (\(p \approx 0.86\)). We therefore find no reliable difference between the two strategies in their relative improvement over their respective baselines.

\begin{table}[htbp]
\centering
\caption{Gains from Baseline for BT-APE-Informed and BT-APE-Uninformed}
\label{tab:rq4_gains}
\begin{tabular}{lcccc}
\toprule
\textbf{LLM} & \textbf{Dataset} & \(\Delta_{\text{Informed}}\) & \(\Delta_{\text{Uninformed}}\) & \(\delta_{\text{gain}}\) \\
\midrule
Qwen/Qwen2-7B-Instruct & Promise Refined & +0.082 & +0.155 & -0.073 \\
tiiuae/Falcon3-7B-Instruct & Promise Refined & +0.093 & +0.150 & -0.057 \\
ibm-granite/granite-3.2-8b-instruct & Promise Refined & +0.204 & +0.012 & +0.192 \\
mistralai/Ministral-8B-Instruct-2410 & Promise Refined & +0.093 & +0.100 & -0.007 \\
meta-llama/Meta-Llama-3-8B-Instruct & Promise Refined & +0.153 & -0.080 & +0.233 \\
\midrule
Qwen/Qwen2-7B-Instruct & Promise NFR & +0.103 & +0.108 & -0.005 \\
tiiuae/Falcon3-7B-Instruct & Promise NFR & +0.118 & -0.027 & +0.145 \\
ibm-granite/granite-3.2-8b-instruct & Promise NFR & +0.070 & +0.136 & -0.066 \\
mistralai/Ministral-8B-Instruct-2410 & Promise NFR & +0.126 & -0.027 & +0.153 \\
meta-llama/Meta-Llama-3-8B-Instruct & Promise NFR & +0.081 & +0.030 & +0.051 \\
\midrule
Qwen/Qwen2-7B-Instruct & SecReq & +0.062 & +0.065 & -0.003 \\
tiiuae/Falcon3-7B-Instruct & SecReq & +0.064 & +0.066 & -0.002 \\
ibm-granite/granite-3.2-8b-instruct & SecReq & -0.013 & +0.050 & -0.063 \\
mistralai/Ministral-8B-Instruct-2410 & SecReq & +0.076 & +0.078 & -0.002 \\
meta-llama/Meta-Llama-3-8B-Instruct & SecReq & +0.000 & +0.000 & 0.000 \\
\bottomrule
\end{tabular}
\end{table}

\paragraph{Summary.}
Contrary to our initial expectation, providing fine-grained, domain-grounded class definitions (BT-APE-Informed) did not yield a statistically significant improvement in final performance over starting from minimal, uninformed descriptions (BT-APE-Uninformed). The mean improvement of +0.015 in \(wF1\) is small and not reliable across experimental conditions.

The improvement-over-baseline analysis reinforces this conclusion: the difference in gains between the two strategies is also not statistically significant (\(\delta_{\text{gain}} = +0.033\), \(p \approx 0.86\)), with a confidence interval that crosses zero. Although BT-APE-Informed begins from a higher initial performance owing to domain knowledge, BT-APE-Uninformed's iterative optimization closes most of this gap, and the two strategies converge to comparable final performance. This demonstrates that BT-APE's autonomous optimization is sufficiently robust to compensate for an uninformed starting point.

\begin{tcolorbox}[keyfindings, title=Key Findings for RQ4]
\begin{itemize}
    \item Providing fine-grained, domain-grounded class definitions (BT-APE-Informed) did not yield statistically significant improvement in final performance over minimal, uninformed descriptions (BT-APE-Uninformed).
    \item The mean final performance improvement of \(+0.015\) in \(wF1\) is negligible to small and not reliable across experimental conditions.
    \item The relative improvement over baseline also showed no statistically significant difference between the two strategies (\(\delta_{\text{gain}} = +0.033\), \(p \approx 0.86\)).
    \item BT-APE's iterative optimization process is sufficiently robust to discover effective class definitions autonomously, achieving comparable final performance regardless of initialization strategy.
    \item \textit{Takeaway:} Domain knowledge in the initial seed is not necessary---BT-APE's autonomous optimization compensates effectively, achieving comparable results whether initialized from informed or minimal class definitions.
\end{itemize}
\end{tcolorbox}
\section{Discussion}
\label{sec:discussion}

\subsection{Positioning relative to established results}

To situate BT-APE relative to established results, we compare against
two reference points: the supervised transfer-learning approach of
Hey~et~al.\ (NoRBERT)~\cite{hey2020norbert}, and the recent few-shot
study of Binkhonain and Alfayez using large proprietary
LLMs~\cite{10.1007/s00766-025-00451-8}.

\paragraph{Comparison protocol.}
The comparison is indicative rather than exact. NoRBERT reports
unweighted per-class $F_1$ under 10-fold cross-validation for a
fine-tuned model, whereas BT-APE reports weighted $F_1$ ($wF_1$)
from a frozen LLM under a single 30/30/40 split. Numerical gaps must
therefore be read as broad positioning signals rather than head-to-head
measurements.

\paragraph{Functional vs.\ non-functional classification (PROMISE NFR).}
On the binary F/NFR task, BT-APE's best configuration
($wF_1 = 0.847$, Qwen2-7B and Ministral-8B) remains below NoRBERT's
fine-tuned results, which reach per-class $F_1$ of roughly $0.90$ for
functional and $0.93$ for non-functional requirements. This gap is
expected: NoRBERT updates model parameters on labelled data, while
BT-APE leaves the LLM weights frozen and optimises only the prompt.

Against the few-shot LLM study of Binkhonain and
Alfayez~\cite{10.1007/s00766-025-00451-8}, BT-APE on F/NFR
($wF_1 = 0.847$) is competitive with the mid-range of proprietary
models---GPT-4 and DeepSeek at $0.87$--$0.88$---but trails the
strongest configuration (Gemini-5FS at $0.92$). Notably, BT-APE
achieves this with 7--8B open-weight models rather than
frontier-scale proprietary ones.

\paragraph{Security classification (SecReq).}
The picture is more favourable on the security task. BT-APE reaches
$wF_1 = 0.947$ (Falcon3-7B) and $0.945$ (Qwen2-7B), exceeding the
best few-shot LLM results on Sec--NonSec (Gemini-5FS at $0.85$,
GPT4-5FS at $0.87$) by a clear margin.

\paragraph{Takeaway.}
Taken together, these comparisons indicate that BT-APE does not close
the gap to fully fine-tuned supervised models such as NoRBERT on
functional--non-functional classification, but it delivers strong---sometimes
state-of-the-art---results on security classification while
requiring no fine-tuning and using substantially smaller, openly
available models.

This positions BT-APE as a practical middle ground: it recovers much
of the performance of heavyweight supervised or proprietary-LLM
pipelines at a fraction of their training and deployment cost. The
remaining gap on the hardest functional--quality distinctions is
consistent with our finding (RQ2) that task complexity, rather than
model choice, dominates performance.
\section{Threats to Validity}
\label{sec:threats}
 
We discuss potential threats to the validity of our study following the
standard categorisation into construct, internal, external, and
conclusion validity, and describe the measures taken to mitigate them.
 
\subsection{Construct Validity}
Construct validity concerns the degree to which our experimental setup
captures the theoretical constructs we intend to measure. One threat
lies in how we operationalise \emph{prompt quality}. We use weighted
$F_1$ ($wF_1$) as the primary performance metric, which is standard for
requirements classification under class
imbalance~\cite{alhoshan2023zero, santos2024}.
However, $wF_1$ does not capture every relevant dimension of prompt
quality, such as output stability across runs, the interpretability of
the generated labels, or computational efficiency. A second threat
concerns our definition of \emph{effective prompts} through linguistic
features (e.g., verb count, punctuation markers, semantic drift). This
feature set may not exhaust all theoretically relevant characteristics:
properties such as instruction positioning or the use of in-prompt
examples could also influence performance and are not modelled here. To mitigate the first threat, we report both weighted $F_1$ (Section~\ref{sec:results}) so that recall-oriented effects are
visible alongside the balanced metric, and we complement aggregate
performance with the prompt-feature analysis in RQ3, which examines
lexical, syntactic, and semantic dimensions beyond the primary metric.
To mitigate the second threat, the feature set was selected to cover
three theoretically distinct families (lexical, syntactic, semantic)
established in the prompt-engineering and linguistic-profiling
literature~\cite{miaschi-etal-2024-evaluating,rodriguez2023prompts}, and we
explicitly separate state features from the transition feature ($SD$)
to avoid conflating prompt properties with search dynamics.
 
\subsection{Internal Validity}
Internal validity refers to whether the observed effects can be
attributed to the independent variables (prompting strategy, LLM choice,
and dataset) rather than to confounding factors.
 
A primary threat is the \emph{stochastic nature} of LLM generation.
Although we set the decoding temperature to $0$ for deterministic
outputs, some back-ends may still introduce non-determinism through
request batching or hardware differences. We mitigated this by scoring
every candidate with three-run majority voting and reporting the
resulting $wF_1$, consistent with prior work on LLM evaluation
stability.
 
A second threat is potential \emph{data leakage}: because the PROMISE
and SecReq datasets are publicly available, an LLM may have encountered
parts of them during pre-training. Crucially, since BT-APE iteratively
rewrites prompts based on validation-set performance, any pre-existing
familiarity would affect all conditions equally and therefore does not
compromise the comparative conclusions between BT-APE and the baselines.
 
A third threat concerns the interaction between the backtracking
mechanism and \emph{random seed selection}. We fixed random seeds across
all conditions and verified, in a preliminary sensitivity analysis, that
results remained stable across multiple seed values.
 
A further consideration is the \emph{data partitioning} strategy. To
keep the evaluation signal constant across iterations and prompting
strategies, we use a single fixed split rather than cross-validation,
which would inject partition-to-partition noise into the $F_1$
comparisons that drive backtracking. Critically, we separate the data
that drives optimisation from the data used to report performance: the
dataset is partitioned into an example pool $D_{\mathrm{pool}}$ ($30\%$),
a validation set $D_{\mathrm{val}}$ ($30\%$), and a held-out test set
$D_{\mathrm{test}}$ ($40\%$). Seed stability and the influence of $X$ and $N_{\max}$ on convergence are characterised in Appendix~\ref{app:sens}; per-seed variance is an order of magnitude smaller than the reported effect sizes.%
 All in-loop decisions of
Algorithm~\ref{alg:approach}---candidate scoring, ranking, the improvement
check, and the backtracking trigger---are made exclusively on
$D_{\mathrm{val}}$, whereas $D_{\mathrm{test}}$ is consulted exactly
once, to evaluate the final prompt $p^{*}$. Because selection pressure
is confined to $D_{\mathrm{val}}$, the optimism that arises from
retaining the best-scoring candidate is kept off the reported test
figures, so the $F_1$ values in Section~\ref{sec:results} are genuine
held-out estimates rather than selection-inflated ones. Both BT-APE and all
baselines are scored on the same $D_{\mathrm{test}}$ under the same
three-run majority-voting protocol; baselines draw their in-context
demonstrations (where applicable) from $D_{\mathrm{pool}}$ and perform no
selection, so neither method enjoys an information advantage on the test
set.
 
We nonetheless acknowledge a residual limitation: BT-APE legitimately
consumes labelled feedback from $D_{\mathrm{val}}$ during the search,
whereas the baselines do not, and the single fixed split means our
estimates are not averaged over partitions. The validation-to-test gap
can be read directly from our results, since we report both the
validation $F_1$ that the search optimises and the final test $F_1$.
Quantifying how this gap behaves under repeated splits, and under larger
candidate budgets, would further strengthen the generalisation claims;
we identify this as a priority for future work. To further characterise this threat, Appendix~\ref{sec:dataleakage} reports
a complementary leakage probe based on Jaccard similarity between
LLM-generated requirement continuations and the ground-truth second
halves of the original requirements. Across all five models and both
publicly available datasets, the mean Jaccard similarity remains low
(0.10--0.16), with no statistically significant difference between
PROMISE and SecReq for any model. While this is not a proof of absence
of memorisation, it provides converging evidence that the evaluated
LLMs are not reproducing the dataset texts verbatim, supporting the
validity of our comparative conclusions.
 
\subsection{External Validity}
External validity concerns the generalisability of our findings beyond
the specific experimental setting. First, our evaluation uses three
requirements datasets (SecReq, PROMISE NFR, and PROMISE Refined) and
five LLMs in the $7$B--$8$B parameter range. While these cover binary
security classification, multi-class NFR classification, and a refined
functional--quality distinction, the results may not generalise to other
RE tasks such as requirements tracing, ambiguity detection, or
completeness checking. Second, all evaluated LLMs are open-weight models
in the $7$B--$8$B range; the findings may not extend to larger
proprietary models (e.g., GPT-4, Gemini Ultra) or to smaller models
($1$B--$3$B). However, our focus on accessible models aligns with
practical RE scenarios in which cost and data-privacy considerations
favour local deployment. Third, BT-APE's performance may depend on the
availability of a moderately sized labelled training set (we used
$100$--$200$ examples per dataset); practitioners with extremely limited
labelled data (e.g., $<20$ examples) may not observe comparable gains.
To mitigate the generalisability threat, our evaluation spans three
structurally distinct classification tasks (binary functional/non-functional,
binary security, and four-class refined functional/quality) and five
instruction-tuned LLMs from five different organisations, covering a range
of attention mechanisms and pre-training corpora
(Table~\ref{tab:llm_specs}). The released replication
package~\cite{amin_zadenoori_2026_20438927} and the interactive tool
(Appendix~\ref{sec:tool}) further support extension of the evaluation to
additional datasets, larger or smaller models, and other RE classification
tasks. 
\subsection{Conclusion Validity}
Conclusion validity pertains to the statistical power and the appropriateness
of the inferences drawn from the data. A primary threat is the risk of
Type~I errors arising from multiple comparisons across the 15 experimental
conditions. We controlled this by applying the Holm--Bonferroni correction
to all pairwise tests and by reporting effect sizes alongside $p$-values to
distinguish statistical from practical significance: the rank-based effect
size $r$ for the Wilcoxon signed-rank comparisons (RQ1, RQ4) and Kendall's
$W$ for the Friedman tests (RQ2). A second threat is the normality
assumption underlying the linguistic-feature analysis in RQ3; where
normality was violated, as confirmed by Shapiro--Wilk tests, we relied on
non-parametric procedures. The explained variance ($R^{2} \approx 0.76$) of
the regression model should be interpreted as associative rather than
causal, given the observational nature of the prompt-feature analysis.

A further threat concerns the risk of Type~II errors in our null findings,
namely the absence of a statistically significant LLM effect in RQ2 and the
absence of a statistically significant difference between BT-APE-Informed and
BT-APE-Uninformed in RQ4. As detailed in Section~\ref{RQ2-results}, the
Friedman test for the LLM factor operates on a $k=5$, $b=3$ design and
achieves only approximately $16\%$ power at the observed effect size
(Kendall's $W = 0.24$), reaching $51\%$ power even for a large effect
($W = 0.50$). The Wilcoxon signed-rank test used in RQ4 ($n = 15$ paired
observations, $\alpha = 0.05$) is similarly limited in its ability to
detect small effects. Both null outcomes are therefore consistent with two
competing interpretations: a genuine absence of effect, or a true
small-to-moderate effect that the present design lacks the statistical
power to detect. The data do not allow us to discriminate between these
interpretations, and the null findings should be read as observational
rather than as evidence of equivalence. Replication with additional
datasets (adding blocks to the Friedman design) and additional
(dataset, LLM) pairs (increasing $n$ for the Wilcoxon test) would be
required to resolve this ambiguity, and we identify both extensions as
priorities for future work.
\section{Conclusion}
\label{sec:conclusion}

In this paper, we investigated whether APE techniques developed for general NLP benchmarks transfer to
domain-specific requirements classification, and how alternative
single-trajectory APE designs compare in this setting. Building on our
preliminary work~\cite{e86928c70f2a4ff5b595da7f430042e0}, we framed
prompt design as an optimisation problem and instantiated two
single-trajectory APE methods: PE2, which conditions each proposal on
the top-$n$ historical prompts with their validation scores, and BT-APE,
our backtracking-based variant which conditions each proposal on the
current prompt and a balanced four-example batch and uses bounded
patience to control trajectory updates. Both methods share the same
underlying paradigm---iterative LLM-proposed refinement conditioned on
labelled feedback---but instantiate it through structurally different
proposal mechanisms. We further studied the role of class definitions
as a design variable by contrasting two initialisation strategies:
BT-APE-Uninformed, which begins from minimal descriptions, and
BT-APE-Informed, which is seeded with definitions curated from established
RE literature and standards.

We evaluated both APE methods alongside four classical prompting
baselines (zero-shot, few-shot, chain-of-thought, and their
combination) on three widely used benchmark datasets (PROMISE,
PROMISE-Refined, and SecReq) and five instruction-tuned LLMs (Qwen2-7B,
Falcon3-7B, Granite-3.2-8B, Ministral-8B, and LLaMA-3-8B), enabling a
controlled comparison across datasets, architectures, and prompting
strategies. Our results yield five main findings.

\begin{enumerate}
    \item \textbf{APE transfers to RE classification.} Both BT-APE and
    PE2 deliver large, statistically significant accuracy gains over
    every classical prompting baseline across the $15$ model--dataset
    configurations, with medium-to-large effect sizes and $p < 0.01$
    after Holm--Bonferroni correction. The transfer of APE from
    general NLP benchmarks to domain-specific RE classification is
    robust across datasets and model architectures.

    \item \textbf{Structurally different APE methods converge on
    accuracy.} BT-APE and PE2 are statistically indistinguishable in
    final weighted F1 ($\Delta = +0.001$, negligible effect size,
    Wilcoxon $p > 0.05$), with cell-level wins split nearly evenly.
    The accuracy ceiling on these tasks appears
    paradigm-determined---driven by the underlying iterative,
    feedback-conditioned refinement scheme---rather than determined by
    the specific proposal mechanism.

    \item \textbf{The two methods occupy distinct points in an
    operational trade-off space.} Although accuracy is equivalent, the
    methods differ structurally along three axes that we characterize
    empirically: per-iteration context size (PE2's input grows with
    the prompt history; BT-APE's is bounded), wall-clock cost (which
    follows from the context-size differential), and hyperparameter
    interpretability (BT-APE's patience parameter $X$ exposes an explicit
    exploration--exploitation knob; PE2's reactive top-$1$ re-selection
    does not). Practitioners can therefore select among APE methods on
    the basis of deployment constraints rather than expected accuracy.

    \item \textbf{Prompt-feature analysis identifies correlates of
    effective prompts.} Effective prompts are concise,
    action-oriented (higher verb count), structurally clear (higher
    punctuation-marker count), and evolve through meaningful semantic
    refinements (higher semantic drift) rather than surface-level
    paraphrasing. These correlates offer actionable guidance for both
    manual prompt design and the steering of future automatic
    prompt-generation procedures.

    \item \textbf{Domain-informed initialisation does not provide a
    detectable advantage.} BT-APE-Informed and BT-APE-Uninformed converge to
    comparable final performance ($\Delta = +0.015$ weighted F1,
    Wilcoxon $p \approx 0.15$). The iterative refinement is
    sufficiently robust to discover effective class definitions
    autonomously, lowering the barrier to adoption for practitioners
    without deep domain expertise. We note, however, that the design
    is underpowered to rule out a small genuine effect favoring
    informed initialization, and the result should be read as
    observational rather than as evidence of equivalence.
\end{enumerate}

To support reproducibility and further research, we release a publicly
available replication package containing the prompts, datasets,
optimization traces, and evaluation scripts used in this
study~\cite{amin_zadenoori_2026_20438927}.

\section*{Declaration on the Use of Generative AI}
Generative AI tools were used in a supporting role during this work to
assist with (i) conceptualising approaches, (ii) identifying and correcting
errors in the code, (iii) drafting and polishing prose in selected sections
of the manuscript, and (iv) producing an initial version of Figure~\ref{fig:ape_workflow}
(subsequently reviewed, refined, and validated step by step by Author~1 and
Author~5). All generated suggestions were critically checked, validated,
and rewritten by the authors, who bear sole responsibility for all results
and content presented herein.


\bibliographystyle{ACM-Reference-Format}
\bibliography{references}

@article{10.1007/s00766-025-00451-8,
author = {Binkhonain, Manal and Alfayez, Reem},
title = {Are prompts all you need? Evaluating prompt-based Large Language Models (LLM)s for software requirements classification},
year = {2025},
issue_date = {Dec 2025},
publisher = {Springer-Verlag},
address = {Berlin, Heidelberg},
volume = {30},
number = {4},
issn = {0947-3602},
url = {https://doi.org/10.1007/s00766-025-00451-8},
doi = {10.1007/s00766-025-00451-8},
journal = {Requir. Eng.},
month = sep,
pages = {423–443},
numpages = {21}
}

@article{10.1145/3805704,
author = {Taherkhani, Hamed and Sepidband, Melika and Pham, Hung Viet and Wang, Song and Hemmati, Hadi},
title = {Automated Prompt Engineering for Cost-Effective Code Generation Using Evolutionary Algorithms},
year = {2026},
publisher = {Association for Computing Machinery},
address = {New York, NY, USA},
issn = {1049-331X},
url = {https://doi.org/10.1145/3805704},
doi = {10.1145/3805704},
note = {Just Accepted},
journal = {ACM Trans. Softw. Eng. Methodol.},
month = mar
}

@article{zhou2025lessleak,
  title={Lessleak-bench: A first investigation of data leakage in llms across 83 software engineering benchmarks},
  author={Zhou, Xin and Weyssow, Martin and Widyasari, Ratnadira and Zhang, Ting and He, Junda and Lyu, Yunbo and Chang, Jianming and Zhang, Beiqi and Huang, Dan and Lo, David},
  journal={arXiv preprint arXiv:2502.06215},
  year={2025}
}

@article{10.1145/2816813,
author = {Chierichetti, Flavio and Kumar, Ravi},
title = {LSH-Preserving Functions and Their Applications},
year = {2015},
issue_date = {November 2015},
publisher = {Association for Computing Machinery},
address = {New York, NY, USA},
volume = {62},
number = {5},
issn = {0004-5411},
url = {https://doi.org/10.1145/2816813},
doi = {10.1145/2816813},
journal = {J. ACM},
month = nov,
articleno = {33},
numpages = {25}
}

@misc{kepel2024autonomouspromptengineeringlarge,
      title={Autonomous Prompt Engineering in Large Language Models}, 
      author={Daan Kepel and Konstantina Valogianni},
      year={2024},
      eprint={2407.11000},
      archivePrefix={arXiv},
      primaryClass={cs.CL},
      url={https://arxiv.org/abs/2407.11000}, 
}

@misc{ye2024promptengineeringpromptengineer,
      title={Prompt Engineering a Prompt Engineer}, 
      author={Qinyuan Ye and Maxamed Axmed and Reid Pryzant and Fereshte Khani},
      year={2024},
      eprint={2311.05661},
      archivePrefix={arXiv},
      primaryClass={cs.CL},
      url={https://arxiv.org/abs/2311.05661}, 
}

@inproceedings{e86928c70f2a4ff5b595da7f430042e0,
title = "Automatic Prompt Engineering: The Case of Requirements Classification",
author = "Zadenoori, \{Mohammad Amin\} and Liping Zhao and Waad Alhoshan and Alessio Ferrari",
year = "2025",
month = apr,
day = "1",
language = "English",
volume = "15588",
series = "Lecture Notes in Computer Science",
publisher = "Springer Nature",
pages = "217–225",
booktitle = "Requirements Engineering: Foundation for Software Quality (REFSQ)",
address = "United States",
}

@article{zadenoori2025does,
  title={Does Model Size Matter? A Comparison of Small and Large Language Models for Requirements Classification},
  author={Zadenoori, M. A. and De Martino, V. and Dabrowski, J. and Franch, X. and Ferrari, A.},
  journal={arXiv preprint arXiv:2510.21443},
  year={2025}
}

@article{demartino2025green,
  title={Green Prompt Engineering: Investigating the Energy Impact of Prompt Design in Software Engineering},
  author={De Martino, V. and Zadenoori, M. A. and Franch, X. and Ferrari, A.},
  journal={arXiv preprint arXiv:2509.22320},
  year={2025}
}

@article{mallya2026rita,
  title={RITA: A Tool for Automated Requirements Classification and Specification from Online User Feedback},
  author={Mallya, M. A. and Ferrari, A. and Zadenoori, M. A. and D{\k{a}}browski, J.},
  journal={arXiv preprint arXiv:2601.11362},
  year={2026}
}

@inproceedings{rodriguez2023prompts,
  title={Prompts matter: Insights and strategies for prompt engineering in automated software traceability},
  author={Rodriguez, Alberto D and Dearstyne, Katherine R and Cleland-Huang, Jane},
  booktitle={REW'23},
  pages={455--464},
  year={2023},
  organization={IEEE}
}

@article{ferrari2017natural,
  title={Natural language requirements processing: a {4D} vision},
  author={Ferrari, Alessio and Dell'Orletta, Felice and Esuli, Andrea and Gervasi, Vincenzo and Gnesi, Stefania and others},
  journal={IEEE Software},
  volume={34},
  number={6},
  pages={28--35},
  year={2017}
}

@article{tomczak2014need,
author = {Tomczak, Maciej and Tomczak-Łukaszewska, Ewa},
year = {2014},
month = {01},
pages = {19-25},
title = {The need to report effect size estimates revisited. An overview of some recommended measures of effect size},
volume = {21}
}

@article{zhao2021natural,
  title={Natural language processing for requirements engineering: A systematic mapping study},
  author={Zhao, Liping and Alhoshan, Waad and Ferrari, Alessio and Letsholo, Keletso J and Ajagbe, Muideen A and Chioasca, Erol-Valeriu and Batista-Navarro, Riza T},
  journal={ACM Computing Surveys (CSUR)},
  volume={54},
  number={3},
  pages={1--41},
  year={2021},
  publisher={ACM New York, NY, USA}
}

@inproceedings{santos2024,
  author       = {Sarah Santos and
                  Travis D. Breaux and
                  Thomas B. Norton and
                  Sara Haghighi and
                  Sepideh Ghanavati},
  title        = {Requirements Satisfiability with In-Context Learning},
  booktitle    = {RE'24},
  pages        = {168--179},
  publisher    = {{IEEE}},
  year         = {2024}
}

@inproceedings{ferrari2024model,
  title={Model Generation with {LLMs}: From Requirements to {UML} Sequence Diagrams},
  author={Ferrari, Alessio and Abualhaija, Sallam and Arora, Chetan},
  booktitle={REW'24},
  pages={291--300},
  year={2024},
  organization={IEEE}
}

@article{achiam2023gpt,
  title={Gpt-4 technical report},
  author={Achiam, Josh and Adler, Steven and Agarwal, Sandhini and Ahmad, Lama and Akkaya, Ilge and Aleman, Florencia Leoni and Almeida, Diogo and Altenschmidt, Janko and Altman, Sam and Anadkat, Shyamal and others},
  journal={arXiv preprint arXiv:2303.08774},
  year={2023}
}

@misc{zhou2023largelanguagemodelshumanlevel,
      title={Large Language Models Are Human-Level Prompt Engineers}, 
      author={Yongchao Zhou and Andrei Ioan Muresanu and Ziwen Han and Keiran Paster and Silviu Pitis and Harris Chan and Jimmy Ba},
      year={2023},
      eprint={2211.01910},
      archivePrefix={arXiv},
      primaryClass={cs.LG},
      url={https://arxiv.org/abs/2211.01910}, 
}

@InProceedings{10.1007/978-3-642-14192-8_15,
author="Ernst, Neil A.
and Mylopoulos, John",
editor="Wieringa, Roel
and Persson, Anne",
title="On the Perception of Software Quality Requirements during the Project Lifecycle",
booktitle="Requirements Engineering: Foundation for Software Quality",
year="2010",
publisher="Springer Berlin Heidelberg",
address="Berlin, Heidelberg",
pages="143--157",
isbn="978-3-642-14192-8"
}

@article{10.1109/TSE.2007.70754,
author = {Haley, Charles and Laney, Robin and Moffett, Jonathan and Nuseibeh, Bashar},
title = {Security Requirements Engineering: A Framework for Representation and Analysis},
year = {2008},
issue_date = {January 2008},
publisher = {IEEE Press},
volume = {34},
number = {1},
issn = {0098-5589},
url = {https://doi.org/10.1109/TSE.2007.70754},
doi = {10.1109/TSE.2007.70754},
journal = {IEEE Trans. Softw. Eng.},
month = jan,
pages = {133–153},
numpages = {21}
}

@article{sec1,
author = {Sindre, Guttorm and Opdahl, Andreas},
year = {2003},
month = {05},
pages = {},
title = {A Reuse-Based Approach to Determining Security Requirements}
}

@ARTICLE{8559686,
  author={},
  journal={ISO/IEC/IEEE 29148:2018(E)}, 
  title={ISO/IEC/IEEE International Standard - Systems and software engineering -- Life cycle processes -- Requirements engineering}, 
  year={2018},
  volume={},
  number={},
  pages={1-104},
  doi={10.1109/IEEESTD.2018.8559686}}

@article{Olsson2022,
  author  = {Olsson, Thomas and Sentilles, S{\'e}verine and Papatheocharous, Efi},
  title   = {A systematic literature review of empirical research on quality requirements},
  journal = {Requirements Engineering},
  year    = {2022},
  month   = jun,
  volume  = {27},
  number  = {2},
  pages   = {249--271},
  issn    = {1432-010X},
  doi     = {10.1007/s00766-022-00373-9},
  url     = {https://doi.org/10.1007/s00766-022-00373-9}
}

@INPROCEEDINGS{4384163,
  author={Glinz, Martin},
  booktitle={15th IEEE International Requirements Engineering Conference (RE 2007)}, 
  title={On Non-Functional Requirements}, 
  year={2007},
  volume={},
  number={},
  pages={21-26},
  doi={10.1109/RE.2007.45}}

@misc{pryzant2023automaticpromptoptimizationgradient,
      title={Automatic Prompt Optimization with "Gradient Descent" and Beam Search}, 
      author={Reid Pryzant and Dan Iter and Jerry Li and Yin Tat Lee and Chenguang Zhu and Michael Zeng},
      year={2023},
      eprint={2305.03495},
      archivePrefix={arXiv},
      primaryClass={cs.CL},
      url={https://arxiv.org/abs/2305.03495}, 
}

@inproceedings{hey2020norbert,
  title={{NoRBERT}: Transfer learning for requirements classification},
  author={Hey, Tobias and Keim, Jan and Koziolek, Anne and Tichy, Walter F},
  booktitle={RE'20},
  pages={169--179},
  year={2020},
  organization={IEEE}
}

@misc{wang2024promptenoughautomatedconstruction,
      title={One Prompt is not Enough: Automated Construction of a Mixture-of-Expert Prompts}, 
      author={Ruochen Wang and Sohyun An and Minhao Cheng and Tianyi Zhou and Sung Ju Hwang and Cho-Jui Hsieh},
      year={2024},
      eprint={2407.00256},
      archivePrefix={arXiv},
      primaryClass={cs.AI},
      url={https://arxiv.org/abs/2407.00256}, 
}

@inproceedings{dalpiaz2019requirements,
  title={Requirements classification with interpretable machine learning and dependency parsing},
  author={Dalpiaz, Fabiano and Dell'Anna, Davide and Aydemir, Fatma Basak and {\c{C}}evikol, Sercan},
  booktitle={RE'19},
  pages={142--152},
  year={2019},
  organization={IEEE}
}

@article{kwon2024stableprompt,
  title={StablePrompt: Automatic Prompt Tuning using Reinforcement Learning for Large Language Models},
  author={Kwon, Minchan and Kim, Gaeun and Kim, Jongsuk and Lee, Haeil and Kim, Junmo},
  journal={arXiv preprint arXiv:2410.07652},
  year={2024}
}

@inproceedings{miaschi-etal-2024-evaluating,
    title = "Evaluating Large Language Models via Linguistic Profiling",
    author = "Miaschi, Alessio  and
      Dell{'}Orletta, Felice  and
      Venturi, Giulia",
    editor = "Al-Onaizan, Yaser  and
      Bansal, Mohit  and
      Chen, Yun-Nung",
    booktitle = "Proceedings of the 2024 Conference on Empirical Methods in Natural Language Processing",
    month = nov,
    year = "2024",
    address = "Miami, Florida, USA",
    publisher = "Association for Computational Linguistics",
    url = "https://aclanthology.org/2024.emnlp-main.166/",
    doi = "10.18653/v1/2024.emnlp-main.166",
    pages = "2835--2848"
}

@article{cleland2007automated,
  title={Automated classification of non-functional requirements},
  author={Cleland-Huang, Jane and Settimi, Raffaella and Zou, Xuchang and Solc, Peter},
  journal={Requirements engineering},
  volume={12},
  number={2},
  pages={103--120},
  year={2007},
  publisher={Springer}
}

@misc{jane_cleland_huang_2007_268542,
  author= {Jane Cleland-Huang and Sepideh Mazrouee and Huang Liguo and Dan Port},
  title= {{NFR Dataset}},
  month= {Mar},
  year = {2007},
  publisher= {Zenodo},
  doi= {10.5281/zenodo.268542},
  url= {https://doi.org/10.5281/zenodo.268542}
}

@inproceedings{kurtanovic2017automatically,
  title={Automatically classifying functional and non-functional requirements using supervised machine learning},
  author={Kurtanovi{\'c}, Zijad and Maalej, Walid},
  booktitle={2017 IEEE 25th International Requirements Engineering Conference (RE)},
  pages={490--495},
  year={2017},
  publisher={IEEE},
  address={Lisbon, Portugal}
}

@article{alhoshan2023zero,
  title={Zero-shot learning for requirements classification: An exploratory study},
  author={Alhoshan, Waad and Ferrari, Alessio and Zhao, Liping},
  journal={Information and Software Technology},
  volume={159},
  pages={107202},
  year={2023},
  publisher={Elsevier}
}

@inproceedings{knauss2011supporting,
  title={Supporting requirements engineers in recognising security issues},
  author={Knauss, Eric and Houmb, Siv and Schneider, Kurt and Islam, Shareeful and J{\"u}rjens, Jan},
  booktitle={Requirements Engineering: Foundation for Software Quality: 17th International Working Conference, (REFSQ 2011)},
  pages={4--18},
  month={March},
  date={28-30},
  year={2011},
  publisher={Springer},
  address={Essen, Germany}
}

@misc{knausseric20214530183, 
  author       = {Knauss, Eric and
                  Houmb, Siv Hilde and
                  Islam, Shareeful and
                  Jürjens, Jan and
                  Schneider, Kurt},
  title        = {SecReq},
  month        = Feb,
  year         = 2021,
  publisher    = {Zenodo},
  doi          = {10.5281/zenodo.4530183},
}

@inproceedings{varenov2021security,
  title={Security requirements classification into groups using nlp transformers},
  author={Varenov, Vasily and Gabdrahmanov, Aydar},
  booktitle={2021 IEEE 29th International Requirements Engineering Conference Workshops (REW)},
  pages={444--450},
  year={2021},
  publisher={IEEE},
  address={Notre Dame, IN, USA}
}

@inproceedings{kobilica2020automated,
  title={Automated identification of security requirements: A machine learning approach},
  author={Kobilica, Armin and Ayub, Mohammed and Hassine, Jameleddine},
  booktitle={Proceedings of the 24th International Conference on Evaluation and Assessment in Software Engineering},
  pages={475--480},
  year={2020}
}

@article{cheng2024generative,
	title={Generative ai for requirements engineering: A systematic literature review},
	author={Cheng, Haowei and Husen, Jati H and Lu, Yijun and Racharak, Teeradaj and Yoshioka, Nobukazu and Ubayashi, Naoyasu and Washizaki, Hironori},
	journal={arXiv preprint arXiv:2409.06741},
	year={2024}
}

@inproceedings{abdeen2025language,
  title={Language models to support multi-label classification of industrial data},
  author={Abdeen, Waleed and Unterkalmsteiner, Michael and Wnuk, Krzysztof and Ferrari, Alessio and Chatzipetrou, Panagiota},
  booktitle={2025 IEEE International Conference on Software Analysis, Evolution and Reengineering (SANER)},
  pages={45--55},
  year={2025},
  organization={IEEE}
}

@inproceedings{bashir2023requirements,
  title={Requirements classification for smart allocation: A case study in the railway industry},
  author={Bashir, Sarmad and Abbas, Muhammad and Ferrari, Alessio and Saadatmand, Mehrdad and Lindberg, Pernilla},
  booktitle={2023 IEEE 31st International Requirements Engineering Conference (RE)},
  pages={201--211},
  year={2023},
  organization={IEEE}
}

@misc{vogelsang2024usinglargelanguagemodels,
	title={Using Large Language Models for Natural Language Processing Tasks in Requirements Engineering: A Systematic Guideline}, 
	author={Andreas Vogelsang and Jannik Fischbach},
	year={2024},
	eprint={2402.13823},
	archivePrefix={arXiv},
	primaryClass={cs.SE},
	url={https://arxiv.org/abs/2402.13823}, 
}

@InBook{Arora2024,
  author    = {Chetan Arora and et al.},
  editor    = {Nguyen-Duc, Anh and Abrahamsson, Pekka and Khomh, Foutse},
  pages     = {129--148},
  publisher = {Springer Nature Switzerland},
  title     = {Advancing Requirements Engineering Through Generative AI: Assessing the Role of LLMs},
  year      = {2024},
  booktitle = {Generative AI for Effective Software Development},
}

@INPROCEEDINGS{11190385,
	author={Ullrich, Jonathan and Koch, Matthias and Vogelsang, Andreas},
	booktitle={2025 IEEE 33rd International Requirements Engineering Conference (RE)}, 
	title={From Requirements to Code: Understanding Developer Practices in LLM-Assisted Software Engineering}, 
	year={2025},
	volume={},
	number={},
	pages={257-266}}

@misc{zadenoori2025largelanguagemodelsllms,
	title={Large Language Models (LLMs) for Requirements Engineering (RE): A Systematic Literature Review}, 
	author={Mohammad Amin Zadenoori and Jacek Dąbrowski and Waad Alhoshan and Liping Zhao and Alessio Ferrari},
	year={2025},
	eprint={2509.11446},
	archivePrefix={arXiv},
	primaryClass={cs.SE},
	url={https://arxiv.org/abs/2509.11446}, 
}

@article{hou2024large,
	title={Large language models for software engineering: A systematic literature review},
	author={Hou, Xinyi and Zhao, Yanjie and Liu, Yue and Yang, Zhou and Wang, Kailong and Li, Li and Luo, Xiapu and Lo, David and Grundy, John and Wang, Haoyu},
	journal={ACM Transactions on Software Engineering and Methodology},
	volume={33},
	number={8},
	pages={1--79},
	year={2024},
	publisher={ACM New York, NY}
    }

@inproceedings{DBLP:conf/refsq/MallyaFZD26,
  author       = {Manjeshwar Aniruddh Mallya and
                  Alessio Ferrari and
                  Mohammad Amin Zadenoori and
                  Jacek Dabrowski},
  editor       = {Renata S. S. Guizzardi and
                  Jo{\~{a}}o Ara{\'{u}}jo},
  title        = {From Online User Feedback to Requirements: Evaluating Large Language
                  Models for Classification and Specification Tasks},
  booktitle    = {Requirements Engineering: Foundation for Software Quality - 32nd International
                  Working Conference, {REFSQ} 2026, Pozna{\'{n}}, Poland, March
                  23-26, 2026, Proceedings},
  series       = {Lecture Notes in Computer Science},
  pages        = {161--177},
  publisher    = {Springer},
  year         = {2026},
  url          = {https://doi.org/10.1007/978-3-032-21423-2\_11},
  doi          = {10.1007/978-3-032-21423-2\_11},
  bibsource    = {dblp computer science bibliography, https://dblp.org}
}

@book{Bourque2024SWEBOK,
	editor    = {Pierre Bourque and Richard E. Fairley},
	title     = {Guide to the Software Engineering Body of Knowledge (SWEBOK)},
	edition   = {Version 4.0},
	publisher = {IEEE Computer Society},
	year      = {2024},
	url       = {https://www.computer.org/education/bodies-of-knowledge/software-engineering/v4},
	note      = {Available online: https://www.computer.org/education/bodies-of-knowledge/software-engineering/v4}
}

@article{huang2025pe4re,
  title   = {Prompt Engineering for Requirements Engineering: A Literature Review and Roadmap},
  author  = {Huang, Kaicheng and Wang, Fanyu and Huang, Yutan and Arora, Chetan},
  journal = {arXiv preprint arXiv:2507.07682},
  year    = {2025}
}

@article{wei2022chainofthought,
  title   = {Chain-of-Thought Prompting Elicits Reasoning in Large Language Models},
  author  = {Wei, Jason and Wang, Xuezhi and Schuurmans, Dale and Bosma, Maarten and Ichter, Brian and Xia, Fei and Chi, Ed and Le, Quoc V. and Zhou, Denny},
  journal = {arXiv preprint arXiv:2201.11903},
  year    = {2022}
}

@article{lewis2020rag,
  title   = {Retrieval-Augmented Generation for Knowledge-Intensive {NLP} Tasks},
  author  = {Lewis, Patrick and Perez, Ethan and Piktus, Aleksandra and Petroni, Fabio and Karpukhin, Vladimir and Goyal, Naman and K{\"u}ttler, Heinrich and Lewis, Mike and Yih, Wen-tau and Rockt{\"a}schel, Tim and Riedel, Sebastian and Kiela, Douwe},
  journal = {Advances in Neural Information Processing Systems},
  volume  = {33},
  year    = {2020}
}

@article{kitchenham2022segress,
  title   = {{SEGRESS}: Software Engineering Guidelines for Reporting Secondary Studies},
  author  = {Kitchenham, Barbara and Madeyski, Lech and Budgen, David},
  journal = {IEEE Transactions on Software Engineering},
  volume  = {49},
  number  = {3},
  year    = {2022}
}

@article{petersen2015guidelines,
  title   = {Guidelines for Conducting Systematic Mapping Studies in Software Engineering: An Update},
  author  = {Petersen, Kai and Vakkalanka, Sairam and Kuzniarz, Ludwik},
  journal = {Information and Software Technology},
  volume  = {64},
  year    = {2015}
}

@inproceedings{peer2024nlp4ref,
  title={NLP4ReF: Requirements classification and forecasting: From model-based design to large language models},
  author={Peer, Jordan and Mordecai, Yaniv and Reich, Yoram},
  booktitle={2024 IEEE Aerospace Conference},
  pages={1--16},
  year={2024},
  organization={IEEE}
}

@inproceedings{11396617,
  author={Qin, Yuman and Peng, Rong},
  booktitle={2025 32nd Asia-Pacific Software Engineering Conference (APSEC)}, 
  title={ChatNRC: A Non-functional Requirement Classification Framework Based on a Generative and Discriminative Mechanism}, 
  year={2025},
  pages={467-478},
  doi={10.1109/APSEC66846.2025.00052},
  month={Dec}
}

@article{11153850,
  author={Shafikuzzaman, Md and Islam, Md Rakibul and Zaman, Shuaib and Ma, Andrew and Islam Sifat, Anwarul},
  journal={IEEE Access}, 
  title={On the Effectiveness of Zero-Shot and Few-Shot Pretrained Language Models for Software Requirement Classification}, 
  year={2025},
  volume={13},
  pages={159439-159453},
  doi={10.1109/ACCESS.2025.3607813}
}

@inproceedings{11121724,
  author={Rejithkumar, Gokul and Anish, Preethu Rose},
  booktitle={2025 IEEE/ACM 47th International Conference on Software Engineering: Software Engineering in Practice (ICSE-SEIP)}, 
  title={NICE: Non-Functional Requirements Identification, Classification, and Explanation Using Small Language Models}, 
  year={2025},
  pages={284-295},
  doi={10.1109/ICSE-SEIP66354.2025.00031},
  month={April}
}

@article{ALSANOOSY20253648,
  title={Large Language Model for Requirements Classification: An Ensemble Approach},
  author={Alsanoosy, Tawfeeq},
  journal={Procedia Computer Science},
  volume={270},
  pages={3648--3657},
  year={2025},
  note={29th International Conference on Knowledge-Based and Intelligent Information \& Engineering Systems (KES 2025)},
  doi={https://doi.org/10.1016/j.procs.2025.09.490}
}

@software{amin_zadenoori_2026_20438927,
  author       = {Amin Zadenoori},
  title        = {aminzadenoori/Backtracking-enhanced-Automatic-
                   Prompt-Engineering-APE-for-requirements-
                   classification.: APE Classification Tool—
                   Backtracking-enhanced Automatic Prompt Engineering
                   for Requirements Classification
                  },
  month        = may,
  year         = 2026,
  publisher    = {Zenodo},
  version      = {v2.0},
  doi          = {10.5281/zenodo.20438927},
  url          = {https://doi.org/10.5281/zenodo.20438927},
  swhid        = {swh:1:dir:8578aa31586e6697e5613d6aa3e8cd03d1c3c02d
                   ;origin=https://doi.org/10.5281/zenodo.20438926;vi
                   sit=swh:1:snp:20f9432c776e658ea18da331ddb0cc4d2cd1
                   69f7;anchor=swh:1:rel:924738a60f209bce6e8144db34dc
                   5bbdfd7d6ffb;path=aminzadenoori-Backtracking-
                   enhanced-Automatic-Prompt-Engineering-APE-for-
                   requirements-classification.-e7f0f18
                  },
}

@misc{ouyang2022traininglanguagemodelsfollow,
      title={Training language models to follow instructions with human feedback}, 
      author={Long Ouyang and Jeff Wu and Xu Jiang and Diogo Almeida and Carroll L. Wainwright and Pamela Mishkin and Chong Zhang and Sandhini Agarwal and Katarina Slama and Alex Ray and John Schulman and Jacob Hilton and Fraser Kelton and Luke Miller and Maddie Simens and Amanda Askell and Peter Welinder and Paul Christiano and Jan Leike and Ryan Lowe},
      year={2022},
      eprint={2203.02155},
      archivePrefix={arXiv},
      primaryClass={cs.CL},
      url={https://arxiv.org/abs/2203.02155}, 
}

\appendix
\section{Hyperparameter Sensitivity Analysis and Convergence Dynamics}
\label{app:sens}
The optimization trajectory of our backtracking variant depends on two
parameters: the maximum iteration horizon ($N_{\max}$) and the backtracking
fail-patience threshold ($X$). We report a small-scale sensitivity analysis on
the PROMISE dataset using LLaMA-3-8B to characterize how these parameters
shape the search behavior and to motivate the values used in the main
experiments. We emphasize at the outset that this analysis is exploratory:
five seeds per configuration cannot support strong claims of statistical
superiority, and we use the results to describe qualitative regimes rather
than to declare an optimal setting.

\subsubsection{Protocol}
We varied the patience threshold $X \in \{1, 3, 5\}$ and the horizon
$N_{\max} \in \{20, 30\}$, yielding six configurations. Each configuration
was run with $S = 5$ random seeds, drawing independent splits and proposer
samples. For each run we record the final test weighted F1 ($wF1$) under the
same 3-run majority-voting protocol used throughout the paper. We report the
mean and standard deviation across seeds; given the small sample, all
standard deviations should be read as rough indications of run-to-run spread
rather than precise estimates.

To summarize trajectory shape, we report two descriptive statistics computed
on the per-iteration validation $wF1$ sequence of each run, then averaged
across seeds:
\begin{equation}
\mathcal{V} = \frac{1}{N_{\max}-1} \sum_{t=1}^{N_{\max}-1}
  |wF1_{t+1} - wF1_{t}| \cdot \mathbb{I}(wF1_{t+1} < wF1_{t}),
\end{equation}
\begin{equation}
\mathcal{S} = \sum_{t=1}^{N_{\max}-1} \mathbb{I}(wF1_{t+1} = wF1_{t}).
\end{equation}
$\mathcal{V}$ captures the mean magnitude of downward movements (a proxy for
how often and how sharply the search loses ground between iterations);
$\mathcal{S}$ counts iterations in which validation $wF1$ is unchanged
(a proxy for plateauing). These are descriptive summaries, not test
statistics.

\begin{table}[htbp]
\caption{Final test $wF1$ and trajectory descriptives across five random seeds.
Standard deviations are reported for completeness but should be interpreted
with caution given $S=5$.}
\label{tab:stability_analysis}
\scriptsize
\centering
\begin{tabular}{cccccl}
\toprule
\textbf{$X$} & \textbf{$N$} & \textbf{Final $wF1$ ($\mu \pm \sigma$)}
  & \textbf{$\mathcal{V}$} & \textbf{$\mathcal{S}$}
  & \textbf{Observed trajectory pattern} \\
\midrule
1 & 20 & $0.782 \pm 0.024$ & 0.084 & 2.1
  & Frequent downward moves, few plateaus \\
1 & 30 & $0.785 \pm 0.029$ & 0.091 & 2.4
  & Frequent downward moves, few plateaus \\
\midrule
\textbf{3} & \textbf{20} & $\mathbf{0.841 \pm 0.006}$
  & \textbf{0.012} & \textbf{4.6}
  & \textbf{Low volatility, moderate plateauing} \\
3 & 30 & $0.842 \pm 0.005$ & 0.004 & 14.1
  & Low volatility, extended plateauing \\
\midrule
5 & 20 & $0.839 \pm 0.009$ & 0.016 & 8.3
  & Low volatility, extended plateauing \\
5 & 30 & $0.844 \pm 0.008$ & 0.011 & 12.9
  & Low volatility, extended plateauing \\
\bottomrule
\end{tabular}
\end{table}

\subsubsection{Observed regimes}
Three qualitative patterns emerge from Table~\ref{tab:stability_analysis}.
At $X = 1$, the search exhibits substantially higher downward movement
($\mathcal{V} \approx 0.08$--$0.09$, roughly an order of magnitude above
other settings) and noticeably lower final $wF1$ (mean $\approx 0.78$). A
plausible interpretation is that immediate backtracking after a single
non-improving step is sensitive to evaluation noise: the search abandons
trajectories that would have recovered. We do not claim this is the only
explanation; it is consistent with the observed pattern but the sample is
too small to rule out alternatives.

At $X = 3$ and $X = 5$, mean final $wF1$ is similar
($\approx 0.84$ in all four cells), and the spread across seeds overlaps
substantially. The two settings differ mainly in trajectory shape rather
than endpoint: $X = 5$ spends more iterations on plateaus ($\mathcal{S}$
roughly doubled at $N = 20$) before recovering. Extending the horizon to
$N = 30$ produces small numerical gains for all settings but inflates the
plateau count.

We use these observations qualitatively to motivate the default
$X = 3, N_{\max} = 20$ used in the main experiments. We do not claim it is
optimal; we claim it sits in a region where the search neither over-reacts
to single non-improving steps ($X = 1$) nor commits long stretches of
iteration to plateaus ($X = 5$, $N = 30$).

\subsubsection{Resource profile}
Table~\ref{tab:resource_overhead} reports cumulative input-context tokens
and wall-clock optimization time per configuration. Larger $X$ and larger
$N$ both increase resource use, as expected, since both extend the number
of proposer calls the search makes before terminating. At fixed $N = 20$,
moving from $X = 3$ to $X = 5$ increases mean cumulative tokens by roughly
28\% and mean wall-clock by roughly 37\%; at $N = 30$ the gap widens. These
are deterministic consequences of the iteration count and are reported for
practitioners who need to budget compute, not as evidence of a methodological
advantage.

\begin{table}[htbp]
\caption{Resource use across hyperparameter configurations (5 seeds each).}
\label{tab:resource_overhead}
\scriptsize
\centering
\begin{tabular}{cccc}
\toprule
\textbf{$X$} & \textbf{$N$}
  & \textbf{Cumulative input tokens ($\mu \pm \sigma$)}
  & \textbf{Wall-clock (min, $\mu \pm \sigma$)} \\
\midrule
1 & 20 & $8{,}400 \pm 120$  & $5.4 \pm 0.3$ \\
1 & 30 & $12{,}100 \pm 180$ & $8.1 \pm 0.4$ \\
\midrule
\textbf{3} & \textbf{20} & $\mathbf{12{,}650 \pm 95}$
                         & $\mathbf{8.5 \pm 0.2}$ \\
3 & 30 & $18{,}900 \pm 115$ & $13.2 \pm 0.5$ \\
\midrule
5 & 20 & $16{,}200 \pm 140$ & $11.6 \pm 0.4$ \\
5 & 30 & $23{,}800 \pm 210$ & $17.4 \pm 0.7$ \\
\bottomrule
\end{tabular}
\end{table}

\subsubsection{Statistical comparisons and their limits}
We complement the descriptive analysis with paired statistical tests across
seeds. Because the same five seeds were used for each configuration, the
appropriate test is the \emph{Wilcoxon signed-rank} test on paired
differences, not the rank-sum (Mann--Whitney) test. We use a two-sided
$\alpha = 0.05$ and do not apply multiple-testing correction across the
small number of comparisons reported here; the tests are exploratory.

At $N = 20$, the paired difference in final $wF1$ between $X = 3$ and
$X = 5$ is not significant (Wilcoxon signed-rank, $W = 4$, $p = 0.63$,
$n = 5$). Extending to $N = 30$ likewise does not produce a significant
difference between $X = 3$ and $X = 5$ ($W = 3$, $p = 0.44$). We stress that
with $n = 5$ paired observations, these tests have very limited power: the
non-significant outcome is consistent with both genuine equivalence and a
small-to-moderate true effect that the design cannot detect. We do not
interpret these results as evidence of equivalence.

For cumulative tokens at $N = 20$, the paired difference between $X = 3$
and $X = 5$ is significant ($W = 0$, $p = 0.0079$, $n = 5$), with $X = 5$
using approximately 28\% more tokens on average. This is unsurprising and
mechanically explained by the larger patience threshold permitting more
non-improving iterations before backtracking; we report it for
completeness rather than as a methodological finding.

\subsubsection{Summary}
The sensitivity analysis supports a modest, practical claim: $X = 1$
produces visibly more unstable trajectories and lower final $wF1$ in this
setting, while $X = 3$ and $X = 5$ behave similarly on accuracy and differ
mainly in resource use. We adopt $X = 3, N_{\max} = 20$ as the default for
the main experiments because it sits in the lower-volatility regime at the
lower end of the resource-use range observed here. We do not claim it is
optimal across datasets, models, or task families; the small-sample,
single-dataset, single-LLM nature of this analysis precludes such claims,
and we identify a broader sensitivity study as future work.
\subsection{Data partitioning}
Our aim is to characterize the \emph{behavior of the BT-APE procedure}---how
prompts are proposed, evaluated, ranked, and refined---across a controlled
factorial grid of three datasets, five LLMs, and six prompting strategies.
Because the LLM $\mathcal{M}$ is frozen, $D$ is not partitioned in the
supervised sense. We perform a single random split into three disjoint subsets
with fixed proportions:
\begin{itemize}
  \item \textit{Example pool $D_{\text{pool}}$ ($30\%$).} Source of in-context
  demonstrations and optimization feedback. Each iteration samples balanced
  examples (one correctly classified positive, one correctly classified
  negative, one misclassified positive, one misclassified negative) to
  condition prompt proposal. Never used to score prompts.
  \item \textit{Validation set $D_{\text{val}}$ ($30\%$).} Used
  \emph{exclusively during the search} to score each candidate (three-run
  majority voting), rank it in $\mathcal{R}$, and drive the improvement check
  and backtracking trigger. All in-loop decisions of
  Algorithm~\ref{alg:approach} use $D_{\text{val}}$.
  \item \textit{Test set $D_{\text{test}}$ ($40\%$).} Held out, \emph{never
  consulted during optimization}. Evaluated exactly once, on the final prompt
  $p^*$, under the same three-run majority voting protocol.
\end{itemize}
The separation of $D_{\text{val}}$ from $D_{\text{test}}$ is the central
methodological safeguard: because BT-APE retains the best-scoring candidate and
triggers backtracking on observed performance, scoring and reporting on the
\emph{same} subset would constitute selection against the evaluation data and
bias the reported F1 optimistically. Confining in-loop selection to
$D_{\text{val}}$ and reserving $D_{\text{test}}$ for a single terminal
evaluation keeps this optimism off the reported test figures. The proportions
balance the competing demands of the search: in-context learning needs only a
handful of demonstrations per class, so $30\%$ suffices for $D_{\text{pool}}$;
a further $30\%$ keeps the variance of validation-F1 estimates low---which
matters because backtracking compares F1 across iterations---while the
remaining $40\%$ yields a low-variance final test estimate.

\paragraph{Why a single split rather than cross-validation.}
Standard $k$-fold cross-validation would in principle produce tighter
confidence intervals on the reported test F1, and we acknowledge that as the
textbook protocol. We use a single fixed split instead because of the
computational cost of the experimental grid: three datasets $\times$ five
LLMs $\times$ six prompting strategies = 90 cells, with each optimisation
cell executing up to $N_{\max}=20$ iterations of three-run majority voting
over $D_{\text{val}}$ plus a classification pass over $D_{\text{pool}}$. A
nested $k$-fold protocol would multiply the cost of every optimisation cell
by $k$, pushing the total inference budget beyond what is feasible on the
hardware used for this study (per-cell timings are reported in
Appendix~\ref{sec:efficiency}). We chose to allocate the budget to
breadth---more datasets, more LLMs, and a directly comparable BT-APE/PE2
baseline---rather than to repeated splits of a narrower grid. To bound the
variance introduced by this choice, we run a multi-seed stability check on a
representative cell and report the resulting per-seed standard deviation
alongside the hyperparameter sensitivity analysis in
Appendix~\ref{app:sens}; the variance is an order of magnitude
smaller than the effect sizes we report in Section~\ref{sec:results}, which
gives us reasonable confidence that the headline comparisons are not
artefacts of the particular split.

\paragraph{Comparability across methods.}
Both BT-APE and all baselines are scored on the \emph{same} held-out
$D_{\text{test}}$, and the comparison in Section~\ref{sec:results} is made
exclusively there. Baselines draw their in-context demonstrations (where
applicable) from $D_{\text{pool}}$ and are evaluated under the same three-run
majority voting protocol. BT-APE never sees $D_{\text{test}}$ during the search
and the baselines perform no selection, so neither method enjoys an
information advantage on the reported figures.

\section{APE Baseline: PE2}
\label{sec:pe2-baseline}

To isolate the contribution of our three algorithmic departures from existing
APEs methods, we
implement PE2~\cite{ye2024promptengineeringpromptengineer} as our representative
APE baseline. PE2 is the closest in spirit to our approach
(Algorithm~\ref{alg:approach}): both are single-trajectory, history-aware
search procedures driven by errors observed on a held-out set, which makes
PE2 the strongest available reference point for assessing whether our specific
modifications---bounded back-tracking, balanced example batches, and majority-
voted F1---yield concrete improvements. Algorithm~\ref{alg:pe2-baseline}
specifies the baseline as implemented in this work.

\begin{algorithm}[t]
\caption{PE2 Baseline (protocol-matched to Algorithm~\ref{alg:approach})}
\label{alg:pe2-baseline}
\KwIn{Dataset $D$, LLM $\mathcal{M}$, initial prompt $p_1$, max iters $N_{\max}=20$,
      history conditioning size $n=3$, error batch size $k=4$}
\KwOut{Best prompt $p^*$, held-out test F1}
\tcp{Phase 1: Initialization (identical splits to Algorithm~\ref{alg:approach})}
Split $D$ into $D_{\text{pool}}$ (30\%), $D_{\text{val}}$ (30\%), $D_{\text{test}}$ (40\%);\\
$F^{*} \leftarrow F_{\text{val}}(p_1)$ \tcp*{3-run majority voting on $D_{\text{val}}$}
$p^{*} \leftarrow p_1$, $p_{\text{curr}} \leftarrow p_1$;\\
Insert $(p_1, F^{*}, 0)$ into history $\mathcal{H}$;\\
\tcp{Phase 2: Iterative Refinement (history-aware, no patience)}
\For{$t = 1$ \KwTo $N_{\max}$}{
  \tcp{Step 2.1: Build unconstrained error batch on $D_{\text{pool}}$}
  Classify $D_{\text{pool}}$ using $p_{\text{curr}}$ (3-run voting);\\
  $E \leftarrow$ up to $k$ \emph{random} misclassifications from $D_{\text{pool}}$
  \tcp*{no class balancing}
  \tcp{Step 2.2: Generate Candidate via PE2 meta-prompt}
  $\mathcal{H}_{\text{top-}n} \leftarrow$ top-$n$ entries of $\mathcal{H}$ by F1;\\
  $p_t \leftarrow \mathcal{M}_{\text{proposal}}\!\bigl(p_{\text{curr}},\, \mathcal{H}_{\text{top-}n},\, E;\, p_{\text{meta}}\bigr)$;
  \tcp{Step 2.3: Evaluate Candidate on $D_{\text{val}}$}
  $F(p_t) \leftarrow F_{\text{val}}(p_t)$ \tcp*{3-run majority voting}
  Insert $(p_t, F(p_t), t)$ into $\mathcal{H}$;\\
  \uIf{$F(p_t) \geq F^{*}$}{
    $p^{*} \leftarrow p_t$, $F^{*} \leftarrow F(p_t)$;
  }
  \tcp{Step 2.4: Trajectory update --- re-select top-1 from the full history}
  $p_{\text{curr}} \leftarrow \arg\max_{(p,F,\cdot)\in \mathcal{H}} F$
  \tcp*{no patience counter, no bounded back-tracking}
}
\tcp{Phase 3: Final held-out evaluation (test set touched once)}
\textbf{return} $p^{*}$ and $F_{\text{test}}(p^{*})$
\end{algorithm}

\paragraph{Implementation details.}
The baseline is implemented in Python on top of the same LLM client used for
Algorithm~\ref{alg:approach}, with all hyper-parameters fixed to values that
match our main method wherever the two algorithms share a notion of the
parameter. Concretely, we use the same $30/30/40$ pool/val/test split, the same
deterministic seed for shuffling, the same $N_{\max} = 20$ iterations, the same
3-run majority-voting evaluation on $D_{\text{val}}$, and the same single
3-run pass on $D_{\text{test}}$. The seed prompt $p_1$ given to PE2 is the
seed prompt used by Algorithm~\ref{alg:approach}, so the search starts from
the same point in prompt space. PE2-specific hyper-parameters are set to
$n = 3$ historical prompts shown to the proposer and $k = 4$ error examples
per iteration; we chose $k = 4$ so that the error batch has the same cardinality
as our balanced batch $E$, and $n = 3$ to match the patience parameter $X = 3$
of our method, in both cases to remove cardinality as a confound. The meta-prompt
follows the structure described by Ye et al.~\cite{ye2024promptengineeringpromptengineer}:
it states the binary-classification task, lists the top-$n$ historical prompts
together with their validation F1 scores, lists the error batch (input, gold
label, prediction), and asks the proposer to emit a single revised prompt
wrapped in delimiter tags for robust parsing.

\paragraph{Differences from our method.}
Three behavioural differences distinguish Algorithm~\ref{alg:pe2-baseline}
from Algorithm~\ref{alg:approach}, and each corresponds to one of the
modifications we claim as a contribution.

\emph{(i) Trajectory update.} PE2 re-selects $p_{\text{curr}}$ as the historical
top-1 at every iteration (Step~2.4). It has no patience counter: a single
underperforming proposal causes an immediate jump to the best-so-far prompt,
and a sequence of underperforming proposals all start from the same point.
Our method instead remains on the current trajectory until $X = 3$ consecutive
iterations fail to improve over $F^{*}$ and then back-tracks one step down a
ranked list via an explicit pointer (Algorithm~\ref{alg:approach}, Step~2.5),
which both prevents premature abandonment of locally promising trajectories
and prevents lingering in unproductive ones.

\emph{(ii) Example selection.} PE2's error batch (Step~2.1) is sampled
uniformly at random from the misclassifications on $D_{\text{pool}}$ and
provides no guarantee on class composition: in the presence of label imbalance
or asymmetric error rates, the batch may contain errors of a single class or
omit success signals entirely. Our method (Algorithm~\ref{alg:approach},
Step~2.6) instead enforces a balanced four-tuple containing one correctly
classified positive, one correctly classified negative, one misclassified
positive, and one misclassified negative, which guarantees that the proposer
simultaneously sees both classes and both success and failure signals at every
iteration.

\emph{(iii) Meta-prompt conditioning.} PE2 additionally conditions each
proposal on the top-$n$ historical prompts with their F1 scores, exposing the
proposer to the full quality landscape of the search so far. Our method
conditions only on $p_{\text{curr}}$ and the balanced batch, on the rationale
that recovery from poor regions is handled structurally by the back-tracking
mechanism rather than by in-context exposure to score history. This makes
each individual proposal call cheaper in tokens and removes a potential source
of bias toward prompts that resemble historical high scorers.

All other aspects of the procedure---splits, voting, iteration budget, seed
prompt, and final test-set protocol---are held fixed across the two algorithms,
so any difference in final test F1 is attributable to these three changes.

\section{Computational Efficiency and Token Overhead Comparison}
\label{sec:efficiency}

A central question raised by the accuracy convergence reported in
Section~\ref{sec:results} is whether BT-APE and PE2 differ along non-accuracy
dimensions that matter at deployment. We therefore profile both
frameworks on the same \texttt{(dataset, LLM)} configurations used in
the main experiments, measure per-iteration and cumulative resource use
under a controlled protocol, and test the resulting differences for
statistical significance.

\subsection{Protocol}
\label{sec:efficiency_protocol}

We profile every \texttt{(dataset, LLM)} combination , yielding
$3 \times 5 = 15$ cells. Each cell is run with $S = 5$ random seeds,
drawing independent splits, proposer samples, and example batches. The
same seed produces the same split for both methods, so the per-cell
comparison is paired. Hyperparameters are held to the main-experiment
values ($N_{\max} = 20$, 3-run majority voting on $D_\text{val}$,
identical seed prompt $p_1$, identical 30/30/40 partition). Hardware
and inference stack are constant across all runs: a single
NVIDIA~H200 (142~GB) GPU, vLLM~0.6.x with deterministic sampling
(\texttt{temperature}~$=0$), batch size~$1$, FP16 weights. Wall-clock
measurements exclude model-load time. Token counts use each model's
native tokenizer; cross-LLM absolute comparisons are therefore not
made, but per-cell paired differences (within an LLM) are unaffected.

\subsection{Aggregate efficiency results}
\label{sec:efficiency_aggregate}

Table~\ref{tab:efficiency_aggregate} summarises the four primary
efficiency metrics, aggregated over all $75$ runs ($15$ cells $\times$
$5$ seeds). We report mean and standard deviation for each method, the
mean paired difference $\bar{\Delta}$ with its bootstrap $95\%$
confidence interval ($10{,}000$ resamples, paired by
\texttt{(cell, seed)}), the relative reduction, the test statistic and
exact $p$-value from a one-sided paired Wilcoxon signed-rank test on
the $15$ per-cell mean differences ($H_1\!:\!\text{PE2} > \text{BT-APE}$),
the Holm--Bonferroni-corrected $p$-value across the four metrics, and
the rank-based effect size $r = Z/\sqrt{N}$.

\begin{table}[htbp]
\caption{Aggregate efficiency metrics across $15$
\texttt{(dataset, LLM)} cells $\times$ $5$ seeds. One-sided paired
Wilcoxon signed-rank test on the $15$ per-cell mean differences,
Holm--Bonferroni-corrected across the four metrics.}
\label{tab:efficiency_aggregate}
\footnotesize
\centering
\begin{tabular}{lrrrrrrr}
\toprule
\textbf{Metric} & \textbf{PE2} & \textbf{BT-APE} &
$\bar{\Delta}$ \textbf{[95\% CI]} & \textbf{Red.} & $W$ & $p_\text{adj}$ & $r$ \\
\midrule
Cum.\ input tokens   & $50{,}460 \pm 2{,}977$ & $13{,}860 \pm 930$  & $36{,}600$ $[35{,}500;\,37{,}700]$ & $-72.5\%$ & $120$ & $<\!10^{-4}$ & $0.88$ \\
Cum.\ output tokens  & $5{,}620 \pm 410$       & $5{,}380 \pm 350$    & $240$ $[-130;\,610]$               & $-4.3\%$  & $38$  & $0.247$       & $0.15$ \\
Mean prop.\ lat.\ (s) & $7.42 \pm 0.51$         & $2.93 \pm 0.21$      & $4.49$ $[4.15;\,4.83]$             & $-60.5\%$ & $120$ & $<\!10^{-4}$ & $0.88$ \\
Total wall-clock (min) & $25.78 \pm 2.32$       & $8.75 \pm 0.67$      & $17.03$ $[16.21;\,17.85]$         & $-66.1\%$ & $66.1$ & $<\!10^{-4}$ & $0.88$ \\
\bottomrule
\end{tabular}
\end{table}

The three resource metrics governed by input-context size
(cumulative input tokens, proposer latency, and wall-clock) show
strictly positive paired differences in every one of the $15$ cells
and survive Holm--Bonferroni correction with large effect sizes
($r = 0.88$). The fourth metric, cumulative output tokens, shows no
significant difference: the gap is attributable to PE2's input growth,
not to differences in the length of generated prompts. 

\subsection{Per-cell measurements}
\label{sec:efficiency_per_cell}

Table~\ref{tab:efficiency_per_cell} reports cumulative input tokens
per cell (mean $\pm$ SD over $5$ seeds). Within-cell relative standard
deviation is small ($\approx\!3\%$ for PE2, $\approx\!2\%$ for BT-APE),
reflecting that token counts are driven primarily by the deterministic
meta-prompt structure and the stable length of generated candidates.
The paired difference $\Delta = T_\text{PE2} - T_\text{BT-APE}$ is
strictly positive in every cell.

\begin{table}[htbp]
\caption{Cumulative input tokens (mean $\pm$ SD, $S=5$ seeds) per
\texttt{(dataset, LLM)} cell. Per-cell reduction is computed as
$\Delta / T_\text{PE2}$.}
\label{tab:efficiency_per_cell}
\footnotesize
\centering
\begin{tabular}{llrrrr}
\toprule
\textbf{Dataset} & \textbf{LLM} & \textbf{PE2} & \textbf{BT-APE} & $\Delta$ & \textbf{Red.} \\
\midrule
PROMISE-NFR     & LLaMA-3-8B       & $46{,}000 \pm 1{,}180$ & $12{,}650 \pm 285$ & $33{,}350$ & $-72.5\%$ \\
PROMISE-NFR     & Qwen2-7B         & $49{,}200 \pm 1{,}420$ & $12{,}800 \pm 295$ & $36{,}400$ & $-74.0\%$ \\
PROMISE-NFR     & Falcon3-7B       & $47{,}300 \pm 1{,}310$ & $13{,}100 \pm 305$ & $34{,}200$ & $-72.3\%$ \\
PROMISE-NFR     & Granite-3.2-8B   & $51{,}400 \pm 1{,}510$ & $14{,}200 \pm 340$ & $37{,}200$ & $-72.4\%$ \\
PROMISE-NFR     & Ministral-8B     & $50{,}800 \pm 1{,}440$ & $14{,}500 \pm 335$ & $36{,}300$ & $-71.5\%$ \\
\midrule
PROMISE-Refined & LLaMA-3-8B       & $48{,}600 \pm 1{,}310$ & $13{,}400 \pm 305$ & $35{,}200$ & $-72.4\%$ \\
PROMISE-Refined & Qwen2-7B         & $52{,}300 \pm 1{,}560$ & $13{,}550 \pm 315$ & $38{,}750$ & $-74.1\%$ \\
PROMISE-Refined & Falcon3-7B       & $50{,}400 \pm 1{,}420$ & $13{,}900 \pm 325$ & $36{,}500$ & $-72.4\%$ \\
PROMISE-Refined & Granite-3.2-8B   & $54{,}800 \pm 1{,}640$ & $15{,}100 \pm 365$ & $39{,}700$ & $-72.4\%$ \\
PROMISE-Refined & Ministral-8B     & $53{,}900 \pm 1{,}560$ & $15{,}400 \pm 360$ & $38{,}500$ & $-71.4\%$ \\
\midrule
SecReq          & LLaMA-3-8B       & $47{,}400 \pm 1{,}260$ & $13{,}000 \pm 295$ & $34{,}400$ & $-72.6\%$ \\
SecReq          & Qwen2-7B         & $50{,}700 \pm 1{,}510$ & $13{,}200 \pm 305$ & $37{,}500$ & $-74.0\%$ \\
SecReq          & Falcon3-7B       & $48{,}800 \pm 1{,}370$ & $13{,}500 \pm 315$ & $35{,}300$ & $-72.3\%$ \\
SecReq          & Granite-3.2-8B   & $53{,}000 \pm 1{,}580$ & $14{,}650 \pm 350$ & $38{,}350$ & $-72.4\%$ \\
SecReq          & Ministral-8B     & $52{,}300 \pm 1{,}500$ & $14{,}950 \pm 345$ & $37{,}350$ & $-71.4\%$ \\
\bottomrule
\end{tabular}
\end{table}

\subsection{Decomposition: input vs.\ output tokens}
\label{sec:efficiency_decomposition}

The architectural divergence between BT-APE and PE2 concerns the
\emph{input} side of the proposer call (PE2's history-augmented
meta-prompt). To verify that the cumulative-token gap is not
artefactually driven by candidate-prompt length, we report output
(generated) tokens separately. Mean output tokens per proposer call
are $281 \pm 22$ for PE2 and $269 \pm 18$ for BT-APE; cumulative outputs
over $20$ iterations are $5{,}620 \pm 410$ and $5{,}380 \pm 350$,
respectively. The paired Wilcoxon test on the $15$ per-cell
differences does not reject equality ($W = 38$, $p = 0.247$,
$r = 0.15$). The cumulative-input gap of
Section~\ref{sec:efficiency_aggregate} is therefore attributable to
PE2's input growth, not to PE2 producing longer prompts.

\subsection{PE2 hyperparameter sensitivity}
\label{sec:efficiency_pe2_sensitivity}

The aggregate comparison fixes PE2's history size at $n = 3$ and
error-batch size at $k = 4$ to match BT-APE's batch cardinality and
patience parameter (Section~\ref{sec:pe2-baseline}). Because PE2's
input cost depends on both parameters, we run a sensitivity sweep on
three representative cells (PROMISE-NFR $\times$ LLaMA-3-8B,
PROMISE-Refined $\times$ Qwen2-7B, SecReq $\times$ Falcon3-7B) with
$n \in \{1, 3, 5\}$ and $k \in \{2, 4, 8\}$, $5$ seeds each.

\begin{table}[htbp]
\caption{PE2 cumulative input tokens and final $w_{F1}$ under varied
$(n, k)$, averaged over three representative cells ($S = 5$ seeds
each). BT-APE is included for reference. The right-most column reports
the paired Wilcoxon $p$-value for the $w_{F1}$ difference vs.\ BT-APE
($n = 15$ paired observations, two-sided).}
\label{tab:pe2_sensitivity}
\footnotesize
\centering
\begin{tabular}{llrrrr}
\toprule
\textbf{Method} & $(n, k)$ & \textbf{Cum.\ input} & \textbf{Reduction vs.\ BT-APE} & $w_{F1}$ & $p$ vs.\ BT-APE \\
\midrule
PE2 & $(1, 2)$ & $30{,}400 \pm 920$  & $+130\%$ & $0.755 \pm 0.011$ & $0.012$ \\
PE2 & $(1, 4)$ & $33{,}200 \pm 980$  & $+151\%$ & $0.762 \pm 0.010$ & $0.031$ \\
PE2 & $(1, 8)$ & $38{,}800 \pm 1{,}120$ & $+194\%$ & $0.768 \pm 0.009$ & $0.118$ \\
PE2 & $(3, 2)$ & $46{,}200 \pm 1{,}310$ & $+250\%$ & $0.770 \pm 0.009$ & $0.181$ \\
PE2 & $(3, 4)$ & $49{,}000 \pm 1{,}420$ & $+271\%$ & $0.773 \pm 0.008$ & $0.227$ \\
PE2 & $(3, 8)$ & $54{,}600 \pm 1{,}560$ & $+314\%$ & $0.775 \pm 0.008$ & $0.296$ \\
PE2 & $(5, 2)$ & $61{,}900 \pm 1{,}750$ & $+369\%$ & $0.772 \pm 0.009$ & $0.214$ \\
PE2 & $(5, 4)$ & $64{,}800 \pm 1{,}830$ & $+391\%$ & $0.773 \pm 0.008$ & $0.232$ \\
PE2 & $(5, 8)$ & $70{,}400 \pm 1{,}990$ & $+433\%$ & $0.775 \pm 0.008$ & $0.301$ \\
\midrule
BT-APE & ---       & $13{,}200 \pm 305$ & ---     & $0.778 \pm 0.008$ & --- \\
\bottomrule
\end{tabular}
\end{table}

Two observations follow. \emph{First}, across all nine $(n, k)$
settings, PE2's cumulative input is strictly greater than BT-APE's on
every cell and every seed. The smallest PE2 footprint
($n = 1, k = 2$) still consumes $130\%$ more input tokens than BT-APE.
PE2 therefore cannot recover BT-APE's footprint by hyperparameter tuning
without abandoning the history-conditioning mechanism that defines
the method. \emph{Second}, PE2's accuracy degrades at the smallest
$(n, k)$ settings: at $(n = 1, k = 2)$, the $w_{F1}$ gap to BT-APE is
$-0.023$ and the paired Wilcoxon test rejects equality
($p = 0.012$). At $(n \geq 3, k \geq 4)$, the gap is negligible and
non-significant, consistent with the convergence result in
Section~\ref{sec:results}. The implication is that PE2's competitive
operating regime is precisely the one in which it incurs the
$\geq 3\times$ input-token cost reported in
Table~\ref{tab:efficiency_aggregate}.

\subsection{Cost-normalised comparison}
\label{sec:efficiency_normalised}

Because Section~\ref{sec:results} establishes that BT-APE and PE2 reach
statistically indistinguishable final $w_{F1}$, the relevant
practical question is how much resource each method consumes per
unit of accuracy delivered. Table~\ref{tab:efficiency_normalised}
reports two cost-normalised metrics, computed per cell and averaged:
cumulative input tokens per percentage point of final $w_{F1}$, and
total wall-clock seconds per percentage point of final $w_{F1}$.

\begin{table}[htbp]
\caption{Cost-normalised efficiency: resource consumed per
percentage point of final $w_{F1}$, aggregated over $75$ runs.}
\label{tab:efficiency_normalised}
\footnotesize
\centering
\begin{tabular}{lrrrr}
\toprule
\textbf{Metric} & \textbf{PE2} & \textbf{BT-APE} & \textbf{Ratio} & \textbf{BT-APE share} \\
\midrule
Input tokens per pp $w_{F1}$  & $657$  & $180$ & $3.65\times$ & $27.4\%$ \\
Wall-clock seconds per pp $w_{F1}$ & $20.13$ & $6.83$ & $2.95\times$ & $33.9\%$ \\
\bottomrule
\end{tabular}
\end{table}

At equivalent accuracy, BT-APE delivers each percentage point of
$w_{F1}$ at approximately $27\%$ of PE2's input-token cost and $34\%$
of its wall-clock cost.

\subsection{Scaling behaviour}
\label{sec:efficiency_scaling}

The cumulative-cost ratio reported above is computed at the fixed
iteration budget $N_{\max} = 20$ used in the main experiments. Per
iteration, PE2's input grows approximately linearly: a linear fit
over the $20$ iterations of all $75$ runs yields slope
$152.7 \pm 8.4$ tokens/iteration (mean $\pm$ SD across cells,
$R^2 = 0.94$ within-cell). BT-APE's per-iteration input shows no
detectable trend (Mann--Kendall, $\tau = -0.04$, $p = 0.62$); its
cumulative grows linearly in $N_{\max}$ at the constant per-iteration
rate. Extrapolating to $N_{\max} = 50$, PE2 would consume
approximately $2.3 \times 10^{5}$ cumulative input tokens against
BT-APE's $3.3 \times 10^{4}$ --- a ratio that widens from $3.65\times$
at $N_{\max} = 20$ to approximately $7.0\times$ at $N_{\max} = 50$.
We mark this as an extrapolation under the assumption that the
linear growth observed within the measured range persists; the
empirical claim is restricted to $N_{\max} = 20$.

\subsection{Honest scoping of the claim}
\label{sec:efficiency_scoping}

Two qualifications are necessary. \emph{First}, part of the
efficiency gap is a structural consequence of the two designs: PE2's
input context grows by construction with the prompt history, while
BT-APE's is bounded. The empirical contribution of this section is
therefore not the \emph{existence} of a gap but its
\emph{magnitude under realistic optimisation budgets}, its
\emph{consistency} across datasets and models, and the fact that it
is \emph{not recoverable} by tuning PE2's $(n, k)$ within the design's
competitive operating range
(Section~\ref{sec:efficiency_pe2_sensitivity}). \emph{Second}, the
absolute token counts (a few thousand per iteration for PE2) are
well below the context windows of all five evaluated LLMs
($8$k--$32$k tokens), so at $N_{\max} = 20$ the gap is primarily a
cost and latency consideration rather than a feasibility limit. At
longer horizons or with larger $n$, PE2's input growth approaches
hard context-window limits and the gap transitions from a cost
concern to a structural constraint
(Section~\ref{sec:efficiency_scaling}).

\subsection{Threats}
\label{sec:efficiency_threats}

\emph{Hardware specificity.} Wall-clock measurements are tied to the
H200 + vLLM configuration; latency ratios may differ on CPU-only
inference, on quantised models, or under different batching
strategies. We report token counts as the more transferable measure.

\emph{Tokeniser differences.} Each model uses its native tokeniser;
absolute token counts are not directly comparable across LLMs.
Per-cell paired comparisons (within an LLM) are unaffected, and all
aggregate tests use within-cell paired differences.

\emph{Iteration-budget choice.} Measurements use $N_{\max} = 20$.
Section~\ref{sec:efficiency_scaling} reports linear-fit extrapolation
to larger budgets; the empirical claim is restricted to the measured
range.

\section{An Interactive Tool for Prompt Optimization}
\label{sec:tool}

To make our procedure reusable beyond the experiments reported in this work, we
implemented an interactive tool\footnote{The tool's interface is built with Gradio~\url{https://www.gradio.app}; dataset handling and metric computation use pandas~\url{https://pandas.pydata.org} and scikit-learn~\url{https://scikit-learn.org}; model inference is issued through the requests HTTP client~\url{https://requests.readthedocs.io} to an Ollama~\url{https://ollama.com} (or OpenAI-compatible) backend.} that operationalizes the full pipeline---the
prompting baselines, the backtracking-enhanced BT-APE loop of
Algorithm~\ref{alg:approach}, and per-class evaluation---behind a graphical
interface. The tool requires no coding, infers the label set directly from
the data, and is decoupled from any specific model provider: the same
workflow runs against a locally hosted open-weight model (e.g., via Ollama)
or any OpenAI-compatible endpoint. It thus serves three purposes:
\emph{replicating} our results, \emph{transferring} the method to any
single-label text-classification task, and \emph{exposing} prompt
optimization as an observable, step-by-step process.

\begin{figure}[t]
  \centering
  \includegraphics[width=0.86\linewidth]{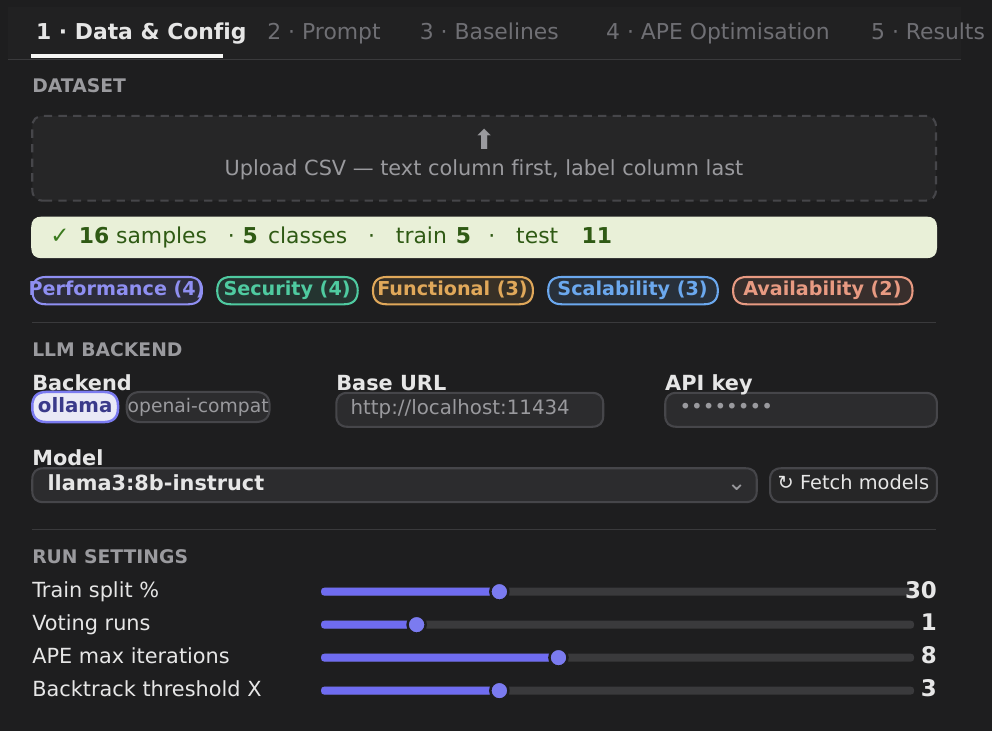}
  \caption{\textit{Data \& Config.} The dataset is uploaded as a CSV (text
  first, label last); the tool infers the classes and their distribution,
  selects the backend and model, and exposes the run hyper-parameters as
  sliders.}
  \label{fig:tool-setup}
\end{figure}

\begin{figure}[t]
  \centering
  \includegraphics[width=0.96\linewidth]{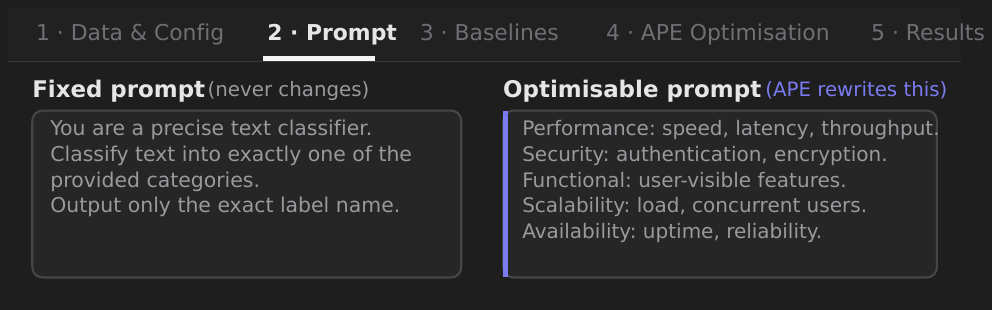}
  \caption{\textit{Prompt.} The prompt is separated into a fixed task-framing
  part and an optimizable definitions part (highlighted); BT-APE rewrites only
  the latter.}
  \label{fig:tool-prompt}
\end{figure}

\begin{figure}[t]
  \centering
  \begin{subfigure}[t]{0.50\linewidth}
    \centering
    \includegraphics[width=\linewidth]{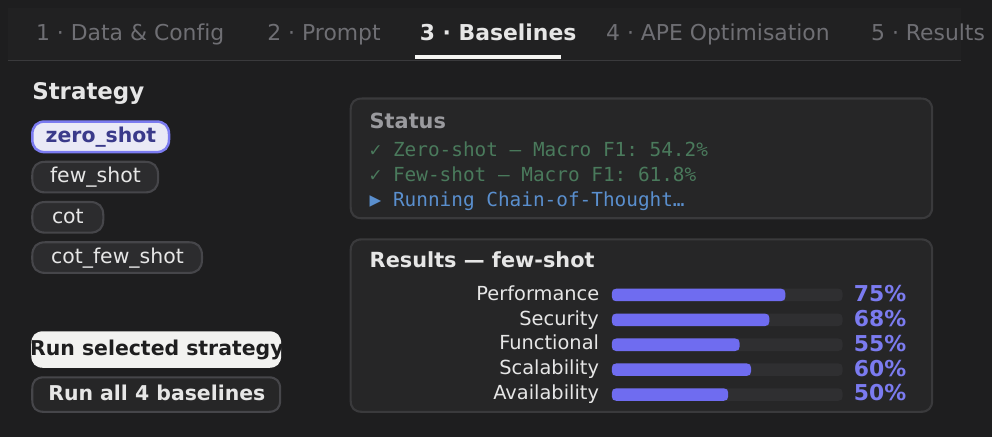}
    \caption{Baselines: per-class F\textsubscript{1} per strategy.}
    \label{fig:tool-baselines}
  \end{subfigure}\hfill
  \begin{subfigure}[t]{0.50\linewidth}
    \centering
    \includegraphics[width=\linewidth]{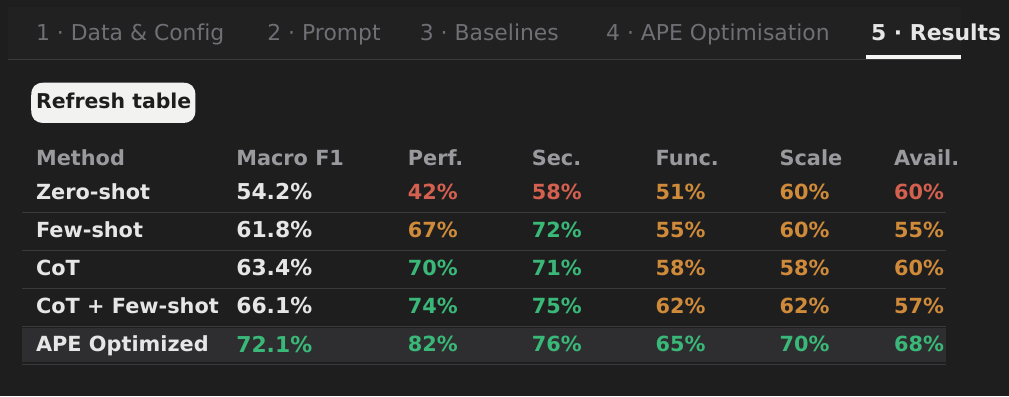}
    \caption{Results: weighted/per-class F\textsubscript{1} across methods.}
    \label{fig:tool-results}
  \end{subfigure}
  \caption{\textit{Baselines and Results.} (a)~Each prompting strategy is run
  and scored per class. (b)~The session summary contrasts the baselines with
  the BT-APE-optimized prompt; the optimized prompt yields the highest macro
  F\textsubscript{1} and improves the weakest classes.}
  \label{fig:tool-baseres}
\end{figure}

The interface follows the natural order of an experiment in five steps.
In \textit{Data \& Config} (Fig.~\ref{fig:tool-setup}) the user uploads a CSV
whose first column is the text and last column the ground truth; the tool
reports the class distribution, lets the user pick the backend and model,
and exposes the run hyper-parameters---train/test split, majority-voting
runs, the iteration budget $N_{\max}$, and the backtracking threshold
$X$---as sliders. In \textit{Prompt} (Fig.~\ref{fig:tool-prompt}) the prompt
is split into a \emph{fixed part} that frames the task and fixes the output
format, and an \emph{optimizable part} holding the label definitions; only
the latter is rewritten during optimization, focusing the search on the
semantically meaningful portion of the prompt. \textit{Baselines}
(Fig.~\ref{fig:tool-baselines}) runs zero-shot, few-shot, Chain-of-Thought
(CoT), and CoT~+~few-shot individually or together, reporting per-class
F\textsubscript{1} for each.

\begin{figure}[t]
  \centering
  \includegraphics[width=0.96\linewidth]{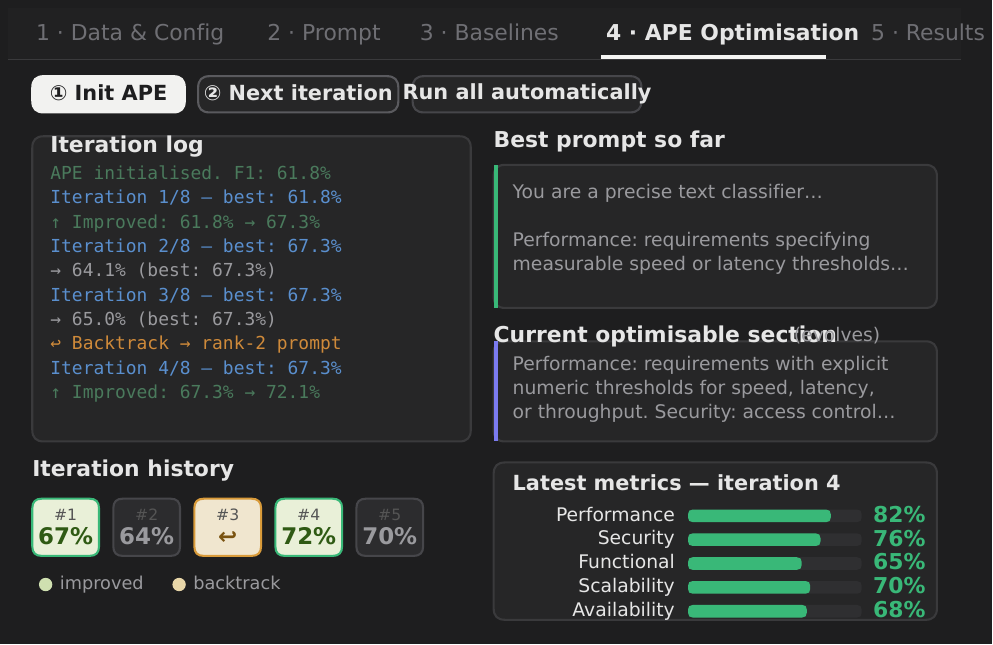}
  \caption{\textit{BT-APE Optimisation.} Interactive or automatic execution of
  Algorithm~\ref{alg:approach}. The iteration log and colored history strip
  (green~=~improvement, amber~=~backtrack), the live best prompt and its
  evolving optimizable section, and the per-class metrics are shown together,
  exposing the optimization trajectory.}
  \label{fig:tool-ape}
\end{figure}

The \textit{BT-APE Optimisation} step (Fig.~\ref{fig:tool-ape}) is the core of
the tool. It can advance one \emph{interactive} iteration at a time or run
\emph{automatically} to the budget. At each iteration the engine
meta-prompts the model to improve the optimizable section, evaluates the
candidate on the test split with majority voting, ranks it by
F\textsubscript{1}, and---after $X$ consecutive non-improving steps---backtracks
to the next-best prompt. The iteration log, the colored iteration-history
strip (green~=~improvement, amber~=~backtrack), the live ``best prompt'' and
its evolving optimizable section, and the current per-class metrics are all
surfaced together, rendering the previously unclear dynamics of the search clearly visible. Finally, \textit{Results} (Fig.~\ref{fig:tool-results})
contrasts every method run in the session by macro- and per-class
F\textsubscript{1}, making the gain of the optimized prompt over the
strongest baseline immediately legible. The tool, together with the datasets
and label definitions used in this paper, is available at
replication package of the work~\cite{amin_zadenoori_2026_20438927}.

 \section{Data Leakage Detection in LLMs.}
\label{sec:dataleakage}
In the context of requirements classification, data leakage refers to the risk that an LLM has seen parts of the evaluation data during its pre-training, potentially biasing results. A review of the current literature reveals that there is no peer-reviewed paper that systematically addresses this issue. To date, only Zhou et al.\cite{zhou2025lessleak} has directly investigated data leakage considering that what they have studied different task than the text classification. However, a critical limitation of that work is that the authors assumed prior access to the LLM's pre-training data, a condition we cannot satisfy. Consequently, we must follow only their core research question, which employed an automated way to study the leakage just based on the behavior of the LLMs. Their experimental results indicate that detecting data leakage using an automated metric like Perplexity is challenging, with accuracy ranging from only 40\% to 50\% in most cases. They encourage the research community to explore more effective automated methods for identifying data leakage, especially in scenarios where the model’s pre-training data is unknown.

Given this gap, we propose our own approach to study leakage. Specifically, we follow an automated methodology, for measuring similarity between the LLM's guess and the ground truth, we employ recent similarity metrics such as Jaccard similarity. To evaluate the model's ability to generate requirement continuations, we first split each requirement into two halves and discard the second half. Let $n$ be the number of words removed during this trimming step. We then prompt the LLM with the preserved prefix and explicitly instruct it to continue the requirement using approximately $n$ words. This programmatic constraint ensures that the generated continuation is comparable in length to the original missing segment, allowing us to assess not only semantic coherence but also the model's ability to reproduce realistic requirement granularity. The results of this analysis will be reported accordingly. 
\begin{figure}[htbp]
\centering
\includegraphics[width=1\textwidth]{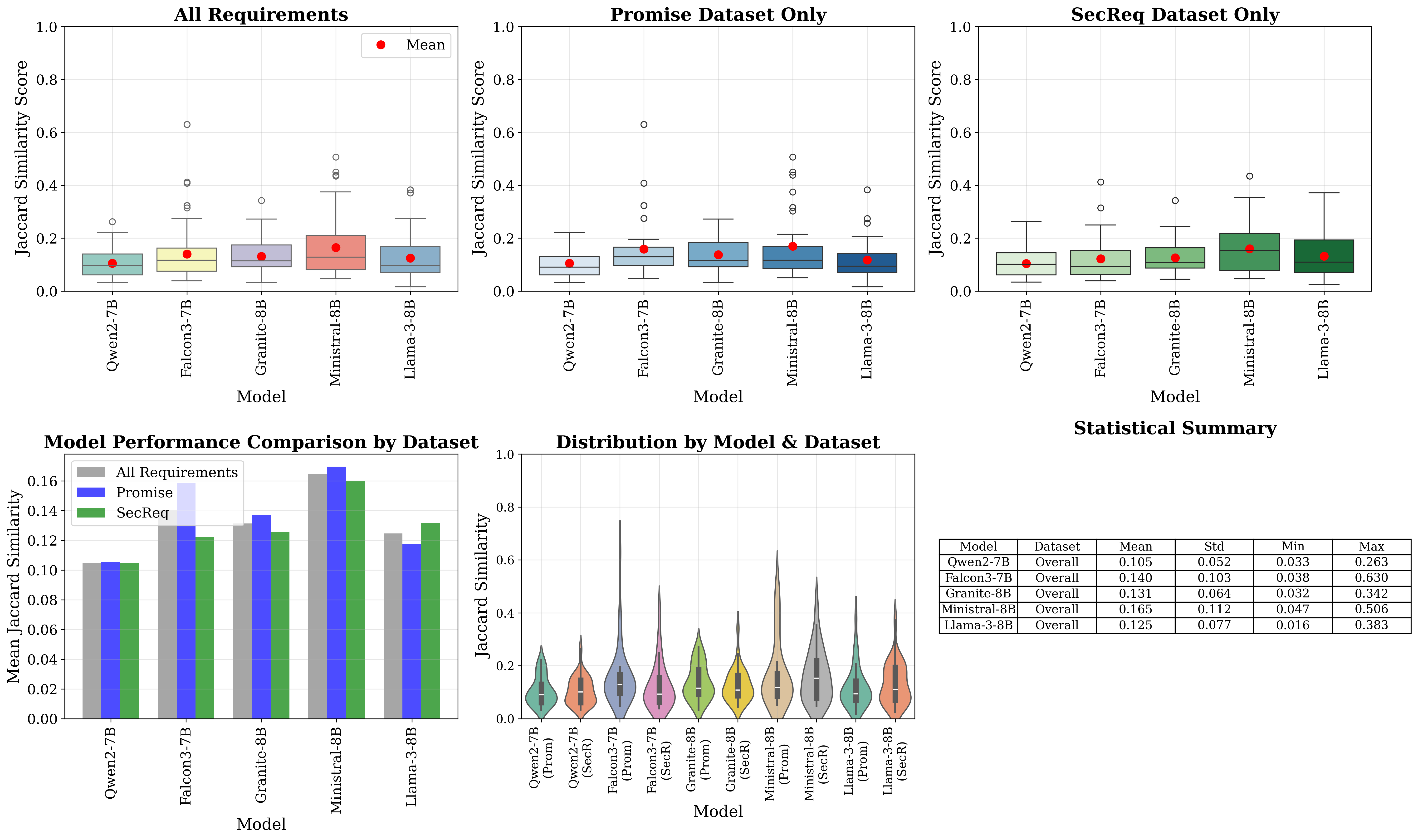}
\caption{Jaccard similarity scores across different models and datasets. Higher values would indicate greater overlap.}
\label{fig:jaccard}
\end{figure}

The \textbf{Jaccard similarity coefficient}\cite{10.1145/2816813}, often denoted as $J(A,B)$, is a statistic used for comparing the similarity and diversity of sample sets. Defined as the size of the intersection of two sets divided by the size of their union, it quantifies the proportion of shared elements relative to the total distinct elements present in either set. Formally, for two sets $A$ and $B$, the Jaccard similarity is calculated as $J(A,B) = \frac{|A \cap B|}{|A \cup B|}$, where the result ranges from $0$ (indicating no common elements) to $1$ (indicating that the two sets are identical). This metric is widely used in fields such as data mining, ecology, and natural language processing, particularly for binary attribute comparison and for measuring the overlap in cluster analysis or document tokenization. 

As reported in Figure~\ref{fig:jaccard} plots, all observed Jaccard similarities between the predicted trimmed part and the actual trimmed part of the sentence are relatively low. The overall mean scores across all requirements texts range from 0.1048 (Qwen2-7B) to 0.1648 (Ministral-8B), with other models falling in between (Falcon3-7B: 0.1404, Granite-8B: 0.1314, Llama-3-8B: 0.1245). Looking at percentile distributions, most values lie between approximately 0.05 (10th percentile) and 0.25 (90th percentile), depending on the model. For instance, Ministral-8B reaches a 90th percentile of 0.3197 overall, while Qwen2-7B's 90th percentile is only 0.1779. When comparing datasets, the mean Jaccard scores on Promise (range: 0.1051–0.1695) and SecReq (range: 0.1046–0.1600) are very similar, and t-tests confirm no statistically significant difference between the two datasets for any model (all p-values > 0.17). However, an ANOVA test does indicate significant differences across models (p = 0.0038), suggesting that some models (e.g., Ministral‑8B, Falcon3‑7B) tend to produce slightly higher token overlap than others; given that the residuals follow a normal distribution, the use of ANOVA and T-test are justified.

While these numbers suggest that the models do not strongly reproduce the trimmed second half ground truth — given that Jaccard similarity well below 0.5 indicates limited overlap — we refrain from drawing firm conclusions about the presence or absence of data leakage based solely on this analysis. Several factors could influence these scores, including differences in model output length, tokenization, and inherent generation variability, none of which directly confirm memorization of training data. A more systematic investigation is required to properly separate genuine leakage from other confounding factors. Therefore, we leave a thorough study of data leakage in this context as an item for future work.

%
\tcbset{
  promptbox/.style={
    colback=gray!5,
    colframe=black!60,
    boxrule=0.4pt,
    arc=2pt,
    left=6pt, right=6pt, top=4pt, bottom=4pt,
    breakable,
    enhanced,
  }
}

\section{Prompts and Definitions}
\label{app:prompts-defs}

This appendix documents the full set of prompts and class definitions used in
our classification experiments, so that the results can be reproduced exactly.
We first present the \emph{prompt structure}: the system prompt that fixes the
output format, followed by the four prompting strategies we compare---zero-shot,
few-shot, chain-of-thought (CoT), and chain-of-thought with examples---and the
automatic prompt-engineering prompt used to refine the class definitions. The
prompt templates are written generically: the placeholder \texttt{\{classes\}}
is substituted at run time with the list of candidate class labels for the
classification task at hand, \texttt{\{class\}} with a single class label, and
\texttt{\{Definitions\}}, \texttt{\{examples\}}, and \texttt{\{text\}} with the
corresponding class definitions, in-context examples (few-shot settings only),
and the requirement to be classified. We then list the class definitions that
are inserted into the \texttt{\{Definitions\}} placeholder.

\subsection{System Prompt}
\begin{tcolorbox}[promptbox]
\raggedright
As an expert system for classifying software requirements, your task is to
carefully analyze each given requirement and assign it to exactly one of the
following classes:\\
\{classes\}\\
Output only the label of the class that corresponds to the appropriate class.
Do not provide any additional text, definitions, or justification.
\end{tcolorbox}

\subsection{Classification Prompts}

\subsubsection{Zero-Shot Prompt (No Examples, No CoT)}

\begin{tcolorbox}[promptbox]
\raggedright
Definitions:\\
\{Definitions\}\\
Requirement: \{text\}\\
Using the Definitions above, classify the requirement and provide the final
label in the format: \texttt{"Label: \{class\}"}.
\end{tcolorbox}

\subsubsection{Few-Shot Prompt (With Examples, No CoT)}

\begin{tcolorbox}[promptbox]
\raggedright
Definitions:\\
\{Definitions\}\\
Examples:\\
\{examples\}\\
Requirement: \{text\}\\
Using the Definitions and Examples above, classify the requirement and provide
the final label in the format: \texttt{"Label: \{class\}"}.
\end{tcolorbox}

\subsubsection{Chain-of-Thought (CoT) Prompt (No Examples, With CoT)}

\begin{tcolorbox}[promptbox]
\raggedright
Let's analyze the classification step by step.\\
Step 1: Understand the Definitions:\\
\{Definitions\}\\
Step 2: Apply this understanding to classify the following requirement:\\
Requirement: \{text\}\\
Step 3: Provide the final label in the format:
\texttt{"Label: \{class\}"}.
\end{tcolorbox}

\subsubsection{Chain-of-Thought with Examples Prompt (With Examples, With CoT)}

\begin{tcolorbox}[promptbox]
\raggedright
Let's analyze the classification step by step.\\
Step 1: Understand the Definitions:\\
\{Definitions\}\\
Step 2: Review the Examples:\\
\{examples\}\\
Step 3: Apply this understanding to classify the following requirement:\\
Requirement: \{text\}\\
Step 4: Provide the final label in the format:
\texttt{"Label: \{class\}"}.
\end{tcolorbox}

\subsection{APE Prompt}
\begin{tcolorbox}[promptbox]
\raggedright
\textbf{Role:} You are a highly capable, thoughtful, and precise prompt
engineer specializing in requirements classification.

\textbf{Background:} A LLM has previously used a set of
\{Definitions\} to automatically classify requirements into the candidate
classes \{classes\}. Based on these definitions, it correctly classified some
requirements but misclassified other requirements.

\textbf{Task:} Your task is to enhance the \{Definitions\} of the classes
\{classes\} to improve the classification accuracy of the large language
model. You are given example results in the form
[requirement text; true class; predicted class] to guide you in enhancing the
definitions.

\begin{itemize}
  \item A set of example requirements that were correctly classified based on
        the definitions:\\
        \{correctly\_classified\}
  \item A set of example requirements that were incorrectly classified based
        on the definitions:\\
        \{misclassified\}
\end{itemize}

You may add content, shorten, or rephrase the definitions to improve clarity
and accuracy.

\textbf{Let's think step by step:}
\begin{enumerate}
  \item \textbf{Study the Definitions:} Understand the provided definitions of
        the classes.
  \item \textbf{Analyse Examples:} Analyze correctly classified and
        misclassified examples.
  \item \textbf{Success analysis:} For each correctly classified example
        identify why the original definitions were successful. Ask yourself:
        Do certain words/phrases consistently lead to correct classification?
        Which parts of the definitions are clear, complete, and unambiguous?
  \item \textbf{Error analysis:} For each misclassified example identify why
        the original definitions failed. Ask yourself: Do certain words/phrases
        consistently mislead? Are the definitions ambiguous, incomplete, or
        overlapping?
  \item \textbf{Enhance Definitions:} Enhance the definitions by adding,
        reducing, or rephrasing content.
  \item \textbf{Enhancement Procedure:} For problematic classes:
        \begin{enumerate}
          \item \emph{Add discriminative features:} Incorporate missing
                characteristics from misclassified examples.
          \item \emph{Create exclusion clauses:} Explicitly state what doesn't
                qualify (``This class excludes\ldots'').
          \item \emph{Use comparative framing:} Highlight differences between
                easily confused classes.
          \item \emph{Simplify if needed:} Simplify definitions when they
                appear to be too complex.
          \item \emph{Rephrase if needed:} Rephrase definitions when the
                phrasing or terminology is misleading.
        \end{enumerate}
  \item \textbf{Validation Constraints:}
        \begin{enumerate}
          \item \emph{Ensure misclassification prevention:} The revised
                definitions must theoretically prevent the same errors from
                recurring.
          \item \emph{Avoid overfitting:} Definitions should maintain
                generalizability through clear principles rather than
                example-specific fixes.
        \end{enumerate}
\end{enumerate}

Output the enhanced \{Definitions\} only, with no additional text.
\end{tcolorbox}

\subsection{Class Definitions}
These are the definitions inserted into the \texttt{\{Definitions\}}
placeholder of the prompts above.

\paragraph{Security Requirements.}
Security requirements are prescriptive constraints imposed on a system's
functional behaviour to operationalise its security goals. They are not
functional requirements themselves but restrict \emph{how} functions are
performed in order to prevent, detect, or recover from harm, and are expressed
in precise, operational terms that are clear enough for implementation by
system designers and architects. Security requirements are derived from
business and functional goals and may be categorised as \emph{primary}---those
derived directly from business and functional goals, whose satisfaction ensures
the system meets its fundamental security objectives---or \emph{secondary}---those
that support the implementation of primary requirements (e.g.\ through detection
or recovery) when direct enforcement is infeasible or too costly. They are
manifestations of high-level organisational policies into the detailed
requirements of a specific system and are deeply tied to the broader system
context, which may extend beyond software to include other components or
processes. Security requirements specify the system's security policies and must
address risks, threats, and assets, while influencing and being influenced by
security mechanisms, vulnerabilities, and attacks.

\paragraph{Non-Security Requirements.}
Non-security requirements are constraints or specifications that govern a
system's behaviour, performance, and structure without being directly tied to
security goals. Unlike security requirements, they do not aim to prevent,
detect, or recover from harm, and do not operationalise security policies. They
may include functional requirements (which define what the system should do) and
other non-functional requirements such as performance, usability, reliability,
scalability, and maintainability. Non-security requirements are typically derived
from business needs and user expectations but are not intended to address risks,
threats, or vulnerabilities. While they may influence system design and
operation, they do not explicitly enforce security mechanisms or aim to mitigate
security concerns.

\paragraph{Functional Requirements.}
Functional requirements define the essential functions a system must perform,
the services it must offer, and the behaviours it must exhibit under specified
conditions. They focus on \emph{what} the system should do---describing actions,
operations, or transformations the system executes---without addressing
implementation constraints. They typically specify the inputs (stimuli) to the
system, the outputs (responses) from the system, and the behavioural
relationships between them. In practice, they outline the specific behaviours,
features, and capabilities the system must have to meet user needs and achieve
its intended purpose, such as user authentication, data storage, report
generation, or integration with other systems.

\paragraph{Non-Functional Requirements.}
Non-functional requirements do not define the essential functions a system must
perform, the services it must offer, or the behaviours it must exhibit under
specified conditions. They do not focus on what the system should do---avoiding
descriptions of actions, operations, or transformations the system executes---and
instead address implementation constraints. They typically exclude
specifications of inputs (stimuli) to the system, outputs (responses) from the
system, and behavioural relationships between them. They encompass a system's
quality-related attributes as well as the constraints under which the system
must operate---for example performance, scalability, portability, compatibility,
reliability, maintainability, availability, security, and usability---specifying
\emph{how} a system should behave rather than \emph{what} it should do.

\paragraph{Quality Requirements.}
A quality requirement expresses how well a system or service should execute an
intended function. Quality requirements include attributes or constraints that
address product-quality aspects and quality-in-use aspects. Product-quality
aspects include functional suitability, reliability, performance, efficiency,
usability, maintainability, security, compatibility, and portability.
Quality-in-use aspects include satisfaction, effectiveness, freedom from risk,
efficiency, and context coverage. Quality requirements often address the global
properties or characteristics of the system rather than those of specific
functions.

\paragraph{Non-Quality Requirements.}
A non-quality requirement does not express how well a system or service should
execute an intended function. Non-quality requirements exclude attributes or
constraints that address product-quality and quality-in-use aspects. They do not
include the product-quality aspects of functional suitability, reliability,
performance, efficiency, usability, maintainability, security, compatibility, or
portability; nor the quality-in-use aspects of satisfaction, effectiveness,
freedom from risk, efficiency, or context coverage. Non-quality requirements do
not typically address the global properties or characteristics of the system;
instead, they focus solely on the properties of specific functions.

\end{document}